\documentclass[11pt]{article}
\usepackage[english]{babel}       
\usepackage{graphicx}                  
\usepackage{amssymb,amsmath}
\usepackage{physics}
\usepackage{simplewick}
\usepackage{commath}
\usepackage{color}
\definecolor{hyperref}{RGB}{026,028,185}
\usepackage[bookmarks=true,colorlinks=true,linkcolor=hyperref,citecolor=hyperref,urlcolor=hyperref,bookmarksnumbered]{hyperref}
\usepackage{enumerate}
\usepackage{bbm}
\usepackage{epsfig}
\usepackage{wrapfig}
\usepackage{yfonts}
\usepackage{geometry}      
\usepackage{verbatim}
\usepackage{titlesec}
\usepackage{subcaption}

\titleformat{\subsubsection}[runin]
        {\normalfont\bfseries}
        {\thesubsection}
        {0.5em}
        {}
        [.]

\def\half{{1\over 2}}

\def\D{\mathcal{D}}
\def\M{\mathcal{M}}
\def\H{\mathcal{H}}
\def\A{\mathcal{A}}
\def\B{\mathcal{B}}
\def\M{\mathcal{M}}

\def\tr{\mathrm{Tr}}

\newcommand{\cl}{\text{cl}}
\newcommand{\cut}{\text{cut}}
\newcommand{\barphi}{\bar{\phi}}
\newcommand{\baromega}{\bar{\omega}}

\usepackage{color}

\newcommand{\diff}[1]{\dif^{\,#1}\!}

\usepackage[sort,compress]{cite}
\usepackage[bulletsep]{collref}

\begin{document}
\renewcommand{\thefootnote}{\arabic{footnote}}
 
\overfullrule=0pt
\parskip=2pt
\parindent=12pt
\headheight=0in \headsep=0in \topmargin=0in \oddsidemargin=0in

\vspace{ -3cm} \thispagestyle{empty} \vspace{-1cm}
\begin{flushright} 
\footnotesize
\end{flushright}%

\begin{center}
\vspace{1.2cm}
{\Large\bf \mathversion{bold}
Entanglement Entropy of Excited States in the Quantum Lifshitz Model}

\vspace{0.8cm} {J.~Angel-Ramelli \footnote{{\tt jfa1@hi.is}}}

\vskip  0.5cm

\small
{\em
University of Iceland,
Science Institute,
Dunhaga 3,  107 Reykjav\'ik, Iceland
}
\normalsize

 \end{center}

\vspace{0.3cm}
\begin{abstract}
In this work we calculate the entanglement entropy of certain excited states of the quantum Lifshitz model. The quantum Lifshitz model is a $2+1$-dimensional bosonic quantum field theory with an anisotropic scaling symmetry between space and time that belongs to the universality class of the quantum dimer model and its generalizations. The states we consider are constructed by exciting the eigenmodes of the Laplace-Beltrami operator on the spatial manifold of the model. We perform a replica calculation and find that, whenever a simple assumption is satisfied, the bipartite entanglement entropy of any such excited state can be evaluated analytically. We show that the assumption is satisfied for all excited states on the rectangle and for almost all excited states on the sphere and provide explicit examples in both geometries. We find that the excited state entanglement entropy obeys an area law and is related to the entanglement entropy of the ground state by two universal constants. We observe a logarithmic dependence on the excitation number when all excitations are put onto the same eigenmode. 

\end{abstract}
\newpage

\tableofcontents
\pagenumbering{arabic}

\setcounter{footnote}{1}
\newpage


\section{Introduction}
In recent years the study of entanglement has brought deep insights across numerous areas of physics such as quantum information, condensed matter physics, holography, quantum gravity, and quantum field theory (QFT). Amongst the many measures of entanglement the entanglement entropy is one of the oldest and most well-studied. For bipartite systems it is defined as follows. Let there be a quantum system on the manifold $M$ with Hilbert space $\H$, and let us prepare a pure state of this system, for example the ground state, described by the density matrix $\rho$. Next, cut $M$ into the two subsystems $A$ and $B$ and assume that the Hilbert space splits accordingly, that is $\H=\H_A\otimes \H_B$. The reduced density matrix on $A$ is obtained by tracing out the degrees of freedom on $B$, that is $\rho_A=\tr_B\rho$, and the entanglement entropy of the subsystem $A$ in the state $\rho$ is given by the von Neumann entropy of $\rho_A$
\begin{equation}
S[A]=-\tr(\rho_A\log\rho_A).
\end{equation}
The entanglement entropy is particularly well-suited for the study of entanglement in bipartite pure systems, where it effectively quantifies the amount of entanglement between the subsystems \cite{Bennett1996,Plenio2007}.

In the context of condensed matter physics and quantum field theory the entanglement entropy has been shown to capture universal properties of critical systems, see for example \cite{Calabrese2006,Amico2008,Eisert2010,Laflorencie2016}. The perhaps most famous example is that of $1+1$-dimensional conformal field theory (CFT), where the entanglement entropy has a leading logarithmic divergence 
\begin{equation}
\label{eq:intro-2d-cft-ee}
S[A]=\frac{c}{3}\log\left(\frac{L_A}{\varepsilon}\right),
\end{equation}
with $L_A$ the size of the subsystem $A$, $c$ the central charge of the CFT \cite{Callan1994,Holzhey1994,Calabrese2004}, and $\varepsilon$ a UV-cutoff. Here, the coefficient of the logarithmic divergence is universal, meaning it is independent of the regularization scheme used and constant within the universality class of the theory. One can thus adopt the point of view that the entanglement entropy is a machinery that extracts universal features from critical theories. However, this machinery is arduous to operate, as entanglement entropy calculations are generally hard, and thus analytic results are scarce. 
It is not surprising that most results for the entanglement entropy have arisen in the context of CFT (and in particular in $2$-dimensional CFT), where powerful methods are available, and for simple states such as the ground state, see \cite{Calabrese2006,Nishioka2009,Rangamani2017}. For ground state one expects the entanglement entropy to obey an area law \cite{Srednicki1993,Hastings2007}. Concretely, for a $d$-dimensional CFT in its ground state one expects the entanglement entropy to have a leading UV-divergence 
\begin{equation}
S[A]= c_{d-2}\frac{\text{Area}(\partial A)}{\varepsilon^{d-2}} +\cdots+c_1 \log\frac{L_A}{\varepsilon}+c_0+\mathcal{O}(\varepsilon),
\end{equation}
where $\partial A$ is the $d-2$-dimensional boundary of $A$ and $c_{d-2}$ is a non-universal coefficient, and $L_A$ a characteristic length associated to $\partial A$. In even dimensions universal information is found in the coefficient of the logarithmic divergence term, whereas in odd dimensions, where the logarithmic divergence is absent, it is found in the constant sub-leading correction. In the ground-state, the entanglement between $A$ and its complement is thus dominated by short-range correlations across the boundary $\partial A$, as opposed to \eqref{eq:intro-2d-cft-ee} where the logarithm indicates that long-range correlations dominate. 

While not as widely studied, the entanglement entropy of excited states has also received some attention over the years. A particularly interesting result is that the entanglement entropy of a pure state picked at random from the Hilbert space will generically obey a volume and not an area law \cite{Page1993,Foong1994,Sen1996,Eisert2010}. The area law behavior of the ground state is consequently quite special within the Hilbert space. Notably, the dominance of extensive states implies that highly excited states are expected to obey volume laws. Another early study of excited states was performed in \cite{Ahmadi2006,Das2006}, where it was found that for free scalars on a lattice  the first excited states of the theory as well as a certain form of coherent state both still obeyed area laws. It was, in fact, argued by \cite{Masanes2009} that low lying excitations in a wide class of theories including many gapped Hamiltonian obey area laws with at most logarithmic corrections.
It was further argued by \cite{Alcaraz2011,Berganza2012} that excitations of CFTs obtained by acting with a primary field on the vacuum continued to obey an area law, albeit with a universal correction with respect to the ground state entanglement determined by the conformal weights of the primary field. As this universal correction was derived from correlation functions of primary fields, one may say that the entanglement entropy of (at least certain) low-lying excitations of a CFT encodes information on the correlation functions of primary fields.
A similar study was performed for finite quasiparticle excitations in a certain limit of a wide class of free theories, where it was again found that these excited states obey area laws with corrections to the ground state entanglement related to the correlation functions of the quasiparticle states \cite{CastroAlvaredo2018,CastroAlvaredo2018a}. This analysis was extended by the same authors in \cite{CastroAlvaredo2019,CastroAlvaredo2019a} to include the treatment of another measure of entanglement, the logarithmic negativity, and they further provided a fairly general proof in the case of a massive free boson showing that the previous results are valid in any dimension and for any entanglement surgery. Thus for a wide range of theories and low excited states, the excited state entanglement entropy has been observed to take the form 
\begin{equation}
\label{eq:intro-SES-expectation}
S_{ES}=S_{GS}+\beta_{ES},
\end{equation}
with $S_{ES}[A]$ and $S_{GS}[A]$ the entanglement entropies of the excited and ground states respectively, and $\beta_{ES}$ a constant depending on the excited state and derived from certain correlation functions in the underlying theory.
Apart from QFTs, some powerful calculations of the excited state entanglement entropy properties of spin chains have been performed, see for example \cite{Alcaraz2008,Alba2009,Berganza2012,Moelter2014}. In particular, it has been observed that generally two types of excited states can be found in the Hilbert space. Excited states showing the same logarithmic behavior typical of the ground states of critical models, see \eqref{eq:intro-2d-cft-ee}, and excited states that exhibit extensive behavior.

In this paper we will perform analytic entanglement calculations for excited states of the quantum Lifshitz model (QLM). The QLM is a $2+1$-dimensional QFT with a scaling symmetry that is anisotropic between time and space and characterized by a so-called dynamical critical exponent $z$. It was introduced in \cite{Ardonne2004} at $z=2$, where it describes the continuum limit of the quantum dimer model \cite{Rokhsar1988} and some of its interacting generalizations \cite{Henley1997,Moessner2001,Castelnovo2005, Freedman2005, Fendley2008}. More importantly for our purposes, the QLM has proven to offer a very fruitful playground for entanglement calculations, partly due to its close connections to CFT. In fact, the ground state  of the QLM can be expressed in terms of the action of a free $2$-dimensional CFT and is thus invariant under spatial conformal transformations \cite{Ardonne2004}. Similarly, its equal-time correlation functions can be expressed in terms of correlation functions in that same CFT \cite{Keraenen2017}. The  ground state entanglement entropy of the QLM and its higher dimensional generalizations \cite{Keraenen2017,Angel-Ramelli2019} have been extensively studied \cite{Fradkin2006,Hsu2009,Hsu2010,Stephan2009,Oshikawa2010,Zaletel2011,Zhou2016}. For the standard $2+1$-dimensional QLM at $z=2$, it has been found to be \cite{Fradkin2006,Hsu2009}
\begin{equation}
S[A]=c_1\frac{L_A}{\varepsilon}+\frac{c}{6}\Delta\chi\log{\frac{L_A}{\varepsilon}}+\gamma_{QCP}+\cdots
\end{equation}
with $L_A$ the length of the boundary $\partial A$ and $c_1$ a non-universal constant. The coefficient of the logarithm is universal and consists of the central charge $c$ of the CFT whose action describes the ground state of the QLM and of the change in Euler characteristic  $\Delta\chi$ due to the surgery. Furthermore, when $\Delta\chi=0$ the sub-leading constant term $\gamma_{\text{QCP}}$, where QCP stands for quantum critical point, is also universal \cite{Hsu2009} and depends on the geometry and topology of the manifolds involved \cite{Hsu2010,Oshikawa2010,Zaletel2011,Zhou2016,Angel-Ramelli2019}. The entanglement of certain mixed states of the QLM has also been studied analytically \cite{AngelRamelli2020}. 

There have been, to our best knowledge, only two works that deal with the excited state entanglement of the QLM. In \cite{Zhou2016a} the entanglement entropy of a local excitation constructed by acting with a vertex and a time evolution operator on the ground state of the QLM is analyzed. The excess entanglement entropy after some time $t$ is found to be related to correlation functions of certain operators on the full and sub-systems with Dirichlet boundary conditions. Finally, in \cite{Parker2017} the authors put the QLM on a compact spatial manifold, and construct states by exciting the eigenmodes of the corresponding Laplacian. Restricting to the case of a single excitation, they show that the trace of the $n$-th power of the reduced density matrix of such a state (whose limit $n\rightarrow 1$ by the replica approach gives the entanglement entropy) can be expressed in terms of correlation functions on a certain $n$-sheeted geometry resulting from a replica calculation. By deriving a form of Wick's theorem on this geometry, the authors then proceed to evaluate these correlation functions and provide expressions for them in terms of quantities related to the Green's functions on the full and sub-systems which they dub entanglement propagator amplitudes (EPA's). A numerical analysis then provides evidence of the universality of these quantities. However, an analytic continuation in $n$ remains elusive in the paper, and thus the authors can only provide expressions for the R\'enyi entropies and a conjecture for the entanglement entropy in the particular case when the two submanifolds resulting from the bipartition of $M$ are equal. 

To summarize, we find that the entanglement entropy of any finitely excited state $S_{ES}[A]$ of the QLM is related to its ground state entanglement entropy $S_{GS}[A]$ by 
\begin{equation}
S_{ES}[A]= \alpha S_{GS}[A] +\beta,
\end{equation}
where evidence points to the universality of the constants $\alpha$ and $\beta$. In the singly excited state and for $m$ excitations in a single mode of the halved $M$ geometry, we further find $\alpha=1$, in agreement with the expectation \eqref{eq:intro-SES-expectation}. In all other cases we provide analytic expressions for both constants. This implies that the entanglement of this class of finitely excited states of the QLM obeys an area law.  When all excitations are put into the same mode, we observe a logarithmic dependence on the excitation number $m$. This simple dependence allows us to consider the highly excited limit, which, contrary to standard expectation, continues to obey an area law. This sub-extensive behavior of the entanglement entropy is a characteristic feature of quantum scars, and suggests this might be such a state. Recently, there has been a lot of interests in these type of states \cite{Bernien2017,Turner2018}, as they exhibit a surprising form of ergodicity breaking and bear some importance to the eigenstate thermalization hypothesis. In particular such states have been observed for quantum dimer models \cite{Lan2017,Wildeboer2020}, which belong to the same universality class as the QLM, and in related one dimensional systems \cite{Iadecola2020}. In \cite{Moudgalya2018,Moudgalya2018a} scar states that exhibit a very similar type of sub-extensive entanglement entropy the our highly-excited state were also observed. In particular, the entanglement entropy of these tower states is determined by a finite correction to the ground state entanglement of the system that, for high excitations, is logarithmic in the excitation number. 

The present work is closely related to \cite{Parker2017}. In section \ref{sec:qlm} we review the basic properties of the QLM as well as the construction of its ground and excited states \`a la \cite{Parker2017} (that is excitations that correspond to the eigenmodes of the Laplacian on the spatial manifold). In section \ref{sec:single-excitation} we concentrate on the special case considered in \cite{Parker2017}, where a single quantum of energy is put into a particular mode. We show that introducing a certain condition for the classical fields the complexity of the problem can be sufficiently reduced such that the previously tricky analytic continuation becomes straight forward, and an explicit expression for the entanglement entropy can be written down. We then show that our assumption is satisfied by all modes on the sphere and rectangle and analyze the entanglement entropy corresponding to the excitation of different modes on both geometries. In section \ref{sec:general-excitation}, we generalize the previous calculation to encompass any excited state by making a slightly stronger assumption than before, and manage to find a closed expression for the entanglement entropy in terms of some complicated tensors of correlation functions. We then show that our assumption is again satisfied for all modes on the rectangle and for almost all modes on the sphere and provide explicit examples when a single mode is excited a finite amount of times. In this case we find that for a wide range of modes the entanglement entropy behaves like the logarithm of the excitation number. In both the singly and general excited state we confirm that the entanglement entropy can be expressed in terms of EPA's, although we define them slightly differently and refer to them as transformed propagators. We calculate these quantities by a spectral approach and find them in agreement with \cite{Parker2017}, providing further evidence for their universality.


\section{The quantum Lifshitz model and its excited states}
\label{sec:qlm}

The quantum Lifshitz model (QLM) is the $(2+1)$ dimensional quantum field theory defined on the spatial manifold $M$ by the Hamiltonian \cite{Ardonne2004}
\begin{equation}
\label{hamiltonian-general-z}
H=\frac{1}{2}\int_{M}\diff{2}x \left(\pi^2+g^2 (\Delta \phi)^2\right),
\end{equation}
where $\phi\sim\phi+2\pi R$ a compactified scalar field with compactification radius $R$, $\pi=-i\frac{\delta}{\delta\phi}$ its conjugate momentum, $\Delta$ the Laplace-Beltrami operator on $M$, and $g$ a free parameter. For the particular value $g=\frac{1}{8\pi}$ the QLM describes the correlations of the Rokhsar-Kievelson quantum dimer model on a square lattice. For simplicity, we will mostly refer to the Laplace-Beltrami operator as the Laplacian on $M$. Let us analyze the ground and excited states of this theory.

\paragraph{Ground state.} In order to find the ground state of the theory we define the operators $A(x)$ and $A^\dagger(x)$ 
\begin{align}
\label{A-in-position-space}
A(x)&:=\frac{1}{\sqrt{2}}\left[i\pi(x)+g\Delta\phi(x)\right], &
A^\dagger(x)&:=\frac{1}{\sqrt{2}}\left[-i\pi(x)+g\Delta\phi(x)\right].
\end{align}
From the commutation relation $[\phi(x),\pi(y)]=i\delta(x-y)$, one can see that $A(x)$ and $A^\dagger(x)$ obey the commutation relation $[A(x),A^\dagger(y)]=g(-\Delta)\delta(x-y)$ and can be interpreted as annihilation and creation operators in position space. We note that $g$ has to be greater than zero, since otherwise the roles of $A$ and $A^\dagger$ are reversed and  $A^\dagger$ becomes the annihilation operator. The Hamiltonian can be written in terms of $A$ and $A^\dagger$ as
\begin{equation}
\label{hamiltonian-factored}
H=\int_M \diff{2}x\  A^\dagger(x)A(x),
\end{equation}
where we, without loss of generality, subtracted the UV-divergent vacuum energy, thereby setting the energy of the ground state to zero. As the Hamiltonian \eqref{hamiltonian-factored} is positive semi-definite,  the ground state $\vert \Psi_0\rangle$ of the theory is found by solving $A(x)\vert\Psi_0\rangle=0$. The solution to this functional differential equation is given
\begin{align}
\label{ground-state}
\vert \Psi_0\rangle &= \frac{1}{\sqrt{Z_M}} \int\D\phi\ e^{-\frac{1}{2}S[\phi]}\vert\phi\rangle, & S[\phi]&=g\int_M\diff{2}x\ \phi (-\Delta)\phi,
\end{align} 
where  we defined the partition function $Z_M:=\int\D\phi\ e^{-S[\phi]}$. Note that the above definition coincides with the one given in \cite{Ardonne2004} after an integration by parts.

\paragraph{Excited states.} Let us concentrate on the case of a compact spatial manifold $M$, so that the spectrum of the Laplacian is discrete. Following \cite{Parker2017}, we can construct excited states of the QLM by exciting the eigenmodes of the Laplacian. Let $L_\lambda(x)$ be the eigenfunction to the eigenvalue $\lambda$ of $(-\Delta)$, that is
\begin{equation}
\label{eigenfunctions}
-\Delta L_\lambda(x)=\lambda L_\lambda(x),
\end{equation}
due to our sign convention, we note that $\lambda\geq 0$.
If $M$ has a boundary, we choose Dirichlet boundary conditions. For $\lambda\neq 0$ we can project the creation operator onto the the $\lambda$ eigenfunction and define
\begin{align}
\label{lambda-creation-operator}
A^\dagger_\lambda&:=\frac{1}{\sqrt{g\lambda}}\int_M\diff{2}x\ L_\lambda(x)A^\dagger(x), &  A_\lambda&:=\frac{1}{\sqrt{g\lambda}}\int_M\diff{2}x\ L_\lambda(x)A(x).
\end{align}
From the commutation relations for $A(x)$, we immediately get the correct commutation relations for a harmonic oscillator for each mode
\begin{gather}
\label{commutation-relations-A}
[A_\lambda, A_\mu^\dagger]=\delta_{\lambda\mu},\quad [A_\lambda^\dagger, A_\mu^\dagger]=[A_\lambda, A_\mu]=0,\\ 
[H, A_\lambda^\dagger]=g\lambda\, A_\lambda^\dagger,
\end{gather}
allowing for interpretation of $A^\dagger_\lambda$ and $A_\lambda$ as the creation and annihilation operators for excitations of the $\lambda$-mode. The Hamiltonian decomposes into the different modes
\begin{equation}
\label{Hamiltonian-lambda}
H=\sum_{\lambda\neq 0} g\lambda\, A_\lambda^\dagger A_\lambda,
\end{equation}
with $g\lambda$ the energy of a single excitation and $A_\lambda^\dagger A_\lambda$ counts the number of excitations in the $\lambda$-mode\footnote{
Before continuing, a comment on the zero-mode is in order. If the Laplacian has an eigenvalue $\lambda=0$, we have to alter the definition \eqref{lambda-creation-operator} for that eigenvalue. A consistent choice is $A_0^\dagger=\int_M L_0(x) A^\dagger(x)$, where $L_0(x)$ is the zero mode. A priori, the expansion of the Hamiltonian then should also contain a term $A^\dagger_0 A_0$, however one can check that the operator $A_0^\dagger A_0$ annihilates any state created by acting with either $A_0^\dagger$ or $A_\lambda^\dagger$ on the vacuum, and can thus be omitted from the Hamiltonian. From a different perspective, one can check that $[A_0, A^\dagger_0]=0$ and thus $A_0$ and $A_0^\dagger$ cannot be considered ladder operators. Hence we cannot excite the $0$-mode by $A^\dagger_0$ and whenever we write $\lambda$ for an eigenvalue it is implied that $\lambda\neq 0$. 
}. We can construct the general excited state by selecting a finite set of non-zero modes $\lbrace\lambda_1,\ldots,\lambda_\nu\rbrace$ and applying the respective creation operators to the ground state a finite amount of times  $\lbrace m_{\lambda_1},\ldots,m_{\lambda_\nu}\rbrace$. The general excited state is then labeled by that set of numbers and, taking into account proper normalization, given by
\begin{align}
\label{eq:excited-state-wave-function}
\ket{(m_{\lambda_1},\ldots,m_{\lambda_r},\ldots,m_{\lambda_\nu})}&= \frac{\left(A_{\lambda_1}^\dagger\right)^{m_{\lambda_1}}}{\sqrt{m_{\lambda_1}!}}\cdots\frac{\left(A_{\lambda_r}^\dagger\right)^{m_{\lambda_r}}}{\sqrt{m_{\lambda_r}!}}\cdots\frac{\left(A_{\lambda_\nu}^\dagger\right)^{m_{\lambda_\nu}}}{\sqrt{m_{\lambda_\nu}!}}\ket{\psi_0}\\
\label{general-excited-state}
&=\frac{1}{\sqrt{Z_M}}\int\D\phi\ \left[\prod_{r=1}^\nu\frac{1}{\sqrt{2^{m_{\lambda_r}} m_{\lambda_r}!}}H_{m_{\lambda_r}}\left(\frac{\phi^{\lambda_r}}{\sqrt{2}}\right)\right]e^{-\half S[\phi]}\vert\phi\rangle, 
\end{align}
where $H_{m}(x)$ is a Hermite polynomial
\begin{equation}
\label{eq:Hermite-polynomials-def}
H_m(x)=m!\sum_{k=0}^{\lfloor\frac{m}{2}\rfloor}\frac{(-1)^k(2x)^{m-2k}}{k!(m-2k)!},
\end{equation}
and where $\phi^\lambda_r$ is defined, similarly to \eqref{lambda-creation-operator}, as the projection of the scalar onto the $\lambda$-eigenmode
\begin{equation}
\label{phi-lambda-def}
\phi^\lambda=\sqrt{2 g \lambda}\int\limits_M \diff{2}x\ L_\lambda(x)\phi(x)
\end{equation}
with the prefactor  $\sqrt{2g\lambda}$ chosen with foresight to simplify later calculations.

\section{A warm up exercise: \\ Entanglement entropy of the singly excited state}\label{sec:single-excitation}
Our final goal is to calculate the bipartite entanglement entropy of the general excited state. Considering that the corresponding calculation is quite involved, it is instructive to first consider the much simpler case of the singly excited. Furthermore, due to its simpler nature, the calculation of the entanglement entropy can be performed under more general assumptions in this case. This problem was first considered by \cite{Parker2017}, where the authors calculated the R\'enyi entropy of a singly excited state by replica approach and provided a conjecture for the entanglement entropy in the special case when the manifold is cut into equal parts. By following a similar procedure in the beginning, yet introducing a simple assumption on the classical fields, we achieve a simplification of the combinatorics of the problem that renders the problem of the analytic continuation straightforward, and thus allows us to find the entanglement entropy exactly. Our findings differ from those in \cite{Parker2017}, but agree with them to leading order in certain quantities. We discuss this at the end of section \ref{sec:replica-single}.

Let us consider the state with one single excitation in the $\lambda$-mode
\begin{equation}
\label{singly-ex-state}
\ket{\psi_\lambda}:=\vert (m_\lambda=1) \rangle =\frac{1}{\sqrt{Z_M}}\int\D\phi\ \phi^\lambda e^{-\frac{1}{2}S[\phi]}\ket{\phi},
\end{equation}
where we used the fact that $H_1(x)=2x$. The action is given in equation \eqref{ground-state}, and after partial integration we choose to rewrite it as
\begin{equation}
\label{action-z=2}
S[\phi]=g\int_M\diff{2}x\ (\nabla\phi)^2.
\end{equation}
The density matrix of the singly excited state can then be written as
\begin{equation}
\label{singly-ex-state-density-matrix}
\rho:=\ket{\psi_\lambda}\bra{\psi_\lambda}=\frac{1}{Z_M}\int\D\phi\D\phi'\ \phi^\lambda\phi'^\lambda e^{-\frac{1}{2}S[\phi]-\frac{1}{2}S[\phi']}\ket{\phi}\langle\phi'\vert.
\end{equation}
We calculate the entanglement entropy by means of the replica method as developed by \cite{Calabrese2004,Calabrese2009}. The ground state entanglement entropy of the QLM has been studied in detail by these methods, see \cite{Fradkin2006,Zaletel2011,Zhou2016,Angel-Ramelli2019}. Our application of the replica method follows \cite{Zhou2016, Angel-Ramelli2019} closely. Let us consider the bipartite geometry obtained by dividing $M$ into the submanifolds $A$ and $B$, and denote by $\Gamma$ the entanglement cut that separates them. We will later consider two specific geometries: a sphere cut at the equator into its two hemispheres and a rectangle cut into two smaller rectangles.

Let us further assume that under the surgery described above the Hilbert space on $M$ splits as $\mathcal{H}=\mathcal{H}_A\otimes\mathcal{H}_B$. For a state in $\mathcal{H}$ we consequently get the splitting $\vert \phi\rangle=\vert\phi_A\rangle\otimes\vert\phi_B\rangle$, where $\vert\phi_{X}\rangle\in\mathcal{H}_{X}$, and $X=A,B$ respectively. The field $\phi(x)$ splits as $\phi(x)=\phi_A(x)+\phi_B(x)$ for $x\in A\cup B$, where
\begin{equation}
\label{phi-A-def}
\phi_A(x)=
\begin{cases}
    \phi(x), \quad x\in A\\
    0,\quad x\in B
\end{cases},
\end{equation}
and analogously for $B$. Whenever we index a field by a submanifold it can be assumed that it only has support there. At the boundary between $A$ and $B$ we have the continuity condition 
\begin{equation}
\label{continuity-condition}
\phi\vert_\Gamma=\phi_A\vert_\Gamma=\phi_B\vert_\Gamma.
\end{equation}
One can easily see that the action splits as $S[\phi]=S_A[\phi_A]+S_B[\phi_B]$, where
\begin{equation}
\label{S-X-def}
S_{X}[\phi_{X}]:=g\int_{X}\diff{2}x\ (\nabla\phi_X)^2,\quad X=A,B
\end{equation}
and similarly $\phi^\lambda=\phi^\lambda_A+\phi^\lambda_B$, where
\begin{equation}
\label{phi-lambda-X-def}
\phi_{X}^\lambda:=\sqrt{2 g\lambda}\int_{X}\diff{2}x\ L_\lambda(x)\phi_{X}(x),\quad X=A,B.
\end{equation}

\subsection{Replica method calculation}
\label{sec:replica-single}
Let $\rho_{A}=\Tr_B\rho$ be the reduced density matrix obtained by tracing out the degrees of freedom on $B$. The gist of the replica method lies in noting that the Von Neumann entropy of the subsystem $A$ can be rewritten as the limit
\begin{equation}
S[A]=-\Tr(\rho_A \log\rho_A)=-\lim_{n\rightarrow1}\partial_n\Tr\rho_A^n,
\end{equation}
as long as one is able to find an analytic continuation in $n$ for $\Tr\rho_A^n$. In the following we construct the trace of the $n$-th power of the reduced density matrix of the singly excited state \eqref{singly-ex-state-density-matrix}, and find that -- under a certain assumption satisfied by all modes in the geometries that we consider -- the analytic continuation is easily found.

\begin{figure}[h]
\centering
\vspace{10pt}
\includegraphics[scale=1.0]{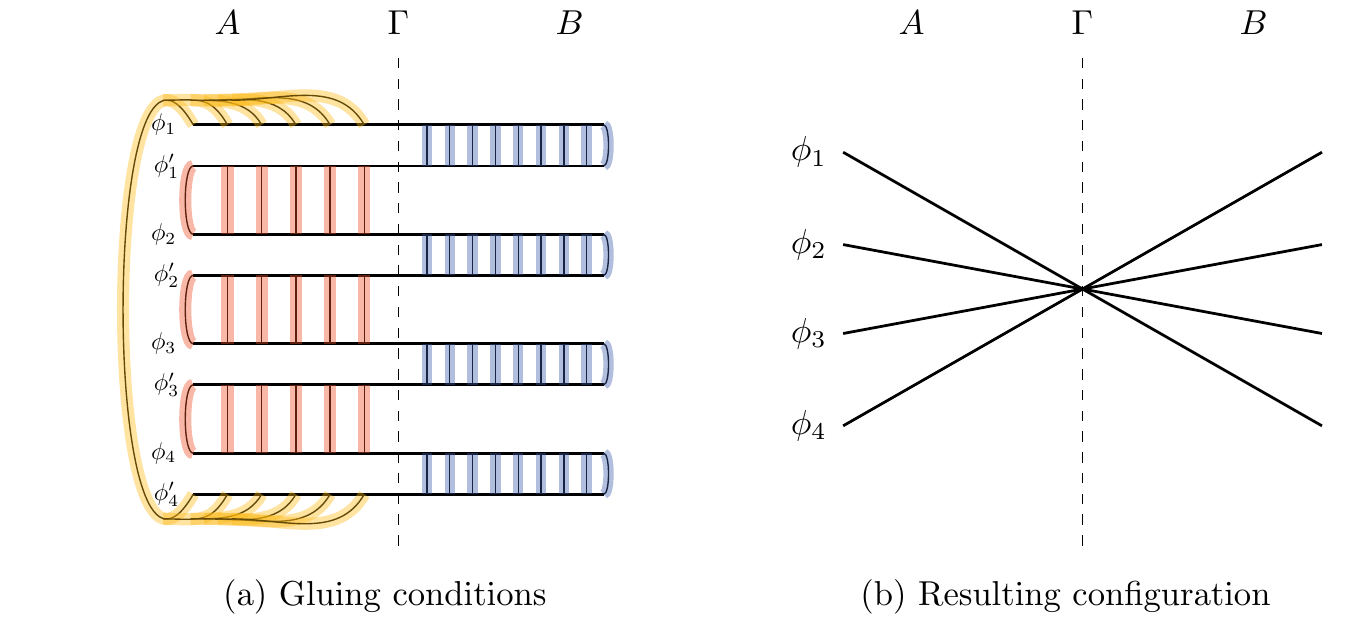}
\caption{The gluing conditions arising from the calculation of $\tr\rho_A^n$ and the geometry that results are depicted. The blue lines represent the gluing conditions due to the partial tracing on the $B$-side of each replica needed to calculate the reduced density matrix $\rho_A$. The red lines represent the gluing conditions resulting from the multiplication of the replicated reduced density matrices. The yellow lines represent the gluing conditions from the final total trace.} 
\label{fig:gluing-conditions}
\end{figure}

 After splitting the Hilbert space into $\H_A$ and $\H_B$, the density matrix takes the form
\begin{multline}
\label{singly-ex-state-density-matrix-after-splitting}
\rho_i=\frac{1}{Z_M}\int\D\phi_{A,i}\D\phi_{B,i}\D\phi_{A,i}'\D\phi_{B,i}'\ (\phi_{A,i}^\lambda+\phi_{B,i}^\lambda)(\phi'^\lambda_{A,i}+\phi'^\lambda_{B,i})\cross\\
 e^{-\frac{1}{2}S_A[\phi_{A,i}]-\frac{1}{2}S_B[\phi_{B,i}]-\frac{1}{2}S_A[\phi'_{A,i}]-\frac{1}{2}S_B[\phi'_{B,i}]}\vert\phi_{A,i}\rangle\otimes\vert\phi_{B,i}\rangle\langle\phi'_{A,i}\vert\otimes\langle\phi'_{B,i}\vert,
\end{multline}
where we introduced the replica index $i=1,\ldots,n$ which serves as an accounting device to distinguish the copies of the reduced density matrix. We note here that all fields in \eqref{singly-ex-state-density-matrix-after-splitting} are integrated over so the replica index doesn't have a physical significance. As discussed in detail in \cite{Angel-Ramelli2019} and depicted in figure \ref{fig:gluing-conditions}, one finds three types of gluing conditions for the fields. 
These arise when we resolve the bras and kets of $\Tr \rho_A^n$ into $\delta$-functions and then integrate over them. The first gluing condition comes from the calculation of the reduced density matrix $\rho_A=\Tr_B\rho$ and tells us to glue the primed and unprimed $B$ fields of the same replica, that is $\phi_{B,i}=\phi'_{B,i}$ for $i=1,\ldots,n$. The second gluing condition is the result of the multiplication between adjacent replicas of the reduced density matrix $\rho_{A,i}\rho_{A,i+1}$ for $i=1,\ldots,n-1$ and, together with the third gluing condition resulting from the total trace $\Tr\rho_{A}^n$, tells us to cyclically glue adjacent primed and unprimed $A$ fields, so $\phi'_{A,i}=\phi_{A,i+1}$ for $i=1,\ldots,n$ and with $n+1\sim 1$. The fact that the gluing conditions on the $A$ side are ``shifted'' with respect to the $B$ side together with the continuity condition \eqref{continuity-condition} forces \emph{all} the fields to agree at the cut $\Gamma$. We denote this boundary condition by $\mathcal{B}$
\begin{equation}
\label{eq:boundary-condition-b}
\mathcal{B}:\quad \phi_{A,i}\vert_\Gamma(x)=\phi_{B,j}\vert_\Gamma(x)=\cut(x)\quad \forall i,j=1,\ldots,n,
\end{equation}
where $\cut(x)$ is a function of the coordinates at the entanglement cut $\Gamma$. After enforcing the gluing conditions, the trace of the product of reduced density matrices takes the form
\begin{multline}
\label{eq:tr-rho-a-n-single-1}
\Tr\rho_A^n=\frac{1}{(Z_M)^n}\int\limits_\mathcal{B}\left[\prod_{i=1}^n\D\phi_{A,i}\D\phi_{B,i}\right] \\
 \cross \left[\prod_{i=1}^n(\phi_{A,i}^\lambda+\phi_{B,i}^\lambda)(\phi^\lambda_{A,i+1}+\phi^\lambda_{B,i})\right]\ e^{-\sum_{i=1}^n (S_{A}[\phi_{A,i}]+S_{B}[\phi_{B,i}])}.
\end{multline}
Since the fields are compactified  $\phi\sim\phi+2\pi R$, the boundary conditions \eqref{eq:boundary-condition-b} only  have to be respected modulo $2\pi R$. A standard way of accounting for this in the calculations, is to split the fields into classical parts and fluctuations, see \cite{Ginsparg1988,DiFrancesco1997}. We write for each field
\begin{equation}
\phi_{X,i}=\phi^\cl_{X,i}+\varphi_{X,i},\quad i=1,\ldots,n.
\end{equation}
From the fluctuations $\varphi$ we only demand that they satisfy Dirichlet boundary conditions
\begin{equation}
\label{fluctuation}
\varphi_{A,i}\vert_\Gamma =\varphi_{B,i}\vert_\Gamma =0,\quad i=1,\ldots,n,
\end{equation}
while the classical fields $\phi^\cl$ obey the equations of motion
\begin{equation}
\label{classical-field-z=2-eom}
\Delta\phi_{A,i}^\cl=\Delta\phi_{B,i}^\cl =0,\quad i=1,\ldots, n,
\end{equation}
and account for the value of the total field $\phi$ at the boundary as well as for its compact behavior. It is not hard to check that the action decouples as
\begin{equation}
\label{action-decouples-z=2}
S[\phi_{A,i}]=S[\phi^\cl_{A,i}]+S[\varphi_{A,i}].
\end{equation}
Finally, we can rewrite the boundary conditions for the classical fields resulting from \eqref{eq:boundary-condition-b} as
\begin{gather}
\label{classical-field-z=2-bc-1-to-(n-1)}
\phi_{A,i}^\cl\vert_{\Gamma}(x)=\phi_{B,i}^\cl\vert_{\Gamma}(x)=\cut(x)+2\pi R\omega_i,\quad i=1,\ldots n-1\\
\label{classical-field-z=2-bc-n}
\phi_{A,n}^\cl\vert_{\Gamma}(x)=\phi_{B,n}^\cl\vert_{\Gamma}(x)=\cut(x), 
\end{gather}
where $\omega_i$ are integers called the winding numbers that reflect the compactness of $\phi^\cl$. Here we used the fact that the cut functions are arbitrary to absorb the $n$-th winding number\footnote{We redefine $\cut\mapsto \cut -2\pi R\omega_n$ for all $n$. This effectively shifts the other winding numbers $\omega_i\mapsto\omega_i-\omega_n$ for $i=1,\ldots,n-1$. Since the winding numbers are just arbitrary integers, we can ignore the shift and rename the shifted winding numbers $\omega_i$.}. It is worth noting at this point that the exact treatment of the winding numbers depends a lot on the geometry, see \cite{Zhou2016} for several examples. The geometries that we will later consider, that is the sphere and the rectangle, in principle would allow for simplifications at this point. However, in order to keep the discussion somewhat more general, we decide to, for the moment, keep the boundary conditions as given above. Furthermore, we note that the zero modes on $A$ and $B$ with the above boundary conditions are in general not trivially related to the zero mode on the complete manifold which doesn't know anything about the cut $\Gamma$. In particular, this means that they are a priori not orthogonal to the eigenmodes of the Laplacian on $M$, so $(\phi^\cl_A)^\lambda+(\phi^\cl_B)^\lambda\neq 0$. We will concentrate on situations where this equation \emph{does} hold, as it leads to a significant simplification of the combinatorics. Thus, let us assume that the following is true
\begin{equation}
\label{eq:assumption-philambdaA+philambdaB-is-zero}
(\phi^\cl_A)^\lambda+(\phi^\cl_B)^\lambda= 0.
\end{equation}
We will later see that this is in fact justified in the geometries we consider.
Defining the field $\phi^\cl$ obtained by stitching together together $\phi_A^\cl$ and $\phi_B^\cl$, that is
\begin{equation}
\label{eq:phi-cl-A-plus-B-def}
\phi^\cl(x):=
\begin{cases}
\phi_{A}^\cl(x),\quad x\in A\\
\phi_{B}^\cl(x),\quad x\in B,
\end{cases}
\end{equation}
we can express the assumption \eqref{eq:assumption-philambdaA+philambdaB-is-zero} in the simpler form $(\phi^\cl)^\lambda =0$. We will later see that this condition is satisfied for all the eigenmodes of the Laplacian on both the sphere and rectangle. As defined above, $\phi^\cl$ is well defined and continuous at $\Gamma$, since $\phi^\cl_A$ and $\phi_B^\cl$ satisfy the same boundary conditions there. In general, however, $\phi^\cl$ is not smooth at the cut. We note here that in the derivation of the generalized Wick's theorem of \cite{Parker2017} this assumption is also implicitly made\footnote{Before eq. (A.11) in the reference, the field $C_a$ is defined by stitching together solutions $0$-modes of the Laplacian on the $A$ and $B$ sides of the manifold with non-trivial boundary conditions in between. Thus, the resulting function is in general \emph{not} a harmonic function over $M$, and the terms denoted $Har$ in (A.11) only vanish when integrated against eigenfunctions of the Laplacian on $M$ iff our assumption \eqref{eq:assumption-philambdaA+philambdaB-is-zero} is fulfilled. }. However, since the authors only consider the rectangular case the assumption is justified. 

At this step, the boundary conditions given in \eqref{eq:boundary-condition-b} can be recast as
\begin{equation}
\mathcal{B}:\quad
\begin{array}{lc}
    \phi_{i}^\cl\vert_{\Gamma}(x)=\cut(x)+2\pi R\omega_i,\quad& i=1,\ldots n-1\\
    \phi_{n}^\cl\vert_{\Gamma}(x)=\cut(x)& \\
    \varphi_{A,i}\vert_\Gamma =\varphi_{B,i}\vert_\Gamma =0,& i=1,\ldots,n.
\end{array}
\end{equation}
Next, we perform the unitary transformation $U_n$ from \cite{Zaletel2011,Zhou2016} on the classical fields. This rotation shifts the dependence on the $\cut$ function to the boundary condition for the $n$-th field, while turning the boundary conditions of the remaining fields into pure winding numbers. Explicitly, we define
\begin{equation}
\label{def-matrix-U}
U_n:= 
\begin{bmatrix}
{1\over \sqrt 2} ~&~ -{1\over \sqrt{2}}~ &~0 & ~&\dots\\
{1\over \sqrt 6} ~& ~ {1\over \sqrt{6}} ~ & ~ - {2\over \sqrt{6}} & ~~0& \dots \\
\vdots 
\\
{1\over \sqrt{n(n-1)}} ~& ~{1\over \sqrt{n(n-1)}} & \dots~ & ~\dots & -\sqrt{1-{1\over n}}\\
{1\over \sqrt{n}} ~& ~{1\over \sqrt{n}} & \dots~ & ~\dots & {1\over \sqrt{n}}\\
\end{bmatrix}
\end{equation}
and rotate the classical fields as
\begin{equation}
\label{classical-field-rotation}
\phi^\cl_i\mapsto \barphi^\cl_i:=(U_n\phi^\cl)_i.
\end{equation}
This results in the altered boundary condition $\mathcal{B}'$
\begin{equation}
\label{eq:boundary-condition-b-prime}
\mathcal{B'}:\quad
\begin{array}{lc}
\barphi_{i}^\cl\vert_{\Gamma}(x)=2\pi R\baromega_i, & i=1,\ldots n-1\\
\barphi_{n}^\cl\vert_{\Gamma}(x)=\sqrt{n}\cut(x)+\frac{2\pi R}{\sqrt{n}}\sum_{i=1}^{n-1}\omega_i,&\\
\varphi_{A,i}\vert_\Gamma =\varphi_{B,i}\vert_\Gamma =0, & i=1,\ldots,n,
\end{array}
\end{equation}
where $\baromega_i=(M_{n-1})_{ij}\omega_j$ and $M_{n-1}$ is the minor matrix resulting from deleting the $n$-th row and column of $U_{n}$. Notice that after the rotation the compactification radius of the $n$-th field becomes $\sqrt{n} R$. 

Finally, we can rewrite equation \eqref{eq:tr-rho-a-n-single-1} for $\Tr\rho_A^n$ in terms of the classical fields and fluctuations as
\begin{multline}
\label{eq:tr-rho-a-n-single-2}
\Tr\rho_A^n =\frac{1}{(Z_M)^n} W(n)\int\limits_\mathcal{B'}\left[\prod\nolimits_{i=1}^{n}\D\varphi_{A,i}\D\varphi_{B,i}\right]\D\barphi_n^\cl\cross\\
\left(\prod_{i=1}^{n}(\varphi_{A,i}^\lambda+\varphi_{B,i}^\lambda)(\varphi^\lambda_{A,i+1}+\varphi^\lambda_{B,i})\right) e^{-S[\barphi^\cl_n]-\sum_{i=1}^{n} (S[\varphi_{A,i}]+S[\varphi_{B,i}])},
\end{multline}
where we used the definition \eqref{phi-lambda-X-def} as well as the assumption \eqref{eq:assumption-philambdaA+philambdaB-is-zero} which together imply
\begin{equation}
\barphi^\lambda_{A,i}+\barphi^\lambda_{B,i}=\varphi_{A,i}^\lambda+\varphi^\lambda_{B,i}
\end{equation}
and allow us to get rid of the classical fields with matching replica indices outside of the exponential. The remaining classical fields in the product, i.e. those coming from $\barphi^\lambda_{A,i+1}+\barphi^\lambda_{B,i}$, don't necessarily vanish by assumption \eqref{eq:assumption-philambdaA+philambdaB-is-zero}. However, they don't contribute to the sum as they always appear multiplied by an odd number of fluctuations. When the path integrals are evaluated, these terms turn into correlation functions, and the correlation functions of an odd number of fluctuations vanish. We can thus safely discard all classical fields outside of the exponential.
We further omitted an $n^{-\frac{\varepsilon}{2L}}$ factor coming from a change of path integral measure, with $\varepsilon$ a UV cut-off and $L$ the length of the entangling cut $\Gamma$, as it only contributes to a non-universal area law term, see \cite{Zaletel2011,Zhou2016} for more details. Furthermore, we separated the contributions from the classical part of the first $n-1$ fields into the so-called winding sector
\begin{align}
W(n)&:=\prod\limits_{i=1}^{n-1}\int\D\bar{\phi}^\cl_{A,i}\D\bar{\phi}^\cl_{B,i}\ e^{-S[\bar{\phi}^\cl_{A,i}]-S[\bar{\phi}^\cl_{B,i}]}\nonumber\\
\label{eq:winding-sector-definition}
&=\prod\limits_{i=1}^{n-1}\int\D\bar{\phi}^\cl_{i}\ e^{-S[\bar{\phi}^\cl_{i}]} =\sum_{\omega\in\mathbb{Z}^{n-1}}e^{-\sum_{i=1}^{n-1}S[\bar{\phi}^\cl_{i}]}.
\end{align}
For clarity, let us make some comments. The assumption \eqref{eq:assumption-philambdaA+philambdaB-is-zero} is formulated for each unrotated replica. However, since $U_n$ is an invertible matrix it is clear that $(\vec{\phi^\cl})^\lambda=\vec{0}$ iff $(U_n\vec{\phi^\cl})^\lambda=\vec{0}$ for $\vec{\phi^\cl}=(\phi^\cl_1,\ldots,\phi^\cl_n)$ and $(\vec{\phi^\cl})^\lambda=((\phi^\cl_1)^\lambda,\ldots,(\phi^\cl_n)^\lambda)$. Furthermore, after our path integral manipulations the $n$-th replica becomes a free field on the complete manifold, such that $\phi^\cl_n$ turns into the $0$-mode on $M$. Thus for the $n$-th rotated field our assumption will always be satisfied after restitching, and we only need to check the other fields\footnote{Alternatively, we could have assumed \eqref{eq:assumption-philambdaA+philambdaB-is-zero} for only the first $n-1$ rotated fields from the beginning. This alters the form of \eqref{eq:tr-rho-a-n-single-2} but also results in \eqref{eq:tr-rho-a-n-single-3} after some manipulations.}. 
In appendix \ref{appendix:path-integral-manipulations-for-single-excitation} we show that, by performing some path integral manipulations, \eqref{eq:tr-rho-a-n-single-2} can be brought into the form
\begin{multline}
\label{eq:tr-rho-a-n-single-3}
\Tr(\rho_{A}^n)=
\frac{1}{(Z_M)^n} W(n)\Bigg[\left(\int_{A_D}\D\varphi_{A}\ (\varphi_{A}^\lambda)^2 e^{-S[\varphi_{A}]}\right)^{n-1}\int\D\bar{\phi}\ \bar{\phi}^\lambda\bar{\phi}^\lambda_A e^{-S[\bar{\phi}]}+\\
+\left(\int_{B_D}\D\varphi_{B}\ (\varphi_{B}^\lambda)^2 e^{-S[\varphi_{B}]}\right)^{n-1}\int\D\bar{\phi}\ \bar{\phi}^\lambda\bar{\phi}^\lambda_B e^{-S[\bar{\phi}]}\Bigg],
\end{multline}
where the subscripts $A_D$ and $B_D$ indicate that those path integrals are performed on either $A$ or $B$ with Dirichlet boundary conditions. By linearity and using the definition \eqref{phi-lambda-X-def}, we can rewrite the path integrals on $X=A,B$ as
\begin{align}
\int_{X_D}\D\varphi_{X}\ (\varphi_{X}^\lambda)^2 e^{-S[\varphi_{X}]}&= 2g\lambda \int\limits_X \diff{2}x\diff{2}x'\ L_\lambda(x)L_\lambda(x')\int_{X_D}\D\varphi_{X}\ \varphi_{X}(x)\varphi_{X}(x') e^{-S[\varphi_{X}]}\\
&=Z_X\ \lambda  \int\limits_X \diff{2}x\diff{2}x'\ L_\lambda(x)L_\lambda(x')G_X(x,x'),
\end{align}
where $Z_X$ is the partition function and $G_X(x,x')$ the Green's function corresponding to the Laplacian on $X$\footnote{Note that the factor of $2g$ cancels, as the two point function actually gives the Green's function corresponding to $2g\Delta$ which is $\frac{1}{2g}G$ with $G$  the Green's function of $\Delta$.}, both with Dirichlet boundary conditions at $\Gamma$. Similarly, we can write the following for the path integrals on the full manifold
\begin{align}
\begin{split}
\int\D\bar{\phi}\ \bar{\phi}^\lambda\bar{\phi}^\lambda_A e^{-S[\bar{\phi}]}&=2g\lambda\int\limits_M \diff{2}x\int\limits_A\diff{2}x' L_\lambda(x)L_\lambda(x')\int\D\bar{\phi}\ \bar{\phi}(x)\bar{\phi}(x') e^{-S[\bar{\phi}]}\\
&=Z_M\ \lambda \int\limits_M \diff{2}x\int\limits_A\diff{2}x' L_\lambda(x)L_\lambda(x')G_M(x,x')
\end{split}
\end{align}
with $G_M(x,x')$ the corresponding Green's function. Let us define the following transformed Green's functions
\begin{align}
\label{eq:g-lambda-definitions}
\begin{split}
& G^\lambda_{X,M}:=\lambda \int\limits_M \diff{2}x\int\limits_X\diff{2}x' L_\lambda(x)L_\lambda(x') G_M(x,x'),\quad X=A,B\\
& G_{X}^\lambda :=\lambda \int\limits_X \diff{2}x\diff{2}x'\ L_\lambda(x)L_\lambda(x')G_{X}(x,x'), \quad X=A,B,M.\\
\end{split}
\end{align}
We emphasize here that $G_M(x,x')$, $G_A(x,x')$, and $G_B(x,x')$ are different objects, each calculated on their corresponding manifold with Dirichlet boundary conditions. In appendix \ref{appendix:transformed-greens-functions} we take a closer look at these transformed propagators and use spectral methods to calculate them in some geometries. In particular we find that
\begin{equation}
\label{eq:g-m-lambda}
G_M^\lambda = 1
\end{equation}
is always true. In \cite{Parker2017}, quantities of this form were dubbed ``Entanglement Propagator Amplitudes'' (EPA's) and some explicit results about them were given. In appendix \ref{appendix:epa-rectangle} we show exactly how the EPA's are related to the transformed propagators and find our results in agreement with \cite{Parker2017}. With all this, we can finally write
\begin{align}
\Tr(\rho_A^n)&=\left(\frac{Z_{A_D}Z_{B_D}}{Z_M}\right)^{n-1}W(n)\left[\left(G_{A}^\lambda\right)^{n-1}G_{A,M}^\lambda+\left(G_{B}^\lambda\right)^{n-1}G^\lambda_{B,M}\right]\nonumber\\
\label{eq:tr-rho-n-final-single}
&=\Tr(\rho_{\text{GS},A}^n)\left[\left(G_{A}^\lambda\right)^{n-1}G_{A,M}^\lambda+\left(G_{B}^\lambda\right)^{n-1}G^\lambda_{B,M}\right],
\end{align}
where $\rho_{GS}$ is the ground state density matrix, see \cite{Zaletel2011,Zhou2016}, meaning that  the ground state R\'enyi entropy $S_{GS}^{(n)}[A]$  factors from the excited state R\'enyi entropies $S_\lambda^{(n)}[A]$
\begin{align}
S_\lambda^{(n)}[A]&:=\frac{1}{1-n}\log(\tr\rho_A^n)\nonumber\\
&=S_{GS}^{(n)}[A]+\frac{1}{1-n}\log\left[\left(G_{A}^\lambda\right)^{n-1}G_{A,M}^\lambda+\left(G_{B}^\lambda\right)^{n-1}G^\lambda_{B,M}\right]
\end{align}
In \cite{Parker2017} an expression for the R\'enyi entropies of the first excited state was also obtained. Our result, however, differs from theirs. In the rectangular case, we will see that $G_{X}^\lambda=G_{X,M}^\lambda-d_\lambda^X$ for $X=A,B$. There, our result agrees with \cite{Parker2017} to leading order in $d_\lambda^A$ and $d_\lambda^B$. As a consequence, our results also disagree with their conjecture for $\tr(\rho_A^n)/\tr(\rho_{GS,A}^n)$ and the resulting entanglement entropy (see eqs. (27) and (75) in \cite{Parker2017}).

The structure of \eqref{eq:tr-rho-n-final-single} offers a simple interpretation. In the above manipulations we effectively separated the $A$ and $B$ sides of the first $n-1$ replicas, while connecting the sides of the $n$-th replica into one complete copy over the whole manifold. The $n-1$-factors of  $G_{X}^\lambda$ encode the free propagation on either the $A$ or $B$ side of the first $n-1$ copies. The $G_{A(B),M}^\lambda$ factor, on the other hand, represents free propagation from either $A$ or $B$ to any point on $M$ for the $n$-th copy. Meanwhile topological effects are factored into the winding sector and the partition functions which give rise to the ground state density matrix. Apart from the winding sector, $n$ only appears in \eqref{eq:tr-rho-n-final-single} as a power. Thus, if an analytic continuation for $W(n)$ is known, it is trivial to analytically continue the rest of the expression. The entanglement entropy of the state with a single excitation in the $\lambda$ mode is thus 
\begin{align}
S_\lambda[A]&=-\lim_{n\rightarrow 1}\partial_n\Tr(\rho_A^n)\\
\label{eq:single-excitation-EE}
&= S_{GS}[A]-\log( G_{A}^\lambda)G_{A,M}^\lambda-\log( G_{B}^\lambda)G^\lambda_{B,M}
\end{align}
where we used that $G^\lambda_{A,M}+G^\lambda_{B,M}=G^\lambda_M=1$ and where $S_{GS}[A]$ is the entanglement entropy of the ground state 
\begin{equation}
\label{eq:ground-state-entanglement-entropy}
S_{GS}[A]=-\lim_{n\rightarrow 1}\partial_n \Tr(\rho_{GS,A}^n).
\end{equation}
Given the ground state entanglement entropy, the calculation of the entanglement entropy of the singly excited state reduces to the calculation of the transformed Green's functions \eqref{eq:g-lambda-definitions}. Let us  consider some explicit cases.

\subsection{Spherical geometry}\label{sec:spherical-geometry-single-excitation}
Let $M=S^2$ be the unit sphere and divide it into two hemispheres $A=B=H^2$ at the equator. The ground state entanglement entropy of a hemisphere is \cite{Zhou2016}  
\begin{equation}
\label{eq:hemisphere-ground-state-EE}
S_{GS}[H^2]=\log(\sqrt{8\pi g}R)-\frac{1}{2}.
\end{equation}
The eigenmodes of the Laplacian on the sphere are the spherical harmonics
\begin{equation}
L_{\lambda}(x)\equiv L_{\ell,m}(x)=Y^m_\ell(x),
\end{equation}
where $x=(\theta,\phi)$ is a point on $S^2$, $\ell$ a non-negative integer, and $m=-\ell,\ldots,\ell$. The eigenvalues of $-\Delta_{S^2}$ are $\lambda=\ell(\ell+1)$ and have a degeneracy of $2\ell+1$. As can be seen above, whenever we use $\lambda$ as an index, we actually mean the two numbers $(\ell,m)$ that uniquely identify a mode. 
The eigenmodes on the hemisphere with Dirichlet boundary conditions at the equator are, up to normalization, the spherical harmonics such that $\ell+m$ is odd. The properly normalized modes on $A$ and $B$ are then
\begin{equation}
L^{H^2}_\lambda(x)=\sqrt{2} Y^m_\ell(x),\quad m+\ell=\text{odd}.
\end{equation}
The first $n-1$ rotated classical fields on the hemisphere with boundary conditions \eqref{eq:boundary-condition-b-prime} is constant and given by
\begin{equation}
\barphi^\cl_i= 2\pi R\baromega_i.
\end{equation}
Since the $0$-mode $L_0=1/(4\pi)$ of the Laplacian on the full sphere is also constant $\barphi^\cl_i$ is proportional to $L_0$, and thus orthogonal to the other eigenmodes. Therefore, our assumption \eqref{eq:assumption-philambdaA+philambdaB-is-zero} is fulfilled in this situation. In appendix \ref{appendix:transformed-propagators-hemisphere} we calculate all the transformed propagators in this geometry and find that
\begin{gather}
\label{eq:g-lambda-sphere}
G_{A,M}^\lambda=G_{B,M}^\lambda=\frac{1}{2} \\
G_A^\lambda=G_B^\lambda=\frac{1}{2} \Sigma_\lambda
\end{gather}
where $\Sigma_\lambda$ is given by
\begin{equation}
\label{eq:Sigma-lambda-hemisphere}
\Sigma_\lambda=
\begin{cases}
\Sigma^\text{e}_\lambda,&\qquad \ell_\lambda+m_\lambda=\text{even}\\
 1, &\qquad \ell_\lambda+m_\lambda=\text{odd},
\end{cases}
\end{equation}
and $\Sigma^\text{e}_\lambda$ is complicated expression involving an infinite sum and an integral over associated Legendre polynomials. It is given explicitly in \eqref{eq:Sigma-lambda-def-appendix} and can be easily evaluated numerically by truncating the sum at a sufficiently large number\footnote{This number must be larger than $\ell_\lambda$ as the summand has a peak there that has to be included.}. We note that $\Sigma_{\ell,-m}=\Sigma_{\ell,m}$ for all modes. The striking difference between the modes with $\ell+m$ even and odd is due to the fact that those with $\ell+m$ odd are orthogonal to the eigenmodes on the hemisphere while the others are not. 
\begin{figure}[h]
\centering
\vspace{10pt}
\includegraphics[scale=0.5]{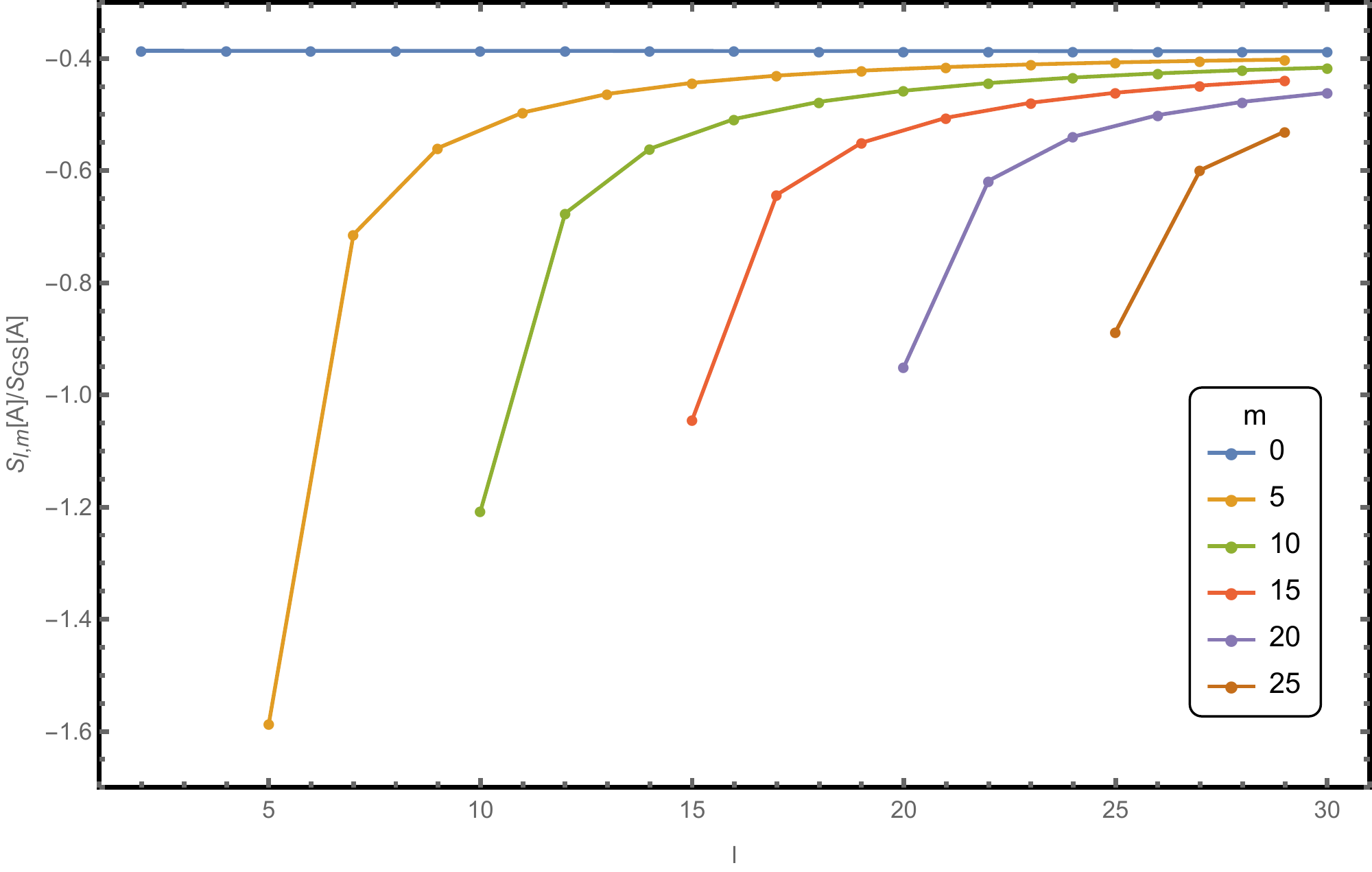}
\caption{The value of the entanglement entropy $S_{\ell,m}[A]$ in units of the ground state entanglement entropy at the RK-point, i.e. $g=\frac{1}{8\pi}$, and for $R=1$ plotted for $\ell$ up to 30 and a selection of values of $m$ such that $\ell+m$ is even.} 
\label{fig:HemispherePlot}
\end{figure}
Using  \eqref{eq:single-excitation-EE} and \eqref{eq:hemisphere-ground-state-EE} we can finally write the entanglement entropy as 
\begin{align}
S_\lambda[A]&=S_{GS}[A]-\log(G_A^\lambda)\\
&=\log\left(\sqrt{2\pi g}R\Sigma_\lambda\right)-\frac{1}{2}
\end{align}
Note that the entanglement entropy of all singly excited modes with $\ell+m$ odd is the same. The behavior of the entanglement entropy for modes with $\ell+m$ even is depicted in Figure \ref{fig:HemispherePlot}. We observe that the entanglement entropy for modes with $m=0$ is the same for all $\ell$. In fact, a numerical evaluation of $\Sigma_\lambda$ seems to indicate that $\Sigma_{\ell,m}\approx 1$. Consequently, the entanglement entropy is the same for \emph{all} modes with $m=0$. Furthermore, we see that for fixed $m$ the entanglement entropy of the modes with $\ell+m$ even and $m\neq 0$ approaches this value with growing $\ell$. Thus, we conclude that
\begin{equation}
\label{eq:sphere-ee-single-simple}
S_{\ell,m}[A]\approx S_{GS}[A]+\log 2
\end{equation}
is valid for all modes with $\ell+m$ odd, all modes with $m=0$, and for all modes in the limit $\ell\gg 1$. In the case $\ell+m$ odd the equation is exact.

\subsection{Rectangular Geometry}\label{sec:rectangular-geometry-single-excitation}
Next, let $M=[0,L_x]\cross[0,L_y]$ be a rectangle and cut it at $x=\ell_x$ into the two smaller rectangles $A=[0,\ell_x]\cross[0,L_y]$ and $B=[\ell_x,L_x]\cross[0,L_y]$. We impose Dirichlet boundary conditions $\phi\vert_{\partial M}=0$. If we go through the replica method calculation, we see that the classical part of the field, after rotation, obeys the boundary conditions \eqref{eq:boundary-condition-b-prime} at the cut. One can check that the only solution to the equation of motion $\Delta \phi^\cl_i=0$ for $i=1,\ldots,n-1$ with boundary conditions \eqref{eq:boundary-condition-b-prime} at the cut and Dirichlet boundary conditions on the other boundaries is the trivial one $\phi^\cl_i(x,y)=0$, which, in turn, is only a solution if the winding numbers vanish. Thus, there cannot be a winding sector for this geometry. The assumption \eqref{eq:assumption-philambdaA+philambdaB-is-zero} is then trivially satisfied for the first $n-1$ fields, since they all vanish.   

The eigenmodes of the Laplacian on $M$ with Dirichlet boundary conditions are given by 
\begin{equation}
\label{eq:rectangle-eigenmodes}
L_\lambda(x)\equiv L_{k_x,k_y}(x,y)= \frac{2}{\sqrt{L_x L_y}}\sin(\frac{\pi k_x}{L_x}x)\sin(\frac{\pi k_y}{L_y}y)
\end{equation}
with eigenvalues
\begin{equation}
\lambda_{k_x,k_y}=\left(\frac{\pi}{L_x}k_x\right)^2+\left(\frac{\pi}{L_y}k_y\right)^2,\quad k_x,k_y\in \mathbb{N}^+.
\end{equation}
In appendix \ref{appendix:det-rectangle} we calculate the $\zeta$-regularized determinant of the Laplacian on the rectangle $M$ and find
\begin{equation}
\label{eq:det-rectangle}
\det \Delta=\frac{1}{\sqrt{2L_y}}\eta\left(i\frac{L_x}{L_y}\right),
\end{equation}
where $\eta$ is the Dedekind $\eta$-function. This coincides with the expression cited in \cite{Parker2017} and calculated by lattice-regularization methods in \cite{Duplantier1988} up to  non-universal terms, and terms that diverge upon taking the continuum limit. We further note that due to the modular properties of the $\eta$-function this expression is invariant under the exchange of $L_x$ and $L_y$. The eigenmodes and determinants on the submanifolds $A$ and $B$ are obtained by inserting $\ell_x$ and $L_x-\ell_x$ respectively instead of $L_x$ into the above expressions. Writing the partition functions as $-\log Z_X=\frac{1}{2}\log\det \Delta_X$ for $X=M,A,B$, we can use the determinant \eqref{eq:det-rectangle} to express the ground state entanglement entropy on the rectangle as 
\begin{equation}
\label{eq:gs-ee-rectangle}
S_{GS}[A]=\frac{1}{2}\log(\frac{\eta\left(i\frac{\ell_x}{L_y}\right)\eta\left(i\frac{L_x-\ell_x}{L_y}\right)}{\sqrt{2L_y}\ \eta\left(i\frac{L_x}{L_y}\right)}).
\end{equation}
In appendix \ref{appendix:G-rectangle} we show that the first  transformed Green's functions on the rectangle are 
\begin{gather}
\label{eq:rectangle-g-a-m-lambda}
G_{A,M}^\lambda = \frac{\ell_x}{L_x}-\frac{1}{2\pi k_x}\sin(2\pi k_x\frac{\ell_x}{L_x}) \\
\label{eq:rectangle-g-b-m-lambda}
G_{B,M}^\lambda =\left(1-\frac{\ell_x}{L_x}\right)+\frac{1}{2\pi k_x}\sin(2\pi k_x\frac{\ell_x}{L_x}).
\end{gather}
and the remaining are given by
\begin{align}
\label{eq:rectangle-g-a-m-lambda}
G_{A}^\lambda &= G_{A,M}^\lambda-\frac{1}{L_x}\frac{2\left(\frac{\pi}{L_y}k_y\right)\coth(\pi k_y\frac{\ell_x}{L_y})\sin(\pi k_x\frac{\ell_x}{L_x})^2-\left(\frac{\pi}{L_x}k_x\right)\sin(2\pi k_x \frac{\ell_x}{L_x})}{\left(\frac{\pi}{L_x}k_x\right)^2+\left(\frac{\pi}{L_y}k_y\right)^2}\\
\label{eq:rectangle-g-b-m-lambda-appendix}
G_{B}^\lambda &= G_{B,M}^\lambda- \frac{1}{L_x}\frac{2\left(\frac{\pi}{L_y}k_y\right)\coth(\pi k_y\frac{L_x-\ell_x}{L_y})\sin(\pi k_x\frac{\ell_x}{L_x})^2+\left(\frac{\pi}{L_x}k_x\right)\sin(2\pi k_x \frac{\ell_x}{L_x})}{\left(\frac{\pi}{L_x}k_x\right)^2+\left(\frac{\pi}{L_y}k_y\right)^2}.
\end{align}
As shown in appendix \ref{appendix:epa-rectangle} our result for the transformed propagators agrees with the approximation found in \cite{Parker2017}. 

\begin{figure}[h]
\begin{subfigure}{.5\textwidth}
\centering
\includegraphics[width=\linewidth]{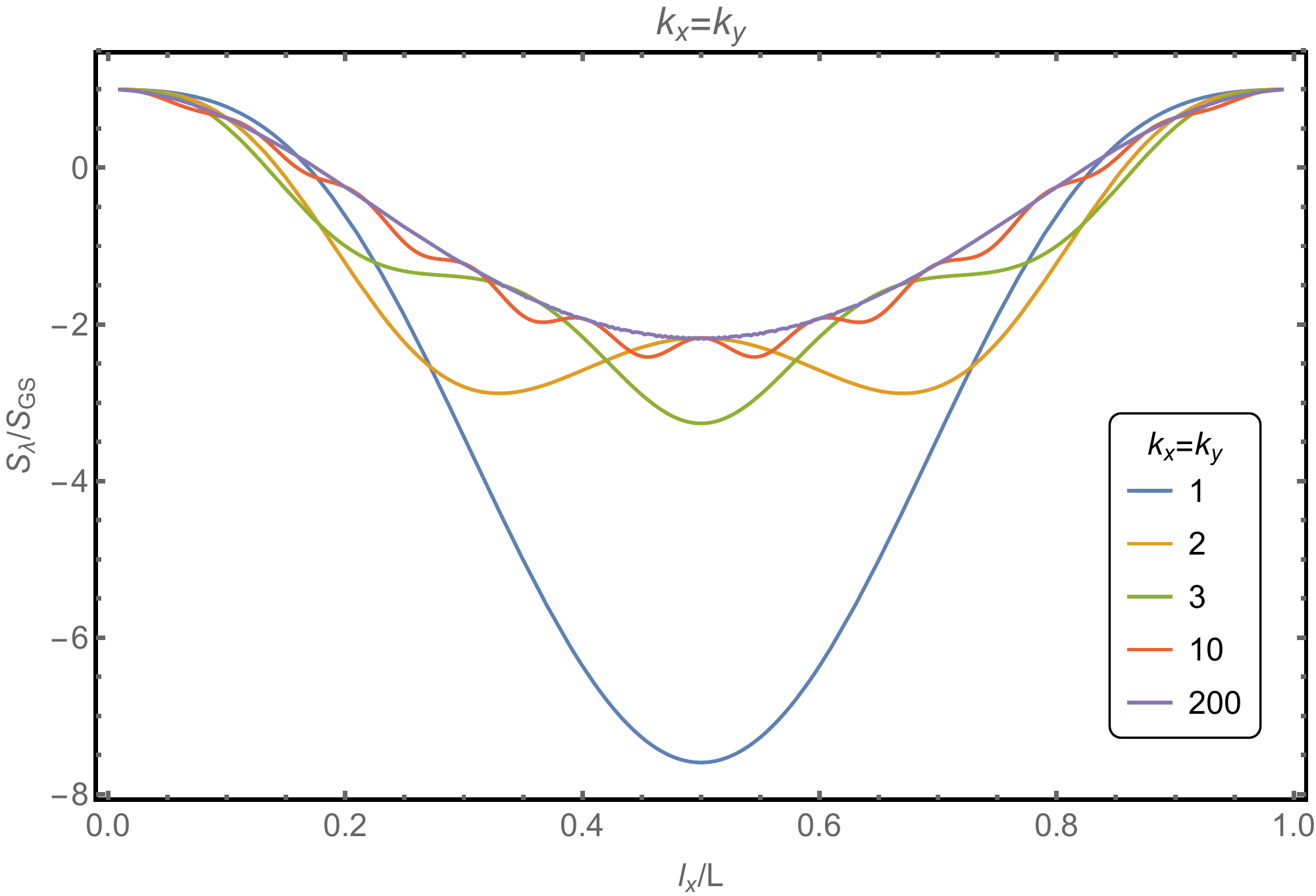} 
\end{subfigure}
\begin{subfigure}{.5\textwidth}
\centering
\includegraphics[width=\linewidth]{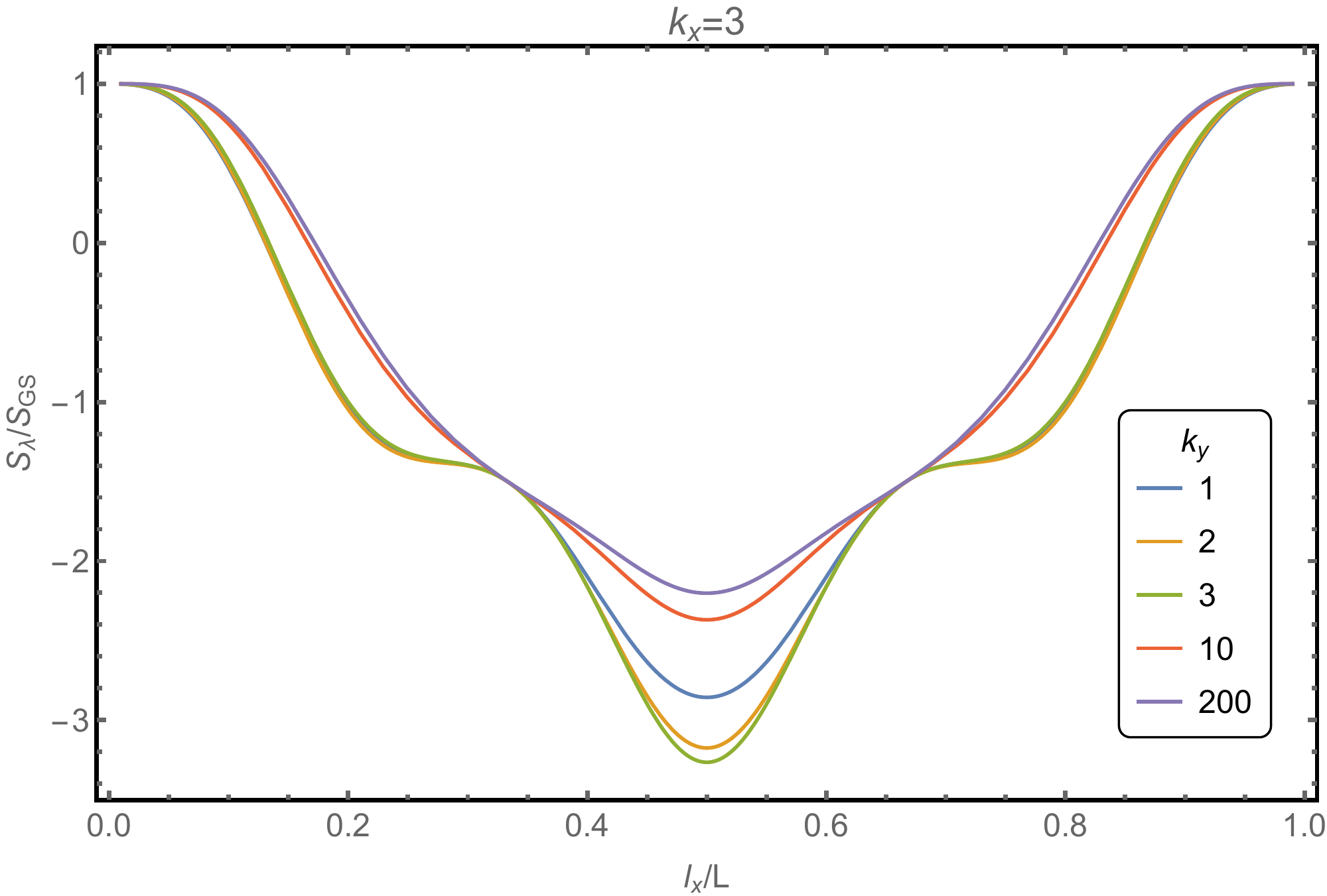} 
\end{subfigure}
\caption{$S_\lambda$ as a function of $\ell_x$ for a selection of modes and for $L_x=L_y$. }
\label{fig:surgery-dependence-rectangle}
\end{figure}

Finally, we find the entanglement entropy for the rectangle cut at $\ell_x$ by inserting the transformed propagators and ground state entanglement entropy into equation \eqref{eq:single-excitation-EE}. In figure \ref{fig:surgery-dependence-rectangle} we show how the entropy depends on the surgery for a selection of modes. We observe that for constant $k_x$ the entanglement entropy is minimal at $k_x=k_y$ and becomes maximal as $k_y\rightarrow\infty$. 
As was the case for the sphere, there are two kinds of modes for a given surgery. Modes that are at the same time eigenmodes of the complete manifold and the submanifolds (up to normalization), and modes that aren't. In the rectangular geometry the first kind of modes are determined by the condition that
\begin{equation}
\label{eq:rect-condition}
k_x\frac{\ell_x}{L_x}=m,\quad m\in\mathbb{N},
\end{equation}
which can only be satisfied whenever $\ell_x/L_x$ is a rational number.  For example, when the rectangle is halved by the surgery, that is $\ell_x=L_x/2$, all modes with even $k_x$ belong to the first category and those with $k_x$ odd to the second. The entanglement entropy is constant for all the modes satisfying  condition \eqref{eq:rect-condition} and given by
\begin{equation}
\label{eq:simple-ee-rectangle}
S_\lambda[A]=S_{GS}[A]-\log\left[\left(\frac{\ell_x}{L_x}\right)^\frac{\ell_x}{L_x}\left(1-\frac{\ell_x}{L_x}\right)^{1-\frac{\ell_x}{L_x}}\right].
\end{equation}
The eigenmodes that do not satisfy the condition have an intricate functional dependence on the $k_x$ and $k_y$. However, it isn't hard to see from the dependence on $k_x$ of the transformed propagators, that for large $k_x$ they converge to the value \eqref{eq:simple-ee-rectangle}. Whenever condition \eqref{eq:rect-condition} cannot be satisfied, that is when $\ell_x/L_x$ is not a rational number, there are no modes of the first kind and the entanglement entropy has a non-trivial dependence on the $k_x$ and $k_y$ for all modes. Nonetheless, equation \eqref{eq:simple-ee-rectangle} is also valid here for $k_x\gg 1$. When the rectangle is halved we can write 
\begin{equation}
S_{k_x,k_y}[A]\approx S_{GS}[A]+\log 2
\end{equation}
for modes with $k_x$ even and for all modes in the limit $k_x\gg1$. In the former case the equation is exact. This result is analogous to that for the halved sphere \eqref{eq:sphere-ee-single-simple}.


\section{General excited states}
\label{sec:general-excitation}
We are now ready to turn our attention to general excitations. From \eqref{eq:excited-state-wave-function} we can directly write the corresponding density matrix as
\begin{align}
\label{eq:es-density-matrix}
\begin{split}
\rho_i&=\ket{(m_{\lambda_1},\ldots,m_{\lambda_r},\ldots,m_{\lambda_\nu})}\bra{(m_{\lambda_1},\ldots,m_{\lambda_r},\ldots,m_{\lambda_\nu})}\\
&= \frac{1}{Z_M}\int\D\phi_i\D\phi'_i\ \left(\prod_{r=1}^\nu\frac{1}{2^{m_{\lambda_r}} m_{\lambda_r}!}H_{m_{\lambda_r}}\left(\frac{\phi_i^{\lambda_r}}{\sqrt{2}}\right)H_{m_{\lambda_r}}\left(\frac{\phi'^{\lambda_r}_i}{\sqrt{2}}\right)\right)e^{-\half S[\phi_i]-\half S[\phi'_i]}\vert\phi_i\rangle\langle\phi'_i\vert,
\end{split}
\end{align}
where we also introduced a replica index. From our discussion about the singly excited state, we know that, in broad terms, computing $\Tr \rho_A^n$ amounts to writing down the product of density matrices, separating the fields into fields with support on the respective submanifolds as $\phi=\phi_A+\phi_B$,  and enforcing the gluing conditions $\phi'_{A,i}=\phi_{A,i+1}$ and $\phi'_{B,i}=\phi_{B,i}$ for $i=1,\ldots,n$ with $n+1\sim 1$, that result in the set of boundary conditions \eqref{eq:boundary-condition-b} on the path integrals. Since the details of the calculation are exactly the same as for the singly excited state we omit them and immediately write down
\begin{multline}
\label{eq:tr-rho-A-n-es-1}
\Tr(\prod_{i=1}^n\rho_{A,i})=\left(\prod_{r=1}^\nu\frac{1}{2^{m_{\lambda_r}} m_{\lambda_r}!}\right)^n\frac{1}{Z_M^n}\int_\mathcal{B}\left[\prod\nolimits_{i=1}^n\D\phi_{A,i}\D\phi_{B,i}\right]e^{-\sum_{i=1}^n(S_{A,i}+S_{B,i})}\cross\\
\prod_{i=1}^n\prod_{r=1}^\nu H_{m_{\lambda_r}}\left(\frac{1}{\sqrt{2}}\left(\phi_{A,i}^{\lambda_r}+\phi_{B,i}^{\lambda_r}\right)\right)H_{m_{\lambda_r}}\left(\frac{1}{\sqrt{2}}\left(\phi^{\lambda_r}_{A,i+1}+\phi^{\lambda_r}_{B,i}\right)\right).
\end{multline}
One can compare this expression with equation \eqref{eq:tr-rho-a-n-single-1} for the singly excited state. As the expression stands, the fields on $A$ and $B$ are still tied together within the Hermite polynomials. We factor them by an identity of Hermite polynomials which we apply on the polynomials with shifted replica indices for $i=1,\ldots,n$ as
\begin{equation}
H_{m_r}\left(\frac{1}{\sqrt{2}}\left(\phi_{A,i+1}^{\lambda_r}+\phi_{B,i}^{\lambda_r}\right)\right)=\frac{1}{2^\frac{m_r}{2}}\sum_{k'_{r,i}=0}^{m_r}\binom{m_r}{k'_{r,i}}H_{k'_{r,i}}(\phi_{A,i+1}^{\lambda_r})H_{m_r-k'_{r,i}}(\phi_{B,i}^{\lambda_r}),
\end{equation}
on the polynomials with matching replica indices for $i=1,\ldots,n-1$
\begin{equation}
\label{eq:hermite-polynomial-identity-1}
H_{m_r}\left(\frac{1}{\sqrt{2}}\left(\phi_{A,i}^{\lambda_r}+\phi_{B,i}^{\lambda_r}\right)\right)=\frac{1}{2^\frac{m_r}{2}}\sum_{k_{r,i}=0}^{m_r}\binom{m_r}{k_{r,i}}H_{m_r-k_{r,i}}(\phi_{A,i}^{\lambda_r})H_{k_{r,i}}(\phi_{B,i}^{\lambda_r}),
\end{equation}
where, with some foresight, we don't apply factorization to the $n$-th copy, as the $A$ and $B$ parts of the $n$-th field will be reunited into a free field on $M$.
Also note that from now on we write $m_r$ instead of $m_{\lambda_r}$ to declutter the notation. Since we are applying the factorization above to each of the Hermite polynomials separately, and they are each labelled by a replica and an $r$ index, there is no way around introducing the, at first overwhelming, amount of indices $k_{r,i}$ and $k'_{r,i}$ for $i=1,\ldots, n$ and $r=1,\ldots,\nu$. 

After factoring the polynomials we can continue as in section \ref{sec:single-excitation}. We first separate the fields into classical parts and fluctuations and then rotate the classical fields.
For the singly excited state it was sufficient to demand the assumption \eqref{eq:assumption-philambdaA+philambdaB-is-zero} to be fulfilled in order to make the calculation tractable. Now, since the fields all appear inside Hermite polynomials, the calculations in appendix \ref{appendix:path-integral-manipulations-for-single-excitation} don't apply, and we must make a new assumption. We demand that the first $n-1$ rotated fields satisfy
\begin{equation}
\label{eq:assumption-philambdaA-and-philambdaB-are-zero}
(\barphi^\cl_{A,i})^\lambda=(\barphi^\cl_{B,i})^\lambda=0,\quad i=1,\ldots,n-1.
\end{equation}
Note that this assumption is stricter than \eqref{eq:assumption-philambdaA+philambdaB-is-zero}, demanding that the integrals on the $A$ and $B$ sides of the manifold vanish separately. We don't require anything from the $n$-th classical field and stitch it back together with the fluctuation to form a complete free field on $M$. As for the singly excited state, the assumption ensures that we don't have to consider correlation functions of the classical fields, which adds a difficult layer of complexity to calculations. While assumption \eqref{eq:assumption-philambdaA-and-philambdaB-are-zero} might seem rather restrictive for general geometries, we will see that it is satisfied by almost all modes on the halved-sphere and by all modes on the rectangle.

Apart from the stronger assumption on the classical fields, there is no difference in the method from the singly excited case and the details go through in an analogous way. With these considerations and some rearranging of the terms we can rewrite \eqref{eq:tr-rho-A-n-es-1} as 
\begin{align}
&\Tr(\prod_{i=1}^n\rho_{A,i})=\frac{1}{Z_M^n}W(n)\sum_{K_1}\sum_{K'_1}\cdots\sum_{K_\nu}\sum_{K'_\nu}\Bigg[ \nonumber\\
&\prod_{i=1}^{n-1}\left(\int_{A_D}\!\D\varphi_{A,i}\left[\prod_{r=1}^\nu \frac{1}{2^{m_{r}}}\sqrt{ \frac{1}{m_r!}\binom{m_r}{k_{r,i}}\binom{m_r}{k'_{r,i-1}}}H_{m_r-k_{r,i}}\!\left(\varphi_{A,i}^{\lambda_r}\right)H_{k'_{r,i-1}}\!\left(\varphi_{A,i}^{\lambda_r}\right)\right]e^{-S[\varphi_{A,i}]}\right) \nonumber\\
&\prod_{i=1}^{n-1}\left(\int_{B_D}\!\D\varphi_{B,i}\left[\prod_{r=1}^\nu \frac{1}{2^{m_{r}}}\sqrt{\frac{1}{m_r!}\binom{m_r}{k_{r,i}}\binom{m_r}{k'_{r,i}}}H_{k_{r,i}}\!\left(\varphi_{B,i}^{\lambda_r}\right)H_{m_r-k'_{r,i}}\!\left(\varphi_{B,i}^{\lambda_r}\right)\right]e^{-S[\varphi_{B,i}]}\right) \nonumber\\
\label{eq:tr-rho-A-n-es-2}
&\int\!\D\bar{\phi_n}\left[\prod_{r=1}^\nu \frac{1}{2^{\frac{3m_{r}}{2}}m_r!} \sqrt{\binom{m_r}{k'_{r,n}}\binom{m_r}{k'_{r,n-1}}}H_{m_r}\!\left(\frac{1}{\sqrt{2}}\bar{\phi_n}^{\lambda_r}\right)H_{k'_{r,n}}\!\left(\bar{\phi}_{A,n}^{\lambda_r}\right)H_{m_r-k'_{r,n-1}}\!\left(\bar{\phi}_{B,n}^{\lambda_r}\right)\right]e^{-S[\bar{\phi}]}\Bigg]
\end{align}
where $\sum_{K_r}\equiv\sum_{k_{r,1}=0}^{m_r}\cdots\sum_{k_{r,n-1}=0}^{m_r}$ and $\sum_{K_r'}\equiv\sum_{k_{r,1}'=0}^{m_r}\cdots\sum_{k_{r,n}'=0}^{m_r}$. Note that for the first $n-1$ replicas there are always two Hermite polynomials for each replica at each side $A$ and $B$ which result from splitting the two Hermite polynomials in \eqref{eq:tr-rho-A-n-es-1}. For the $n$-th replica we only split one of the Hermite polynomials in \eqref{eq:tr-rho-A-n-es-1}, thus we see only three in the resulting expression: the unsplit Hermite polynomial of a full field, and the two factors of the one we split.  Although at first glance the resulting situation seems quite desperate, there are a couple of simplifications we can make. First of all, we note that all the replica indices for the fields are now dummy indices, and that the path integral expressions are just correlation functions. For example the path integral over $A$ is just 
\begin{multline}
\label{eq:A-correl-def}
\int_{A_D}\!\D\varphi_{A}\ \left[\prod_{r=1}^\nu \frac{1}{2^{m_{r}}}\sqrt{\frac{1}{m_r!}\binom{m_r}{k_{r,i}}\binom{m_r}{k'_{r,i-1}}}H_{m_r-k_{r,i}}\!\left(\varphi_{A}^{\lambda_r}\right)H_{k'_{r,i-1}}\!\left(\varphi_{A}^{\lambda_r}\right)\right]e^{-S[\varphi_{A}]}\\
=Z_A\left(\prod_{r=1}^\nu \frac{1}{2^{m_r}}\sqrt{\frac{1}{m_r!}\binom{m_r}{k_{r,i}}\binom{m_r}{k'_{r,i-1}}}\right)\left\langle \prod_{r=1}^\nu H_{m_r-k_{r,i}}\!\left(\varphi_{A}^{\lambda_r}\right)H_{k'_{r,i-1}}\!\left(\varphi_{A}^{\lambda_r}\right)\right\rangle_{A_D},
\end{multline}
and similarly for the path integrals over $B$ and $M$. We further note that the quantity above is labelled by the $2\nu$ indices $k_{r,i}$ and $k'_{r,i-1}$ with the replica index constant. Hence, we package them into rank $2\nu$ tensors by defining 
\begin{align}
\label{eq:def-tensor-A-B-M}
\mathcal{A}^{k'_{1},\ldots, k'_{\nu}}_{\ k_1,\ldots,k_\nu}&:=\left(\prod_{r=1}^\nu \frac{1}{2^{m_r}}\sqrt{\frac{1}{m_r!}\binom{m_r}{k_{r}}\binom{m_r}{k'_{r}}}\right)\left\langle \prod_{r=1}^\nu H_{m_r-k_{r}}\!\left(\varphi_{A}^{\lambda_r}\right)H_{k'_{r}}\!\left(\varphi_{A}^{\lambda_r}\right)\right\rangle_{A_D}\\
\mathcal{B}^{k'_{1},\ldots, k'_{\nu}}_{\ k_1,\ldots,k_\nu}&:=\left(\prod_{r=1}^\nu \frac{1}{2^{m_r}}\sqrt{\frac{1}{m_r!}\binom{m_r}{k_{r}}\binom{m_r}{k'_{r}}}\right)\left\langle \prod_{r=1}^\nu H_{m_r-k_{r}}\!\left(\varphi_{B}^{\lambda_r}\right)H_{k'_{r}}\!\left(\varphi_{B}^{\lambda_r}\right)\right\rangle_{B_D}\\
\mathcal{M}^{k'_{1},\ldots, k'_{\nu}}_{\ k_1,\ldots,k_\nu}&:=\left(\prod_{r=1}^\nu \frac{1}{2^{\frac{3m_{r}}{2}}m_r!}\sqrt{\binom{m_r}{k_{r}}\binom{m_r}{k'_{r}}}\right)\left\langle \prod_{r=1}^\nu H_{m_r}\!\left(\frac{1}{\sqrt{2}}\bar{\phi}^{\lambda_r}\right) H_{k'_{r}}\!\left(\bar{\phi}_{A}^{\lambda_r}\right)H_{m_r-k_{r}}\!\left(\bar{\phi}_{B}^{\lambda_r}\right)\right\rangle_{M}
\end{align}
where the $k_r$ and $k'_r$ indices run from $0$ to $m_r$. Using the Einstein summation convention we can define multiplication and a trace operation for these tensors in the standard way
\begin{align}
\label{eq:multiplication-trace-definition}
(\mathcal{A}\mathcal{B})^{k'_{1},\ldots, k'_{\nu}}_{\ k_1,\ldots,k_\nu}&:=\mathcal{A}^{k'_{1},\ldots, k'_{\nu}}_{\ j_1,\ldots,j_\nu}\mathcal{B}^{j_{1},\ldots, j_{\nu}}_{\ k_1,\ldots,k_\nu},\\
\Tr\mathcal{A}&:=\mathcal{A}^{k_{1},\ldots, k_{\nu}}_{\ k_1,\ldots,k_\nu}.
\end{align}
These definitions take care of the index structure of \eqref{eq:tr-rho-A-n-es-2}, which can then be simply written as
\begin{align}
\label{eq:tr-rho-A-n-es-2}
\begin{split}
\Tr(\rho_{A}^n)&= \left(\frac{Z_A Z_B}{Z_M}\right)^{n-1}W(n)\Tr[(\mathcal{AB})^{n-1}\mathcal{M}]\\
&=\Tr(\rho_{\text{GS},A}^n)\Tr[(\mathcal{AB})^{n-1}\mathcal{M}].
\end{split}
\end{align}
In order to find an analytic continuation in $n$ for this expression, we first note that the tensors above can all be written as square matrices $A^i_{j}$ with $i,j=1,\ldots,\prod_{r=1}^\nu (m_r+1)$ such that matrix multiplication and the standard trace correspond to the operations on the tensors defined in \eqref{eq:multiplication-trace-definition}\footnote{This is satisfied if we, for example, define $A^i_j$ by the map 
\begin{equation}
\mathcal{A}^{k'_{1},\ldots, k'_{\nu}}_{\ k_1,\ldots,k_\nu}\mapsto A^{k'_1+\sum_{r=2}^\nu (m_{r-1}+1+k'_r)}_{k_1+\sum_{r=2}^\nu (m_{r-1}+1+k_r)}
\end{equation}}. 
As a matrix, the tensor $\mathcal{AB}$ is invertible which provides a sufficient condition for the existence of its matrix logarithm.  Later, we will see that at least in the special case where all the excitations are on the same mode, the matrix $\A\B$ is in fact diagonal and with real eigenvalues, making the matrix logarithm unique and well-defined. For now, we are satisfied knowing it exists as it provides us with the desired analytic continuation. We can next differentiate and take the limit $n\rightarrow 1$ to find
\begin{equation}
\label{eq:excited-state-entanglement-entropy}
S_{m_{\lambda_1},\ldots,m_{\lambda_\nu}}[A]=S_{GS}[A]\Tr(\mathcal{M})+\Tr(\log(\mathcal{AB})\mathcal{M}).
\end{equation}
It is not hard to check that this agrees with the entanglement entropy of the singly excited state \eqref{eq:single-excitation-EE}. The trace of $\mathcal{M}$ in the first term, is easy to evaluate using the identity of Hermite polynomials \eqref{eq:hermite-polynomial-identity-1} and gives
\begin{equation}
\Tr(\mathcal{M})\equiv \mathcal{M}^{k_{1},\ldots, k_{\nu}}_{\ k_1,\ldots,k_\nu}=\left(\prod_{r=1}^\nu \frac{1}{2^{m_r}m_r!}\right)\left\langle \prod_{r=1}^\nu H_{m_r}\!\left(\frac{1}{\sqrt{2}}\bar{\phi}^{\lambda_r}\right)H_{m_r}\!\left(\frac{1}{\sqrt{2}}\bar{\phi}^{\lambda_r}\right)\right\rangle_{M}.
\end{equation}
As was the case for the singly excited state, the coefficient of the ground state entanglement entropy encodes information about correlation functions on the complete manifold. The second term, also as before, includes information about the correlation functions in the submanifolds as well as correlation functions that communicate between them. Using our expression for $\tr\rho_A^n$ \eqref{eq:tr-rho-A-n-es-2} it is also straight forward to write down an explicit form for the R\'enyi entropies
\begin{equation}
S^{(n)}_{m_{\lambda_1},\ldots,m_{\lambda_\nu}}[A]=S^{(n)}_{GS}[A]+\frac{1}{1-n}\log\left(\Tr[(\mathcal{AB})^{n-1}\mathcal{M}]\right),
\end{equation}
where we observe, as for the singly excited state, that the ground state R\'enyi entropy factors from the excited state. While it is quite nice that an analytic expression can be written down for the entropy of an arbitrarily excited state, its form is quite involved. In the next section we will consider an interesting subcase, and see how the result behaves in some specific geometries.

\subsection{An interesting subcase: only one mode is excited}
\label{sec:interesting-subcase}
The density matrix corresponding to the state with $m$ excitations in the $\lambda$-mode is follows directly from \eqref{eq:excited-state-wave-function} with $\nu=1$
\begin{align}
\rho&=\ket{(m_\lambda)}\bra{(m_\lambda)}\\
&=\frac{1}{Z_M}\frac{1}{2^{m_{\lambda}} m_{\lambda}!}\int\D\phi\D\phi'\ H_{m_{\lambda}}\!\left(\frac{\phi^{\lambda}}{\sqrt{2}}\right)H_{m_{\lambda}}\!\left(\frac{\phi'^{\lambda}}{\sqrt{2}}\right)e^{-\half S[\phi]-\half S[\phi']}\vert\phi_i\rangle\langle\phi'_i\vert
\end{align}
The replica calculation goes through as described in the previous section, leading to tensors \eqref{eq:def-tensor-A-B-M} that are square $(m+1)$-dimensional matrices with elements
\begin{align}
\label{eq:tensor-A-single-mode}
\mathcal{A}^{k'}_{\ k}&=\frac{1}{2^m}\sqrt{\frac{1}{m!}\binom{m}{k}\binom{m}{k'}}\left\langle H_{m-k}\!\left(\varphi_A^\lambda\right)H_{k'}\!\left(\varphi_A^\lambda\right)\right\rangle_A, \\
\label{eq:tensor-B-single-mode}
\mathcal{B}^{k'}_{\ k}&=\frac{1}{2^m}\sqrt{\frac{1}{m!}\binom{m}{k}\binom{m}{k'}}\left\langle H_{m-k}\!\left(\varphi_B^\lambda\right)H_{k'}\!\left(\varphi_B^\lambda\right)\right\rangle_B,\\
\label{eq:tensor-M-single-mode}
\mathcal{M}^{k'}_{\ k}&=\frac{1}{2^\frac{3m}{2}m!}\sqrt{\binom{m}{k}\binom{m}{k'}}\left\langle H_{m}\!\left(\frac{1}{\sqrt{2}}\bar{\phi}^{\lambda}\right) H_{k}\!\left(\bar{\phi}_{A}^{\lambda}\right)H_{m-k'}\!\left(\bar{\phi}_{B}^{\lambda}\right)\right\rangle_{M}.
\end{align}
The correlation functions in the tensors are straight-forward to calculate. In appendix \ref{sec:correlation-functions-appendix}  we use the definition of Hermite polynomials and the linearity of the correlation functions to express the matrix elements above as sums of simple correlation functions of transformed fields. These can then be evaluated by Wick's theorem and simplified by considering that the Green's functions appearing in the expansion are all integrated against the same eigenfunctions of the Laplacian, turning the space variables into dummy integration variables. Thus, all Green's functions in Wick's expansion that are integrated over the same domain, are the same. For example, we have  
\begin{align}
\label{eq:correl-one-field}
\left\langle(\varphi_X^\lambda)^{\beta}\right\rangle_X&=\delta_{\beta,\text{even}}(\text{\# of full Wick contractions})(G_X^\lambda)^\frac{\beta}{2}\nonumber\\
&=\delta_{\beta,\text{even}}(\beta-1)!!(G_X^\lambda)^\frac{\beta}{2},
\end{align}
which together with \eqref{eq:Hermite-polynomials-def} is used to express the tensors $\mathcal{A}$ and $\mathcal{B}$ as polynomials in $G_A^\lambda$ and $G_B^\lambda$. Explicitly, we have
\begin{align}
\label{eq:tensor-A-polynomial}
\mathcal{A}^{k'}_{k}&= F^m_{k,k'}\left(G_A^\lambda\right), &
\mathcal{B}^{k'}_{k}&= F^m_{k,k'}\left(G_B^\lambda\right),
\end{align}
where $F$ is a polynomial given in equation \eqref{eq:F-polynomial-definition-appendix} of appendix  \ref{sec:correlation-functions-appendix}. We note that $F^m_{k,k'}\propto \delta_{m-k+k',\text{even}}$.
The evaluation of the correlation functions in $\mathcal{M}$ is combinatorically more involved, as three different projections of the field appear in it. This means that there are three distinct transformed Green's functions appearing in Wick's expansion. We find it convenient to work with the following partial projections
\begin{equation}
\label{eq:g-x-y-def}
G_{X,Y}^\lambda:=\lambda \int_X \diff{2}x\int_Y\diff{2}x' L_\lambda(x)L_\lambda(x') G_M(x,x'),
\end{equation}
where $X,Y=A,B$. Since $G_{A,B}^\lambda=G_{B,A}^\lambda$ these are three distinct objects. By definition, they are related to the transformed Green's functions we dealt with before, see \eqref{eq:g-lambda-definitions}, by 
\begin{gather}
G^\lambda_{X,M}=G^\lambda_{A,B}+G^\lambda_{X,X},\quad X=A,B\\
G^\lambda_{M}=G^\lambda_{A,A}+2G^\lambda_{A,B}+G^\lambda_{B,B}.
\end{gather}
As for $\mathcal{A}$ and $\mathcal{B}$, the correlation functions in $\mathcal{M}$ can be expressed as polynomials in the three transformed Green's functions by linearity of the correlation functions, identities of Hermite polynomials, and some combinatorics. The result is 
\begin{equation}
\label{eq:tensor-M-polynomial}
\mathcal{M}^{k'}_k= T^m_{k,k'}\left(G_{A,B}^\lambda,G_{A,A}^\lambda,G_{B,B}^\lambda\right),
\end{equation}
where $T$ is a polynomial in three variables defined in equation \eqref{eq:T-polynomial-def-appendix} of appendix \ref{sec:correlation-functions-appendix}. Note that for a fixed surgery and eigenmode $G_{A,B}^\lambda$, $G_{A,A}^\lambda$, and $G_{B,B}^\lambda$ are just real numbers. Furthermore, we note that $T^m_{k,k'}\propto \delta_{k+k',\text{even}}$.

In the following sections we will apply this to the same geometries we studied when considering singly excited states in section \ref{sec:single-excitation}. We will concentrate on the modes that are simultaneously eigenmodes on the manifold and submanifolds, and observe the effect that adding excitations to them has on the entanglement entropy.

\subsection{Spherical geometry} 
Let us start by considering again the geometry of section \ref{sec:spherical-geometry-single-excitation}, that is the sphere $M$ cut at the equator into the hemispheres $A$ and $B$. The first thing we need to do, is identify the modes for which the assumption \eqref{eq:assumption-philambdaA-and-philambdaB-are-zero}, that is $(\barphi_{A_i}^\cl)^\lambda=(\barphi_{B,i}^\cl)^\lambda=0$ for $i=1,\ldots,n-1$, is satisfied. As the classical field is constant, the condition is satisfied iff
\begin{equation}
\int_A\diff{2}{x}\ Y^m_\ell(x)=\int_B\diff{2}{x}\ Y^m_\ell(x)=0,
\end{equation}
where $Y^m_\ell$ is a spherical harmonic. The spherical harmonics are proportional to $e^{im\varphi}$, and $\varphi$ is integrated from $0$ to $2\pi$, so the condition is automatically satisfied for all spherical harmonics with $m\neq 0$. It isn't hard to see that, by the properties of Legendre polynomials, the remaining integrals vanish iff $\ell$ is even. Thus our results from section \ref{sec:general-excitation} are valid for all eigenmodes of the Laplacian on the sphere \emph{except} those with $m=0$ \emph{and} $\ell$ odd.

\begin{figure}[h]
\begin{subfigure}{.5\textwidth}
\centering
\includegraphics[width=\linewidth]{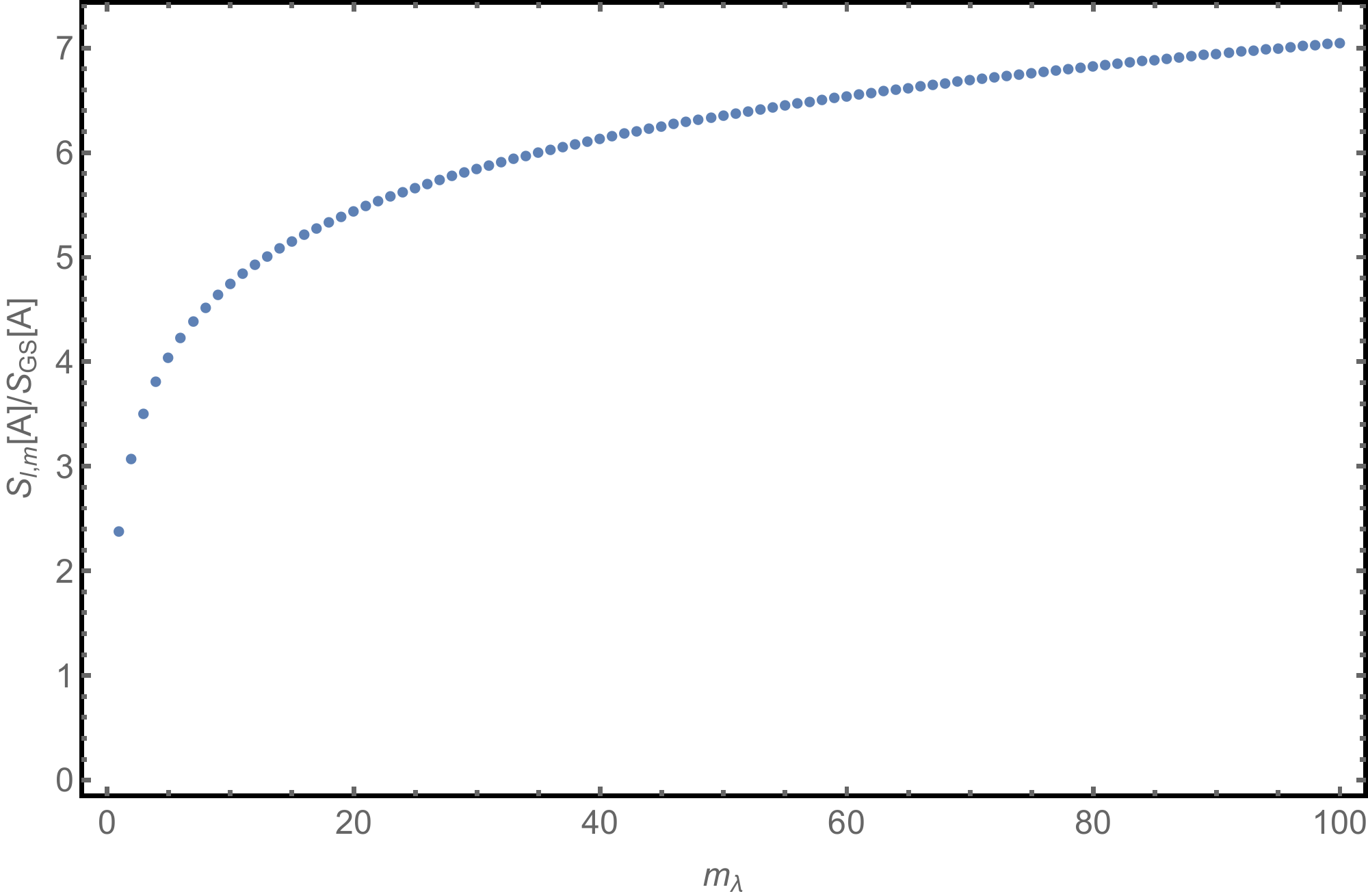} 
\end{subfigure}
\begin{subfigure}{.5\textwidth}
\centering
\includegraphics[width=\linewidth]{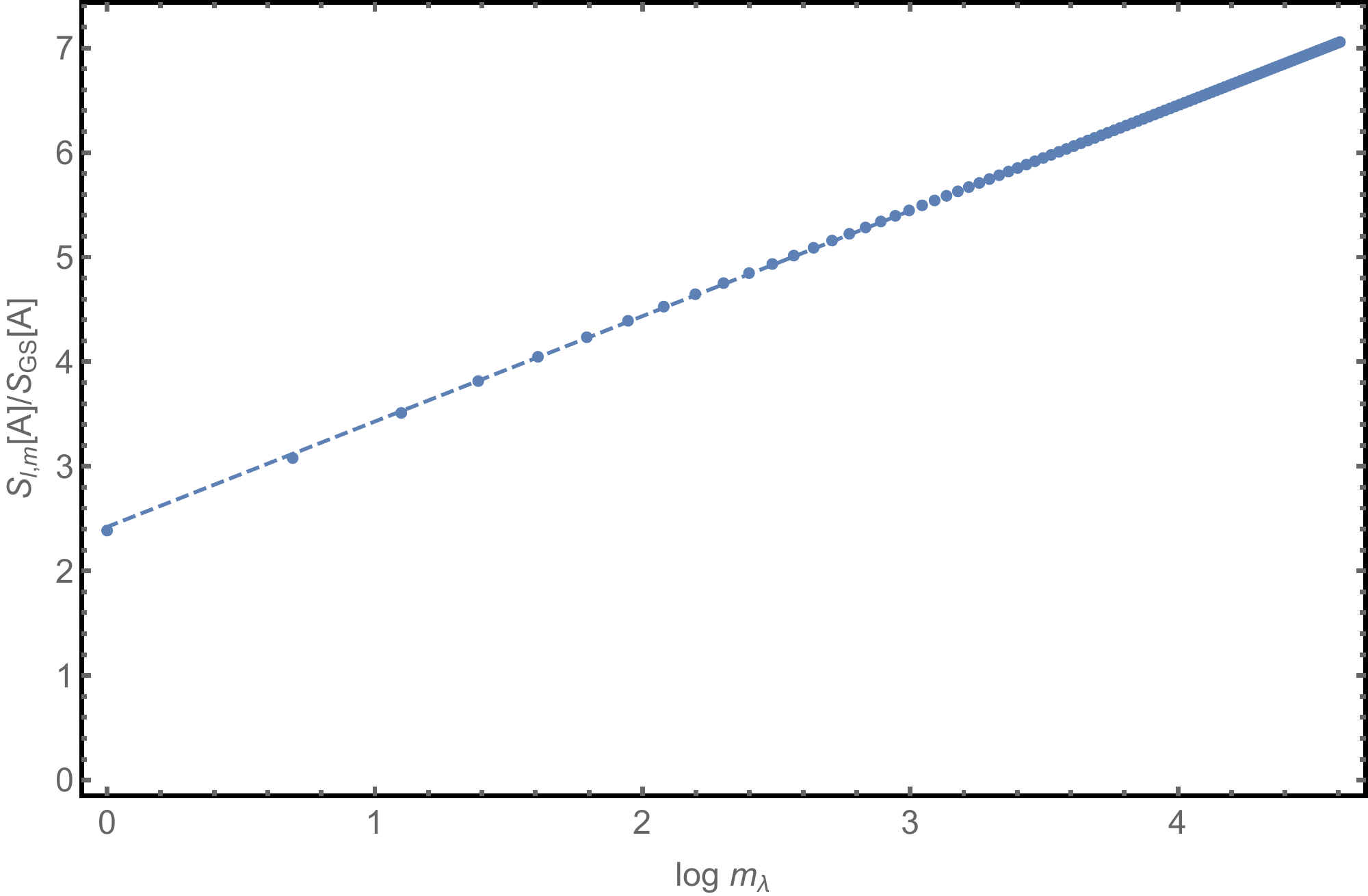} 
\end{subfigure}
\caption{In the first image $S_{m_\lambda}$ is plotted as a function of the number of excitations $m_\lambda$ for any mode satisfying $\ell+m=$odd and $m\neq 0$. We evaluate it at the RK-point $g=\frac{1}{8\pi}$ and for $R=1$. The second image is a log-linear plot of the same data, the dashed line represents the linear fit $y=1.008 x+2.421$ with an $R^2$ value above $0.999$.}
\label{fig:m-dependence-sphere-first-kind}
\end{figure}

During our analysis of the singly excited state, we observed that the transformed propagators, and thus the entanglement entropy, where highly dependent on whether the eigenmode on the complete manifold was also a eigenmode on the submanifold. In particular, all modes that were simultaneous eigenmodes on the manifold and submanifold had the same entanglement entropy. On the halved-sphere geometry these modes are characterized by having $\ell+m$ odd, and, as before, all of them have the same entanglement entropy. The calculations of the relevant transformed propagators are found in appendix \ref{appendix:transformed-propagators-hemisphere}, they are given by
\begin{align}
\label{eq:g-all-lambda-subspectrum-sphere}
\ell+m=\text{odd and }m\neq 0\qquad \Longrightarrow \qquad
\begin{cases}
&G_{A}^\lambda=G_{B}^\lambda=\frac{1}{2}\\
&G_{A,A}^\lambda=G_{B,B}^\lambda=1/4+\sigma_\lambda\\
&G_{A,B}^\lambda=1/4-\sigma_{\lambda}
\end{cases}
\end{align}
where we exclude the modes with $m=0$ to fulfill the assumption on the classical fields, and $\sigma_\lambda$ is defined in \eqref{eq:sigma-alpha-def-appendix}. We will see that the entanglement entropy is independent of $\sigma_\lambda$, so its exact value is irrelevant for our purposes. As we discussed for the singly excited state, the above expressions are also approximately valid for high angular momentum modes, that is modes in the limit $\ell\gg 1$. 
We can thus use the polynomial expressions \eqref{eq:tensor-A-polynomial} and \eqref{eq:tensor-M-polynomial} of the tensors appearing in the entanglement entropy for the state with $m_\lambda$ excitations in one of the allowed modes by
\begin{align}
\label{eq:tensors-halved-sphere}
\mathcal{A}^{k'}_{\ k}&=\mathcal{B}^{k'}_{\ k}= F^m_{k,k'}\!\left(\frac{1}{2}\right),&
\mathcal{M}^{k'}_{\ k}&=T^m_{k,k'}\!\left(\frac{1}{4}-\sigma_\lambda,\frac{1}{4}+\sigma_\lambda,\frac{1}{4}+\sigma_\lambda\right),
\end{align}
with $k,k'=0,\ldots,m_\lambda$. One can check, that for these values the product matrix $\A \B$ is diagonal and has the following simple form in terms of binomial coefficients
\begin{equation}
\A\B=\frac{1}{2^m}\mathrm{diag}\left[\binom{m}{0},\ldots,\binom{m}{k},\ldots,\binom{m}{m}\right].
\end{equation}
The matrix $M$ has lower triangular form and shares the same diagonal elements as $\A\B$, such that $\tr(\M) =1$. In particular, this shows that $\sigma_\lambda$ does not affect the entanglement entropy. Using the ground state entanglement entropy for this configuration, see \eqref{eq:hemisphere-ground-state-EE}, as well as our general result for the excited state entanglement entropy \eqref{eq:excited-state-entanglement-entropy}, we can write the entanglement entropy for $m_\lambda$ quanta of energy in an eigenmode with $\ell+m$ odd and $m\neq 0$ in the halved-sphere geometry as
\begin{align}
S_{m_\lambda}[H^2]&=S_{GS}[H^2]\Tr(\mathcal{M})+ \Tr(\log(\A\B)\mathcal{M})\nonumber\\
&=\log(\sqrt{8\pi g}R)-\frac{1}{2}+\sum_{k=0}^m\frac{1}{2^m}\binom{m}{k}\log\left(\frac{1}{2^m}\binom{m}{k}\right).
\end{align}
The equation above is also valid as an approximation for all modes with $\ell+m$ even in the limit of high angular momentum $\ell\gg 1$. We further note that this type of correction to the ground state entanglement was also observed in \cite{Moudgalya2018} for a similar class of states \cite{Moudgalya2018a} thought to be examples of many-body quantum scars.
In figure \ref{fig:m-dependence-sphere-first-kind} the value of the entanglement entropy is shown for different excitation levels.
As can be seen in the right image of the figure, a logarithmic behavior in $m$ dominates. We can thus provide the following approximation for the entanglement entropy
\begin{equation}
\label{eq:s-is-log-m}
S_{m_\lambda}[H^2]\approx S_{GS}[H^2]-a_1\log m_\lambda -a_2
\end{equation}
with $a_1\approx 0.504$ and $a_2\approx 0.711$. In particular, we learn that for highly excited states, that is for $m_\lambda\gg 1$, the entanglement entropy for this type of modes develops a logarithmic divergence in $m_\lambda$.

\subsection{Rectangular geometry}
\label{sec:excited-rectangle}
Let us turn again to the rectangular geometry of section \ref{sec:rectangular-geometry-single-excitation}, with $M$ a rectangle of side lengths $L_x$ and $L_y$ cut at $\ell_x$ into the two smaller rectangles $A$ and $B$. As was the case for the single excitation, the vanishing of the classical part of the field ensures that our assumption \eqref{eq:assumption-philambdaA-and-philambdaB-are-zero} is trivially satisfied for all modes. 

The transformed propagators needed to write down the explicit expression for the entanglement entropy for a general surgery are calculated in appendix \ref{appendix:G-rectangle}. As far as we know, there are no simple closed forms to be written down in this case, as both the transform Green's function and the entanglement entropy have an intricate functional dependence on the surgery parameter $\ell_x$. We thus refrain from repeating what general formulas we have stated up to now, and concentrate on a specific surgery.

\begin{figure}[h]
\begin{subfigure}{.5\textwidth}
\centering
\includegraphics[width=\linewidth]{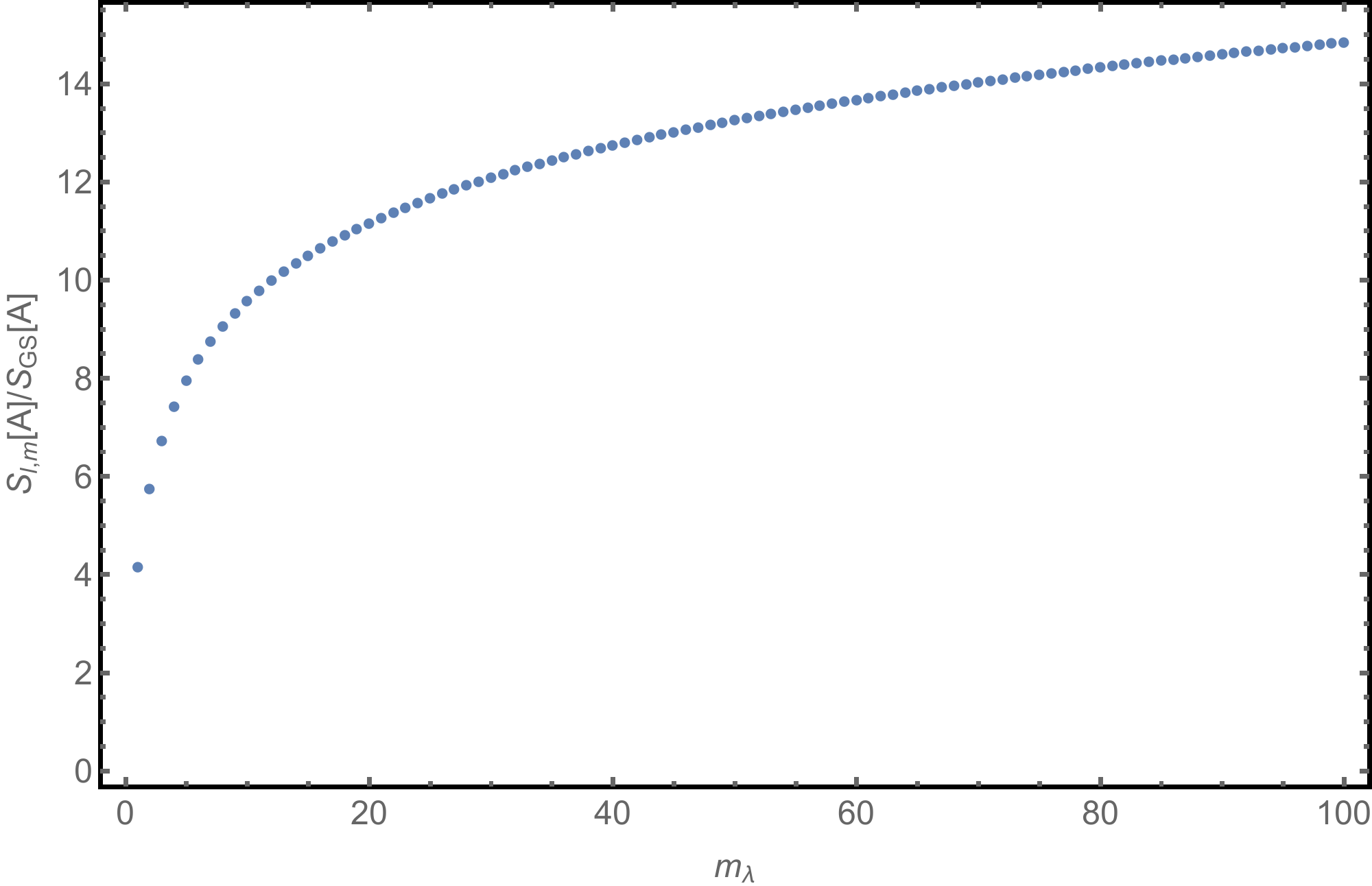} 
\end{subfigure}
\begin{subfigure}{.5\textwidth}
\centering
\includegraphics[width=\linewidth]{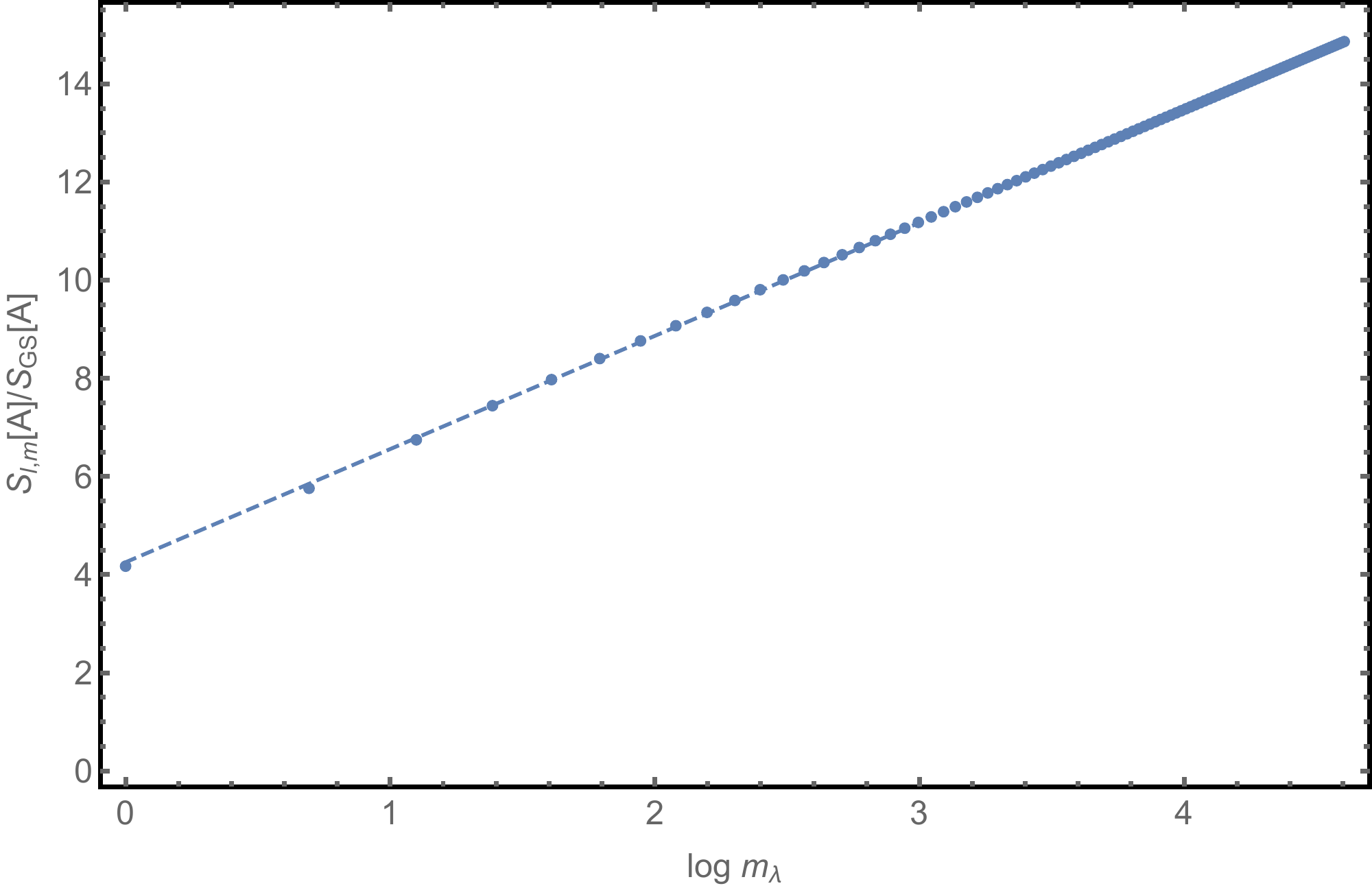} 
\end{subfigure}
\caption{In the first image $S_{m_\lambda}$ is plotted as a function of the number of excitations $m_\lambda$ for any mode satisfying $k_x$ even in the halved rectangle geometry with $L_x=L_y$. The second image is a log-linear plot of the same data. The dashed line represents the linear fit $y=2.306 x+4.252$ with an $R^2$ value above $0.999$.}
\label{fig:m-dependence-rectangle-first-kind}
\end{figure}

Let us choose the halved rectangle geometry, that is $\ell_x=L_x/2$. Here, for $k_x$ even, the eigenmodes on the complete rectangle are also eigenmodes on $A$ and $B$, and from the expressions for the propagators given in \ref{appendix:G-rectangle}, we get 
\begin{align}
\label{eq:g-all-lambda-subspectrum-rectangle}
k_x\text{ even}\qquad \Longrightarrow \qquad
\begin{cases}
&G_{A}^\lambda=G_{B}^\lambda=\frac{1}{2}\\
&G_{A,A}^\lambda=G_{B,B}^\lambda=1/4+\sigma_\lambda\\
&G_{A,B}^\lambda=1/4-\sigma_{\lambda}.
\end{cases}
\end{align}
From our discussion about the single excitation, we know the above equations to be approximately true for all modes with high wave number perpendicular to the entanglement cut, that is $k_x\gg 1$. From the point of view of the tensors, the situation here is exactly the same as on the halved sphere\footnote{Note that the value of $\sigma_\lambda$ on the rectangle a priori differs from that on the sphere. This is however not relevant, as $\sigma_\lambda$ doesn't contribute to the entanglement entropy.}, and the entanglement entropy only differs in the contribution from the ground state entanglement entropy which is now given by \eqref{eq:gs-ee-rectangle}. Thus, for even $k_x$  the excited state entanglement entropy with $m_\lambda$ quanta of energy in the mode labeled by $k_x$ and $k_y$ is given by
\begin{equation}
S_{m_\lambda}[H^2]= \frac{1}{2}\log(\frac{\eta\left(\frac{i}{2}\frac{L_x}{L_y}\right)^2}{\sqrt{2L_y}\eta\left(i\frac{L_x}{L_y}\right)})+\sum_{k=0}^m\frac{1}{2^m}\binom{m}{k}\log\left(\frac{1}{2^m}\binom{m}{k}\right).
\end{equation}
We note that we don't expect the contribution from the excited state to the entanglement entropy to be independent on the geometry in general. Here, this is a feature of the type of modes that we chose and also of the surgery. We do, however, expect the excited states corresponding to eigenmodes that are simultaneously eigenmodes on the full and halved submanifold to have an entanglement entropy of the above form for any similar ``halved'' geometry (for example halved cylinders or tori).
In figure \ref{fig:m-dependence-rectangle-first-kind} we show the dependence of $S_{m_\lambda}$ on $m_\lambda$. As expected, we see the same type of logarithmic behavior as for the sphere. In particular, equation \eqref{eq:s-is-log-m} is also valid here for all modes with $k_x$ even or $k_x\gg 1$, and replacing the ground state entanglement entropy corresponding to the sphere by that corresponding to the rectangle.


\section{Discussion}
We calculated the entanglement entropy of the states of the QLM obtained by exciting the eigenmodes of the Laplace-Beltrami on a compact spatial manifold. These are natural states for the model, as the ground state is given in terms of a conformal action determined by that same operator. Assuming that certain integrals of the classical fields vanish, see \eqref{eq:assumption-philambdaA+philambdaB-is-zero} and \eqref{eq:assumption-philambdaA-and-philambdaB-are-zero}, which can be interpreted as its partial projections onto the excited eigenmodes, we show that the replica calculation becomes tractable for even the most general excited state. This allows us to provide the elusive analytic continuations needed to find the entanglement entropy, which we do separately for the singly excited state -- where the assumption on the classical fields is a little less stringent -- and for the general case. 

We find that the ground state R\'enyi entropy factors from the excited state R\'enyi entropy in agreement with \cite{Parker2017}, and further confirm that these can be written down in terms of the EPA's from \cite{Parker2017}, although we write them slightly differently and refer to them as transformed propagators. We further find that, though the leading terms in the R\'enyi entropies of the singly excited states agree to those in \cite{Parker2017}, the other terms do not. We believe the differences to origin in the derivation of the generalized Wick's theorem by \cite{Parker2017}, although at the time of writing we have been unable to locate their root. In the general case, we find that the entanglement entropy is again written in terms of transformed propagators, though here they are packaged inside complicated tensors. We calculate the transformed propagators by spectral methods and provide explicit formulae in both the spherical and rectangular geometry. In the latter our results are in agreement with the expressions found in \cite{Parker2017} which adds to the evidence given by \cite{Parker2017} for the universality of the objects.

An interesting observation about these transformed propagators is that they distinguish between two classes of modes. The first class are modes that are, up to normalization, simultaneously eigenmodes on the full manifold and the submanifold. In the spherical geometry, for example, this condition is satisfied by the spherical harmonics with $\ell+m$ odd, which are both eigenfunctions of the Laplacian on the sphere and on the hemisphere with Dirichlet boundary conditions. The transformed propagators are the same for all of these modes. In particular, this implies that the entanglement entropy is constant within this class of modes. The second class of modes are those, for which the above condition is not satisfied. There we find a non-trivial dependence of the transformed propagators on the chosen mode, and thus also a dependence of the entanglement entropy on the modes. Looking back at equation \eqref{eq:excited-state-entanglement-entropy}, we see that the excited state entanglement entropy is related to the ground state entanglement entropy by two universal constants. Notably, this implies that the excited state entanglement entropy still obeys an area law. Keeping in mind that the two universal constants are derived from correlation functions on the different subsystems, our findings are similar to those in \cite{Alcaraz2011,Berganza2012,CastroAlvaredo2018,CastroAlvaredo2018a,Zhou2016a}, where excited states are constructed that obey area laws with corrections with respect to the ground state entanglement determined by certain correlation functions of the underlying theories. For the halved sphere and rectangle we observe that when $m$ quanta of energy are put into one mode, the entanglement entropy behaves logarithmically in $m$, that is $S_m\sim \log m$. For a highly excited state, we thus observe a logarithmic divergence in the excitation number, instead of the extensive behavior one would generically expect.

Our results and calculation can be applied to any bipartite compact geometry on which the spectrum of the Laplace-Beltrami operator on both the full manifold and the submanifolds is known. A generalization to higher dimensional geometries as in \cite{Angel-Ramelli2019} and any even critical exponent $z$ should also be relatively straight forward. The key element here being the replacement of the Laplace-Beltrami operator by the higher dimensional conformal generalization of one of its powers (for example a GJMS operator in the spherical case), whose eigenmodes would be excited. Up to some subtleties that are treated in \cite{Angel-Ramelli2019}, we expect our replica calculation to hold up also in these cases. While not as straight forward, a generalization to non-compact geometries would also be desirable. Even in that case, we would still expect our results to hold up for any excited state that can be written in the form \eqref{eq:excited-state-wave-function}. We would further like to find ways to relax our assumptions on the classical fields. As was already noted in \cite{Parker2017} the combinatorics of the problem become prohibitively difficult in the most general case. However a slight relaxation of our assumptions -- such as demanding the assumption \eqref{eq:assumption-philambdaA+philambdaB-is-zero} for all excited states instead of only the singly excited ones --  should be possible at least in some situations. Finally, a deeper exploration of the highly excited states of the model is desireable. In particular, we would like to better understand whether the area law behavior that we observe in the highly excited state with all excitations in one mode is in fact connected to a quantum scar, and, if this is the case, if this is a property of only that type of states. This also indicates that an analysis of the eigenstate thermalization hypothesis in the QLM might provide interesting results.

\section{Acknowledgements}
First of all I would like to thank Valentina Giangreco M. Puletti for her help and patience during this project. I also thank Watse Sybesma, Lukas Schneiderbauer, and L\'arus Thorlacius for their valuable comments on the manuscript, as well as Erik Tonni for his comments on an early version of the project, and Daniel E. Parker for his comments and in particular for pointing out the possible connection to quantum scars. I would also like to thank Cl\'ement Berthiere for providing me with an excellent template for figure \ref{fig:gluing-conditions}. Finally, I thank our research group at the University of Iceland for enduring more than one presentation on the topic with keen interest, and providing some helpful comments along the way. This research was supported in part by the Icelandic Research Fund under contract 163419-053, and by grants from the University of Iceland Research Fund.


\appendix 

\section{Rewriting $\Tr\rho_A^n$}
\label{appendix:path-integral-manipulations-for-single-excitation}

In this appendix we perform the path integral manipulations needed to get from \eqref{eq:tr-rho-a-n-single-2} to \eqref{eq:tr-rho-a-n-single-3}.
We start with the following factor that appears in \eqref{eq:tr-rho-a-n-single-2}
\begin{multline}
\label{eq:path-int-single-1-appendix}
I:=\int\limits_\mathcal{B'}\left[\prod\nolimits_{i=1}^{n}\D\varphi_{A,i}\D\varphi_{B,i}\right]\D\barphi_n^\cl\cross\\
\left(\prod_{i=1}^{n}(\varphi_{A,i}^\lambda+\varphi_{B,i}^\lambda)(\varphi^\lambda_{A,i+1}+\varphi^\lambda_{B,i})\right) e^{-S[\barphi^\cl_n]-\sum_{i=1}^{n} (S[\varphi_{A,i}]+S[\varphi_{B,i}])}.
\end{multline}
Let us take a closer look at the product of fields in the path integral. Keeping in mind the calculations of \cite{Zhou2016,Angel-Ramelli2019}, we want to separate the $n$-th field from the rest, and find out which terms survive the integration. Remembering that our assumption \eqref{eq:assumption-philambdaA+philambdaB-is-zero} implies $\phi_n^\lambda=\varphi_{A,n}^\lambda+\varphi_{B,n}^\lambda$ the product can be written as
\begin{multline}
\left(\prod_{i=1}^{n-1}(\varphi_{A,i}^\lambda+\varphi_{B,i}^\lambda)(\varphi^\lambda_{A,i+1}+\varphi^\lambda_{B,i})\right)(\varphi^\lambda_{A,1}+\varphi^\lambda_{B,n})\phi_n^\lambda \\
=\left(\prod_{i=1}^{n-1}(\varphi_{A,i}^\lambda)^2\varphi^\lambda_{A,n}\phi_n^\lambda\right)+\left(\prod_{i=1}^{n-1}(\varphi_{B,i}^\lambda)^2\varphi^\lambda_{B,n}\phi_n^\lambda\right)\\+(\text{at least one field appears alone}).
\end{multline}
When we perform the path integrals only the leading terms survive, since the other terms will contain something proportional to at least one one-point function of the fluctuations or the full field, which vanish. We can also glue the integration of the $n$-th field back together by
\begin{equation}
\int_{\mathcal{B}'}\D \barphi^\cl_n\D\varphi_{A,n} \D \varphi_{B,n}\ (\cdots)=\int \D\bar{\phi}_n\ (\cdots),
\end{equation}
to form an integral over a free field on the complete manifold (see \cite{Fradkin2006,Zaletel2011,Zhou2016}). This then leads us to 
\begin{multline}
\label{eq:path-int-single-2-appendix}
I=\left(\prod_{i=1}^{n-1}\int_{A_D}\D\varphi_{A,i}\ (\varphi_{A,i}^\lambda)^2 e^{-S[\varphi_{A,i}]}\right)\int\D\bar{\phi}_n\ \bar{\phi}_n^\lambda\bar{\phi}^\lambda_{A,n}e^{-S[\bar{\phi}_n]}+\\
+\left(\prod_{i=1}^{n-1}\int_{B_D}\D\varphi_{B,i}\ (\varphi_{B,i}^\lambda)^2 e^{-S[\varphi_{B,i}]}\right)\int\D\bar{\phi}_n\ \bar{\phi}_n^\lambda\bar{\phi}^\lambda_{B,n} e^{-S[\bar{\phi}_n]},
\end{multline}
where $A_D$ and $B_D$ remind us that those path integrals are taken over $A$ and $B$ respectively and with Dirichlet boundary conditions, and where we also used the fact that
\begin{align}
&\int\D\bar{\phi}_n\  \bar{\phi}_n^\lambda\barphi^\lambda_{A,n}e^{-S[\bar{\phi}_n]}\nonumber\\
\quad&=\int\D\varphi_{A,n}\D\varphi_{B,n}\D\barphi_n^\cl \left[(\phi^\cl_{A,n})^\lambda+\varphi_{A,n}^\lambda\right]\left[\varphi_{A,n}^\lambda+\varphi_{B,n}^\lambda\right]e^{-S[\barphi^\cl_n]-S[\varphi_{A,n}]-S[\varphi_{B,n}]}\nonumber\\
\quad&=\int\D\bar{\phi}_n\  \bar{\phi}_n^\lambda\varphi^\lambda_{A,n}e^{-S[\bar{\phi}_n]}+\nonumber\\
&\quad\quad\quad\quad \underbrace{\int\D\varphi_{A,n}\D\varphi_{B,n}\D\barphi_n^\cl(\phi^\cl_{A,n})^\lambda\left[\varphi_{A,n}^\lambda+\varphi_{B,n}^\lambda\right]e^{-S[\barphi^\cl_n]-S[\varphi_{A,n}]-S[\varphi_{B,n}]}}_{=0}\nonumber\\
\quad&=\int\D\bar{\phi}_n\  \bar{\phi}_n^\lambda\varphi^\lambda_{A,n}e^{-S[\bar{\phi}_n]}.
\end{align}
In the second to last line the term vanishes because it is proportional to a one-point function of the fluctuations on either $A$ or $B$. If we furthermore notice that all the fields in the path integrals are  dummy fields that are integrated over, we can get rid of the replica indices and write
\begin{multline}
\label{eq:path-int-single-final-appendix}
I=
\left(\int_{A_D}\D\varphi_{A}\ (\varphi_{A}^\lambda)^2 e^{-S[\varphi_{A}]}\right)^{n-1}\int\D\bar{\phi}\ \bar{\phi}^\lambda\bar{\phi}^\lambda_A e^{-S[\bar{\phi}]}+\\
+\left(\int_{B_D}\D\varphi_{B}\ (\varphi_{B}^\lambda)^2 e^{-S[\varphi_{B}]}\right)^{n-1}\int\D\bar{\phi}\ \bar{\phi}^\lambda\bar{\phi}^\lambda_B e^{-S[\bar{\phi}]}, 
\end{multline}
which is precisely the factor that appears in \eqref{eq:tr-rho-a-n-single-3}.


\section{Transformed Green's functions}
\label{appendix:transformed-greens-functions}
We calculate the transformed Green's functions $G^\lambda_{X,M}$ and $G^\lambda_{X}$ by means of their eigenvalue expansions as well as the orthogonality and completeness of the eigenmodes of the Laplacian. We denote the eigenmodes of the Laplacian on $M$ by $L_\lambda(x)$, and on $X=A,B$ as $L_{\alpha}^X(x)$, where we always take Dirichlet boundary conditions. In order to avoid confusion, we index eigenmodes and eigenvalues on $M$ by $\lambda$ and $\mu$, and eigenmodes and eigenvalues on $A$ and $B$ by $\alpha$ and $\beta$. The Green's function $G_M(x,x')$ has an eigenvalue expansion of the form
\begin{equation}
\label{eq:eigenvalue-expansion-greens-appendix}
G_M(x,x')=\sum_{\lambda} \frac{L_{\lambda}(x) L_{\lambda}(x')}{{\lambda}},
\end{equation}
where the sum runs over the complete set of eigenmodes of the Laplacian on $M$, and similarly we have the eigenvalue expansion on $X=A,B$
\begin{equation}
\label{eq:eigenvalue-expansion-greens-A,B-appendix}
G_X(x,x')=\sum_{\alpha} \frac{L^{X}_{\alpha}(x) L^X_{\alpha}(x')}{{\alpha}},
\end{equation}  
where the sum now runs over the complete set of eigenmodes of the Laplacian on either $A$ or $B$.
Let us take a look at the different transformed Green's functions in \eqref{eq:g-lambda-definitions}. 
The integral in $G_M^\lambda$ is easily evaluated by the orthonormality of the eigenfunctions leading to 
\begin{equation}
\label{eq:g-m-lambda-appendix}
G_M^\lambda =\lambda\sum_\mu \frac{1}{\mu}\left(\int_M\diff{2}{x}\ L_\lambda(x)L_\mu(x)\right)^2=\lambda\sum_\mu\frac{\delta_{\lambda\mu}}{\mu}=1.
\end{equation}
We can proceed with $G_{X,M}^\lambda$ in a similar fashion and write
\begin{align}
G_{X,M}^\lambda&=\lambda \int_M \diff{2}x\int_X\diff{2}x' L_\lambda(x)L_\lambda(x') G_M(x,x')\nonumber\\
&=\lambda \int_M \diff{2}x\int_X\diff{2}x' L_\lambda(x)L_\lambda(x')\sum_{\mu} \frac{L_{\mu}(x) L_{\mu}(x')}{{\mu}}\nonumber\\
&=\lambda\int_X\diff{2}x' L_\lambda(x')\sum_\mu\frac{L_\mu(x')}{\mu}\int_M \diff{2}x L_\lambda(x)L_\mu(x)\nonumber\\
&=\lambda\int_X\diff{2}x' L_\lambda(x')\sum_\mu\frac{L_\mu(x')}{\mu}\delta_{\lambda\mu}\nonumber\\
\label{eq:g-x-m-lambda-general-appendix}
&=\int_X\diff{2}x' (L_\lambda(x'))^2.
\end{align}
Using the expansion \eqref{eq:eigenvalue-expansion-greens-A,B-appendix} we can rewrite $G_X^\lambda$ as
\begin{align}
G_X^\lambda&=\lambda\int_X\diff{2}{x}\diff{2}{x'}L_\lambda(x)L_\lambda(x')G_X(x,x')\nonumber\\
&= \lambda\int_X\diff{2}{x}\diff{2}{x'}L_\lambda(x)L_\lambda(x')\sum_{\beta} \frac{L^{X}_{\beta}(x) L^X_{\beta}(x')}{{\beta}}\nonumber\\
\label{eq:g-x-lambda-general-appendix}
&=\lambda\sum_\beta\frac{1}{\beta}\left(\int_X\diff{2}{x}\ L_\lambda(x)L^X_\beta(x)\right)^2.
\end{align}

Let us take a look at the specific situation, where the eigenmodes $\alpha$ on the submanifold are simultaneously eigenmodes on the full manifold (i.e. $\exists \lambda$ s.t. $\alpha=\lambda$). In other words,
this is the case where some of the eigenmodes on $M$ coincide, up to normalization, with the eigenmodes on $A$ and $B$. Here, we have a notion of orthogonality between $L_\lambda(x)$ and $L^X_\alpha(x)$. In particular we can write $c_\alpha^X L^X_\alpha(x)= L_\alpha(x)$, where $c_\alpha^X$ accounts for the different normalization of the eigenmodes. For this type of mode we can, for example, evaluate the following integral exactly
\begin{equation}
\label{eq:c-x-lambda-subspectrum-def}
\int_X\diff{2}x\ L_\alpha(x)L_\beta(x)=\delta_{\alpha\beta} \left(c_\alpha^X\right)^2,
\end{equation}
allowing us to evaluate \eqref{eq:g-x-m-lambda-general-appendix}
\begin{equation}
\label{eq:g-x-m-lambda-subspectrum-appendix}
G_{X,M}^\alpha=(c_\alpha^X)^2.
\end{equation}
We can also continue the calculation in \eqref{eq:g-x-lambda-general-appendix} and write
\begin{equation}
\label{eq:g-x-lambda-subspectrum-appendix}
G_X^\alpha=\alpha\sum_\beta \frac{(c_\beta^X)^2}{ \beta}\left(\int_X\diff{2}{x}\ L^X_\alpha(x)L^X_\beta(x)\right)^2=(c_\alpha^X)^2,
\end{equation}
Finally, we can rewrite $G_{A,B}^\lambda$ using the same techniques
\begin{align}
G_{A,B}^\alpha&=\alpha \int_A \diff{2}x\int_B\diff{2}x' L_\alpha(x)L_\alpha(x') G_M(x,x')\nonumber\\
&=\alpha \int_A \diff{2}x\int_B\diff{2}x' L_\alpha(x)L_\alpha(x')\sum_{\mu} \frac{L_{\mu}(x) L_{\mu}(x')}{{\mu}}\nonumber\\
\label{eq:g-a-b-lambda-general-appendix}
&=\alpha\sum_\mu \frac{1}{\mu} \left(\int_A \diff{2}x\  L_\alpha(x)L_\mu(x)\right)\left(\int_B \diff{2}x\  L_\alpha(x)L_\mu(x)\right)\nonumber\\
&= \left(c_\alpha^A\right)^2\left(c_\alpha^B\right)^2+\sigma_\alpha^{A,B},
\end{align}
where $\sigma_\alpha^{A,B}$ is defined as
\begin{equation}
\sigma_\alpha^{A,B}:=\alpha\sum_{\mu\in S}\frac{1}{\mu} \left(\int_A \diff{2}x\  L_\alpha(x)L_\mu(x)\right)\left(\int_B \diff{2}x\  L_\alpha(x)L_\mu(x)\right).
\end{equation}
Here, $S$ is the set of eigenmodes on $M$ that are \emph{not} eigenmodes on $A$ and $B$. 
Similarly for $X=A,B$
\begin{equation}
\label{eq:g-x-x-lambda-general-appendix}
G_{X,X}^\alpha=\left(c_\alpha^X\right)^4+\sigma_\alpha^{X,X}
\end{equation}
with 
\begin{equation}
\sigma_\alpha^{X,X}:=\alpha\sum_{\mu\in S}\frac{1}{\mu} \left(\int_X \diff{2}x\  L_\alpha(x)L_\mu(x)\right)^2.
\end{equation}
Note that 
\begin{align}
\int_A\diff{2}x\ L_\alpha(x)L_\mu(x)&=\int_M\diff{2}x\ L_\alpha(x)L_\mu(x)-\int_B\diff{2}x\ L_\alpha(x)L_\mu(x)\nonumber\\
&=-\int_B\diff{2}x\ L_\alpha(x)L_\mu(x),
\end{align}
where in the second line we used the fact that the integral over $M$ is proportional to $\delta_{\alpha\mu}$, but $\mu\neq\alpha$ since we are only summing over modes $\mu$ that are not eigenmodes on $A$ and $B$. In particular, we can define 
\begin{equation}
\label{eq:sigma-alpha-def-appendix}
\sigma_\alpha:=\sigma_\alpha^{A,A}=\sigma_\alpha^{B,B}=-\sigma_\alpha^{A,B}.
\end{equation}
Next we discuss the calculation of the transformed Green's functions for specific geometries.

\subsection{Spheres and hemispheres}
\label{appendix:transformed-propagators-hemisphere}
Let us consider the situation when a sphere $M=S^2$ is cut into the northern hemisphere $A=H_{N}^2$ and the southern hemisphere $B=H^2_S$ along the equator. The relevant eigenfunctions are then $L^{S^2}_{\ell,m}(x)=Y^m_l(x)$ on the sphere and $L^{H^2}_{\ell,m}(x)=\sqrt{2}Y^m_\ell(x)$ with $m+\ell$ odd on the hemispheres with Dirichlet boundary conditions. Thus the eigenmodes on the hemisphere are also eigenmodes of the sphere, and in our notation $\lambda=(\ell,m)$ and $\alpha=(\ell,m)$ with $\ell+m$ odd. We note that in this case $c_\alpha^{H^2}=\frac{1}{\sqrt{2}}$. Using equations \eqref{eq:g-m-lambda-appendix}, \eqref{eq:g-x-m-lambda-subspectrum-appendix}, \eqref{eq:g-x-lambda-subspectrum-appendix}, \eqref{eq:g-a-b-lambda-general-appendix}, \eqref{eq:g-x-x-lambda-general-appendix}, and \eqref{eq:sigma-alpha-def-appendix} we find
\begin{align}
\label{eq:g-all-lambda-subspectrum-sphere-appendix}
\ell+m=\text{odd}\qquad \Longrightarrow \qquad
\begin{cases}
&G_M^\alpha=1\\
&G_{A,M}^\alpha=G_{B,M}^\alpha=G_{A}^\alpha=G_{B}^\alpha=\frac{1}{2}\\
&G_{A,A}^\alpha=G_{B,B}^\alpha=1/4+\sigma_\alpha\\
&G_{A,B}^\alpha=1/4-\sigma_{\alpha}
\end{cases}
\end{align}
Let us take a look at the other modes, that is those with $\ell+m$ even. We remind the reader that the spherical harmonics transform as $Y^m_l(\pi-\theta,\phi)=(-1)^{l+m} Y^m_l(\theta,\phi)$ under reflection over the equatorial plane. In particular this implies that 
\begin{equation}
\int_A\diff{2}{x}\ (L_\lambda(x))^2 = \int_B\diff{2}{x}\ (L_\lambda(x))^2 = \frac{1}{2}\int_M\diff{2}{x}\ (L_\lambda(x))^2=\frac{1}{2} 
\end{equation}
and thus from equation \eqref{eq:g-x-m-lambda-general-appendix} we see that $G_{A,M}^\lambda=G_{B,M}^\lambda=\frac{1}{2}$ is actually true for all modes. Furthermore it isn't hard to see from the reflection properties and normalization that and that $G_{X,X}^\lambda=G_X^\lambda/2$ and $G_{A,B}^\lambda=-G_X^\lambda/2$. Thus finding $G_X^\lambda$ provides us with the remaining propagators. The transformed propagators $G_X^\lambda$ are given by
\begin{align}
G_X^\lambda&=\lambda\sum_\alpha \frac{1}{\alpha}\left(\int_X\diff{2}{x}\ L_\lambda(x)L^X_\alpha(x)\right)^2\nonumber\\
&=2\ \ell_\lambda(\ell_\lambda+1) \sum_{\substack{\ell_\mu,m_\mu \\ \ell_\mu+m_\mu=\text{odd}}}\frac{1}{\ell_\mu(\ell_\mu+1)}\left(\int_X\diff{2}{x}\ Y^{m_\lambda}_{\ell_\lambda}(x)\left(Y^{m_\mu}_{\ell_\mu}(x)\right)^*\right)^2,
\end{align}
where $\ell_\lambda+m_\lambda=$even. Using the definition of the spherical harmonics 
\begin{equation}
Y^m_\ell(\theta,\phi)=N(\ell,m) e^{im\phi} P^m_\ell(\cos(\theta)),
\end{equation}
 where $N$ is the normalization
\begin{equation}
N(\ell,m)=\sqrt{\frac{2\ell+1}{4\pi}\frac{(\ell-m)!}{(\ell+m)!}},
\end{equation}
and $P^m_\ell(x)$ are associated Legendre polynomials, we can evaluate part of the integral and write
\begin{multline}
\int_A\diff{2}{x}\ Y^{m_\lambda}_{\ell_\lambda}(x)\left(Y^{m_\mu}_{\ell_\mu}(x)\right)^*= -\int_B\diff{2}{x}\ Y^{m_\lambda}_{\ell_\lambda}(x)\left(Y^{m_\mu}_{\ell_\mu}(x)\right)^*=\\
 2\pi \delta_{m_\lambda,m_\mu} N(\ell_\lambda,m_\lambda)N(\ell_\mu,m_\lambda) \int_0^{1}\diff{}{x}\ P^{m_\lambda}_{\ell_\lambda}(x)P^{m_\lambda}_{\ell_\mu}(x).
\end{multline}
Unfortunately, to our best knowledge, there is no simple closed form for the integral over the associated Legendre polynomials. We can however write the transformed propagator as
\begin{equation}
\label{eq:G-A-lambda-hemisphere-value-appendix}
G_A^\lambda=G_B^\lambda = \frac{1}{2}\Sigma_\lambda,
\end{equation}
with 
\begin{equation}
\label{eq:Sigma-lambda-hemisphere-appendix}
\Sigma_\lambda=
\begin{cases}
\Sigma^\text{e}_\lambda,&\qquad \ell_\lambda+m_\lambda=\text{even}\\
 1, &\qquad \ell_\lambda+m_\lambda=\text{odd},
\end{cases}
\end{equation}
and where for $\ell_\lambda+m_\lambda=$ even we defined
\begin{multline}
\label{eq:Sigma-lambda-def-appendix}
\Sigma_\lambda^e\equiv \Sigma^e_{\ell_\lambda,m\lambda}:=(2\ell_\lambda+1)\ell_\lambda(\ell_\lambda+1)\cross \\ \frac{(\ell_\lambda-m_\lambda)!}{(\ell_\lambda+m_\lambda)!}\sum_{\substack{\ell_\mu\geq m_\lambda\\ \ell_\mu+m_\lambda=\text{odd} }}^\infty\frac{(2\ell_\mu+1)}{\ell_\mu(\ell_\mu+1)}\frac{(\ell_\mu-m_\lambda)!}{(\ell_\mu+m_\lambda)!}\left( \int_0^{1}\diff{}{x}\ P^{m_\lambda}_{\ell_\lambda}(x)P^{m_\lambda}_{\ell_\mu}(x)\right)^2.
\end{multline}
Note that we used the fact that $P^{m_\lambda}_{\ell_\mu}=0$ for $\ell_\mu<m_\lambda$ to alter the starting point of the sum. We also note that the conditions that $m_\lambda+\ell_\mu$ be odd is equivalent to demanding that $\ell_\mu$ be odd whenever $\ell_\lambda$ is even and vice-versa. We further note that $\Sigma^{e}_{\ell,-m}=\Sigma^{e}_{\ell,m}$, due to the transformation behavior of associated Legendre polynomials when $m\mapsto -m$:
\begin{equation}
P^{-m}_\ell(x)=(-1)^m\frac{(\ell-m)!}{(\ell+m)!}P^m_\ell(x).
\end{equation}
The value of $\Sigma_\lambda$ can be determined numerically from this expression by truncating the sum at some sufficiently large value and performing a numerical integration at each step. We note that the sum converges for all desired values.

\subsection{Rectangles}
\label{appendix:G-rectangle}
We now consider the case described in \ref{sec:rectangular-geometry-single-excitation}, that is $M=[0,L_x]\cross [0,L_y]$, $A=[0,\ell_x]\cross[0,L_y]$, and $B=[\ell_x,L_x]\cross[0,L_y]$. As before, $G_M^\lambda=1$. We start with $G_{X,M}^\lambda$
\begin{equation}
G_{X,M}^\lambda=\int_X\diff{2}x' (L_\lambda(x'))^2.
\end{equation}
Using \eqref{eq:rectangle-eigenmodes} for the eigenmodes, the integral can be easily evaluated and gives
\begin{align}
\label{eq:rectangle-g-a-m-lambda-appendix}
G_{A,M}^\lambda &= \frac{\ell_x}{L_x}-\frac{1}{2\pi k_x}\sin(2\pi k_x\frac{\ell_x}{L_x}), \\
\label{eq:rectangle-g-b-m-lambda-appendix}
G_{B,M}^\lambda &=\left(1-\frac{\ell_x}{L_x}\right)+\frac{1}{2\pi k_x}\sin(2\pi k_x\frac{\ell_x}{L_x}).
\end{align}
Now we can tackle $G_X^\lambda$. We take, as before, $L_\lambda$ to be eigenfunctions on $M$ and $L^X_\alpha$ eigenfunctions on $X=A,B$, and start by considering $G_A^\lambda$. In order to calculate
\begin{equation}
G_A^\lambda=\lambda\sum_\alpha \frac{1}{\alpha}\left(\int_A\diff{2}{x}\ L_\lambda(x)L^A_\alpha(x)\right)^2,
\end{equation}
we first need to evaluate the integral. Due to the surgery we perform on $M$ the eigenmodes on $M$ and $A$ have the same $y$ dependence. In particular, this means that the $y$ integral only contributes a $\delta$-function and we can write
\begin{align}
\int_A\diff{2}{x}\ L_\lambda(x)L^A_\alpha(x) &\equiv \int_A\diff{2}{x}\ L_{k_x,k_y}(x)L^A_{n_x,n_y}(x)\nonumber\\
&=\frac{2}{\sqrt{\ell_x L_x}}\delta_{k_y,n_y}\int_0^{\ell_x}\diff{}{x} \sin(\pi k_x \frac{x}{L_x})\sin(\pi n_x \frac{x}{\ell_x})\nonumber\\
&= \frac{2}{\sqrt{\ell_x L_x}} (-1)^{n_x}\sin(\pi k_x \frac{\ell_x}{L_x})\frac{\frac{\pi}{\ell_x}n_x}{\left(\frac{\pi}{\ell_x}n_x\right)^2-\left(\frac{\pi}{L_x}k_x\right)^2}.
\end{align}
With this integral we can conclude the evaluation of $G_{A}^\lambda$
\begin{align}
G_A^\lambda &= \frac{4}{\ell_x L_x} \left(\left(\frac{\pi}{L_x}k_x\right)^2+\left(\frac{\pi}{L_y}k_y\right)^2\right)\sin(\pi k_x \frac{\ell_x}{L_x})^2\cross\nonumber\\
&\qquad\qquad \qquad\qquad \qquad\sum_{n_x=1}^\infty \frac{1}{\left(\frac{\pi}{\ell_x}n_x\right)^2+\left(\frac{\pi}{L_y}k_y\right)^2}\frac{\left(\frac{\pi}{\ell_x}n_x\right)^2}{\left(\left(\frac{\pi}{\ell_x}n_x\right)^2-\left(\frac{\pi}{L_x}k_x\right)^2\right)^2}\nonumber\\
\label{eq:G-a-rectangle-appendix}
&= \frac{\ell_x}{L_x}-\frac{1}{2\pi k_x} \sin(2\pi k_x \frac{\ell_x}{L_x})- d_\lambda^A,
\end{align}
where we defined 
\begin{equation}
\label{eq:d-lambda-A-appendix}
d_\lambda^A:=\frac{1}{L_x}\frac{2\left(\frac{\pi}{L_y}k_y\right)\coth(\pi k_y\frac{\ell_x}{L_y})\sin(\pi k_x\frac{\ell_x}{L_x})^2-\left(\frac{\pi}{L_x}k_x\right)\sin(2\pi k_x \frac{\ell_x}{L_x})}{\left(\frac{\pi}{L_x}k_x\right)^2+\left(\frac{\pi}{L_y}k_y\right)^2}.
\end{equation}
We can proceed similarly with $G^\lambda_B$ and get
\begin{equation}
\label{eq:G-b-rectangle-appendix}
G^\lambda_B=\left(1-\frac{\ell_x}{L_x}\right)+\frac{1}{2\pi k_x}\sin(2\pi k_x \frac{\ell_x}{L_x})-d_\lambda^B,
\end{equation}
with 
\begin{equation}
\label{eq:d-lambda-B-appendix}
d_\lambda^B:=\frac{1}{L_x}\frac{2\left(\frac{\pi}{L_y}k_y\right)\coth(\pi k_y\frac{L_x-\ell_x}{L_y})\sin(\pi k_x\frac{\ell_x}{L_x})^2+\left(\frac{\pi}{L_x}k_x\right)\sin(2\pi k_x \frac{\ell_x}{L_x})}{\left(\frac{\pi}{L_x}k_x\right)^2+\left(\frac{\pi}{L_y}k_y\right)^2}.
\end{equation}


\section{EPA's and transformed Green's functions on the rectangle} 
\label{appendix:epa-rectangle}
\begin{figure}[h]
\begin{subfigure}{.5\textwidth}
\centering
\includegraphics[width=.9\linewidth]{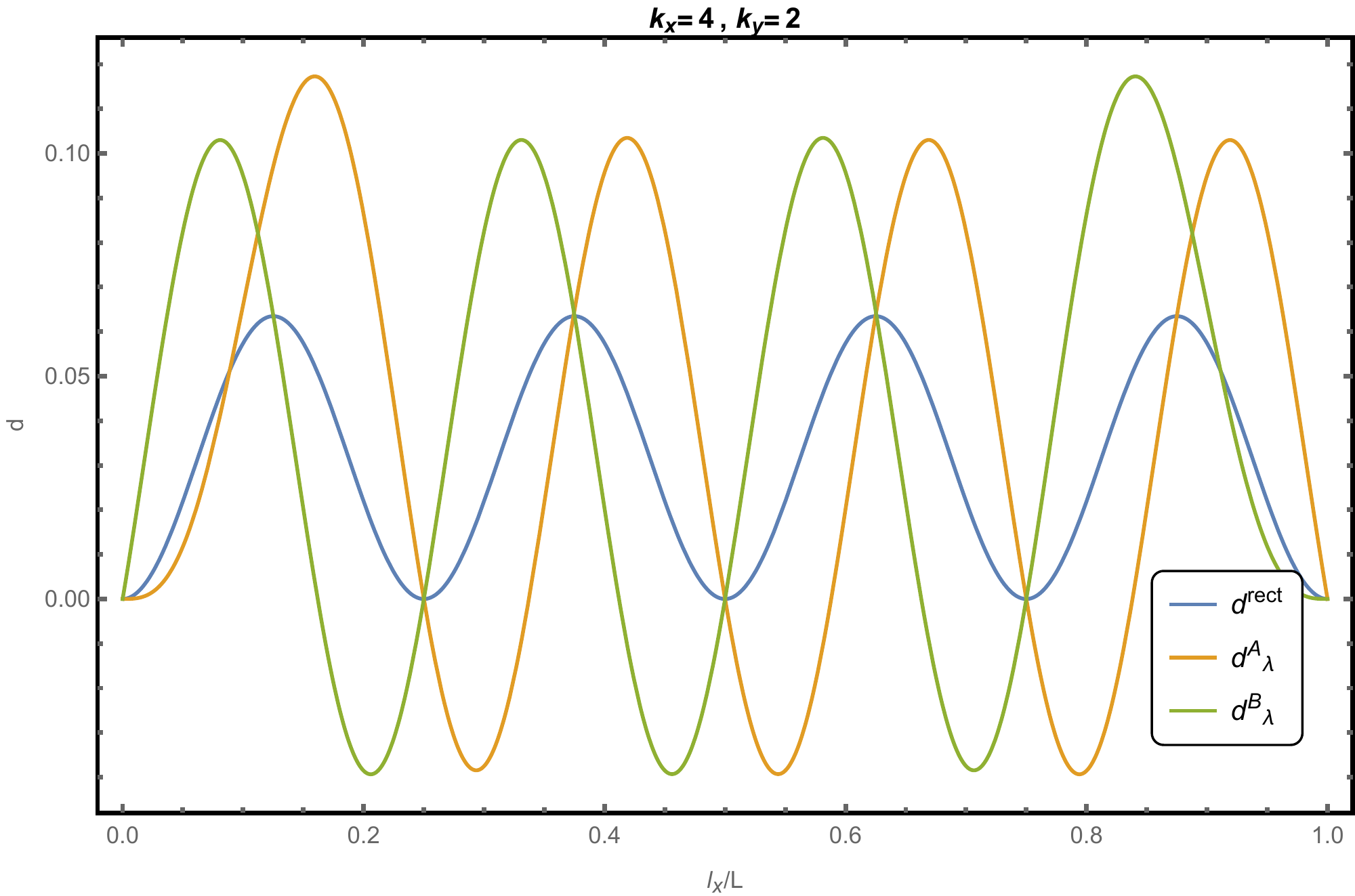} 
\end{subfigure}
\begin{subfigure}{.5\textwidth}
\centering
\includegraphics[width=.9\linewidth]{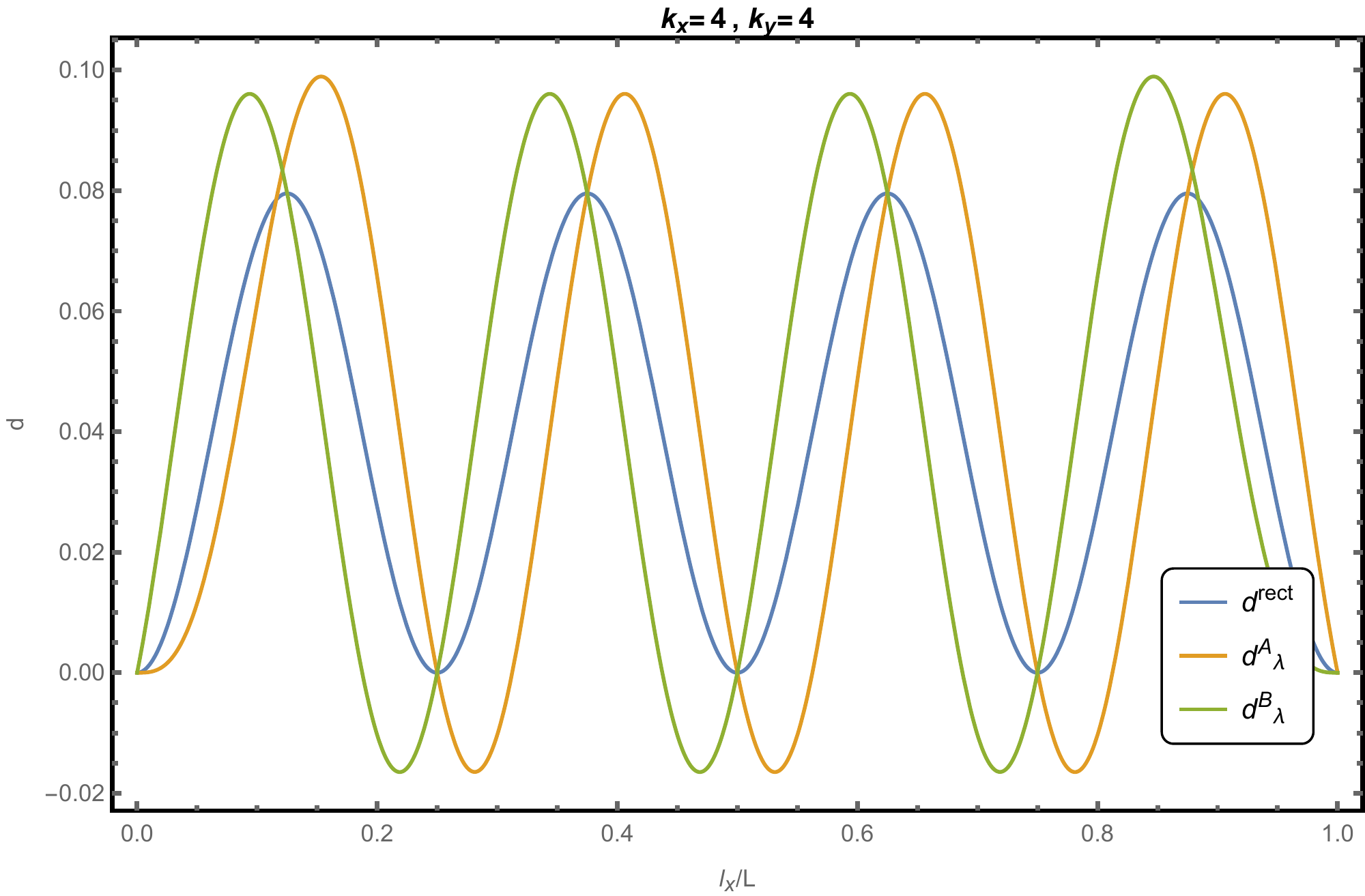} 
\end{subfigure}
\begin{subfigure}{.5\textwidth}
\centering
\includegraphics[width=.9\linewidth]{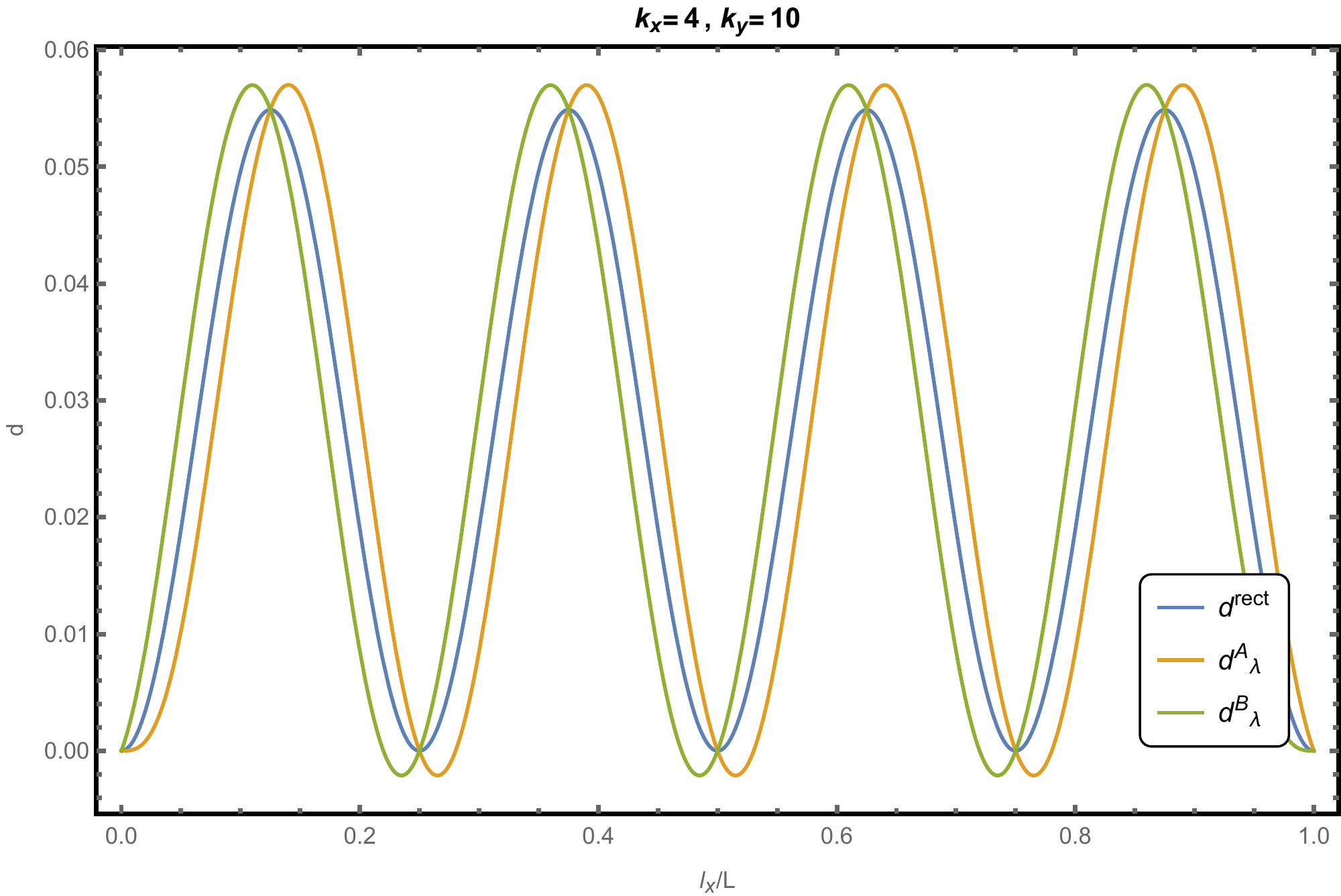} 
\end{subfigure}
\begin{subfigure}{.5\textwidth}
\centering
\includegraphics[width=.9\linewidth]{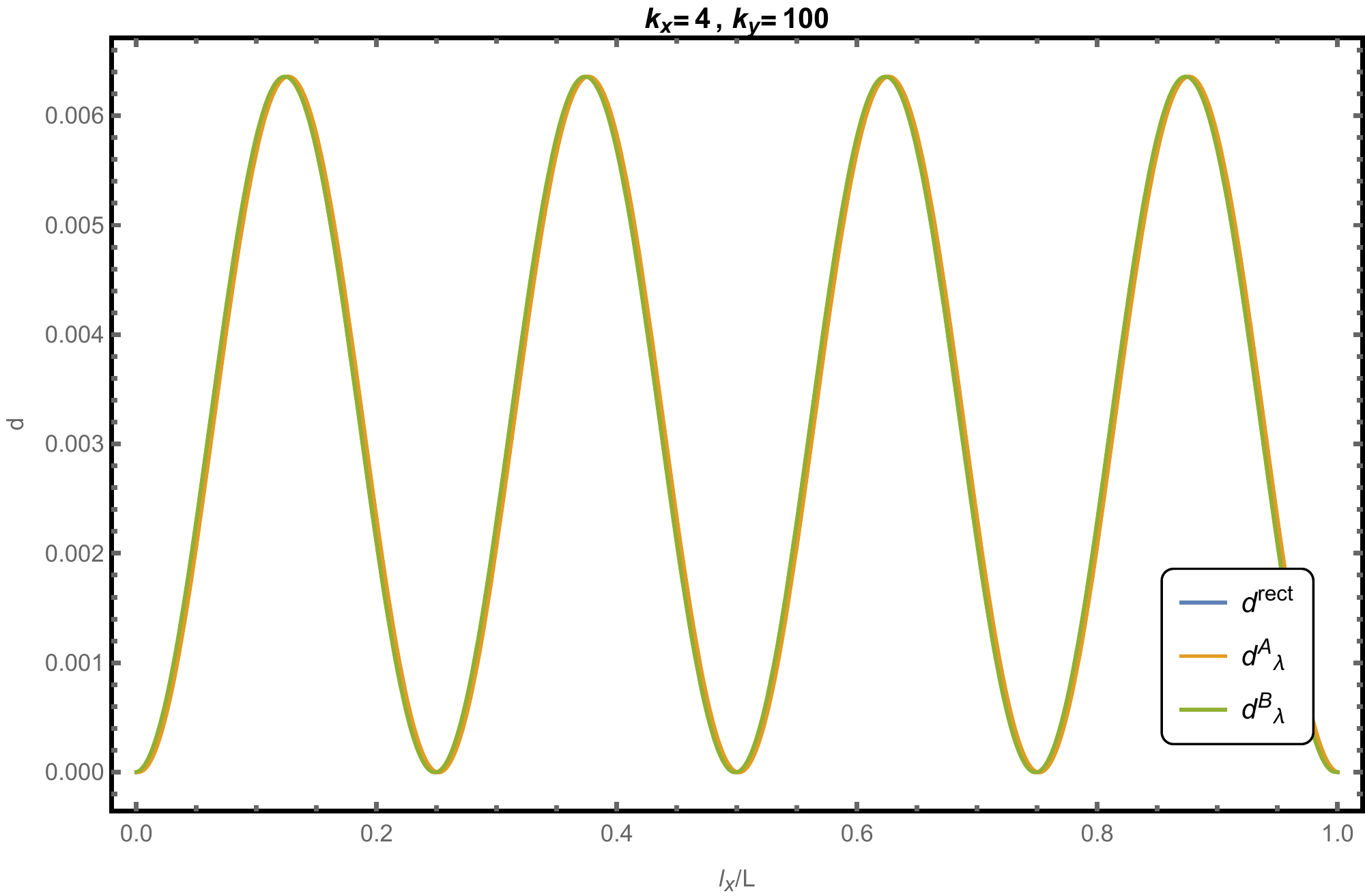} 
\end{subfigure}
\caption{The quantity $d^\text{rect}$ calculated by \cite{Parker2017} is compared to our exact results $d^A_\lambda$ and $d^B_\lambda$ for constant $k_x=4$ and $k_y=2,4,10,100$. The approximation used to calculate $d^\text{rect}$ is only valid at large $k_y$, there it agrees with our result. }
\label{fig:comparison-d-parker}
\end{figure}
In \cite{Parker2017} quantities dubbed Entanglement Propagator Amplitudes (EPA's) are introduced. These are closely related to our transformed propagators. In fact, the first two EPA's are, up to a different normalization, just $G^\lambda_A$ and $G^\lambda_B$
\begin{align}
\alpha&=G^\lambda_A\\
\beta&=G^\lambda_B.
\end{align}
The remaining EPA can be expressed in terms of our quantities as 
\begin{equation}
\gamma=G^\lambda_M-G^\lambda_A-G^\lambda_B=1-G^\lambda_A-G^\lambda_B.
\end{equation}
As an example these quantities are calculated explicitly on a rectangle with side lengths $L_x$ and $L_y$ and an entanglement cut at $\ell_x$, where the the eigenmodes of the Laplacian are labeled by the positive integers $k_x$ and $k_y$ and correspond to the Fourier modes on the rectangle. In \cite{Parker2017} the following expression is given for the transformed propagators
\begin{align}
G_A^\lambda&=\frac{\ell_x}{L_x}-\frac{1}{2\pi k_x}\sin(2\pi k_x \frac{\ell_x}{L_x})-d^\text{rect},\\
G_B^\lambda&=\left(1-\frac{\ell_x}{L_x}\right)-\frac{1}{2\pi k_x}\sin(2\pi k_x \frac{\ell_x}{L_x})- d^\text{rect},
\end{align}
where the following approximation\footnote{Note that in \cite{Parker2017} there is a an extra factor of $2$  in equation (72) with respect to their calculation of $\gamma$ in appendix C.2 of the paper.}, valid for large $k_y$ and $L_x=L_y$, is given for $d^\text{rect}$
\begin{equation}
d^\text{rect}\approx\frac{4}{\pi^3}\frac{2\pi k_y \int_{0}^{2\pi k_y}\frac{\sin t}{t}\diff{}{t}-\pi k_y \int_{0}^{\pi k_y}\frac{\sin t}{t}\diff{}{t}-(-1)^{k_y}+1}{k_x^2+k_y^2}\sin(\pi k_x\frac{\ell_x}{L_x})^2.
\end{equation}
In figure \ref{fig:comparison-d-parker} we compare this result to our analytic results for $d^A_\lambda$ and $d^B_\lambda$ at $L_x=L_y$ from section \ref{appendix:G-rectangle}, see equations \eqref{eq:d-lambda-A-appendix} and \eqref{eq:d-lambda-B-appendix} in particular. We find the results in good agreement where the approximation for $d^\text{rect}$ is valid.

\section{Determinant of the Laplacian on a rectangle}
\label{appendix:det-rectangle}
In this section we calculate the zeta regularized determinant of the Laplacian on a rectangle. The determinant of the Laplacian can be written as
\begin{equation}
\det \Delta = e^{-\zeta_S'(0)},
\end{equation}
where $\zeta_S$ is the spectral $\zeta$-function corresponding to the Laplacian. On the rectangle, where the eigenvalues of the Laplacian are
\begin{equation}
\lambda_{k_x,k_y}=\left(\frac{\pi}{L_x}k_x\right)^2+\left(\frac{\pi}{L_y}k_y\right)^2,\quad k_x,k_y\in \mathbb{N}^+,
\end{equation}
the spectral $\zeta$-function takes the form
\begin{equation}
\zeta_S(s)=\sum_{k_x,k_y=1}^\infty\left(\left(\frac{\pi}{L_x}k_x\right)^2+\left(\frac{\pi}{L_y}k_y\right)^2\right)^{-s}.
\end{equation}
Our goal is thus to find an analytic continuation of this expression to $s=0$ and calculate the value of its derivative at that point. Let us first recall some identities. The integral representation of the $\Gamma$ function can be used to write
\begin{equation}
\ell^{-s}=\frac{1}{\Gamma(s)}\int_{0}^\infty \diff{}{t}\ t^{s-1} e^{-\ell t}.
\end{equation}
The Poisson summation formula is given by 
\begin{equation}
\sum_{n=1}^\infty e^{-a n^2}=-\frac{1}{2}+\frac{1}{2}\sqrt{\frac{\pi}{a}}+\sqrt{\frac{\pi}{a}}\sum_{n=1}^\infty e^{-\frac{n^2\pi^2}{a}}.
\end{equation}
The Riemann $\zeta$-function is defined as
\begin{equation}
\zeta(s)=\sum_{n=1}^\infty \frac{1}{n^s},
\end{equation}
and finally the integral representation of a modified Bessel function is 
\begin{equation}
K_\nu(z)=\frac{1}{2}\int_0^\infty\diff{}{u}\ e^{-\frac{z}{2}\left(u+\frac{1}{u}\right)} u^{\nu-1}.
\end{equation}
We can now use these identities to perform some manipulations on the spectral $\zeta$-function
\begin{align}
\zeta_S(s)&=\frac{1}{\Gamma(s)}\sum_{k_x,k_y=1}^\infty\int_{0}^\infty\diff{}{t}\ t^{s-1} e^{-t\left(\left(\frac{\pi}{L_x}k_x\right)^2+\left(\frac{\pi}{L_y}k_y\right)^2\right)}\nonumber\\
&=\frac{1}{\Gamma(s)}\sum_{k_y=1}^\infty\int_{0}^\infty\diff{}{t}\ t^{s-1} e^{-t\left(\frac{\pi}{L_y}k_y\right)^2} \sum_{k_x=1}^\infty e^{-t\left(\frac{\pi}{L_x}k_x\right)^2}\nonumber\\
&=\frac{1}{\Gamma(s)}\sum_{k_y=1}^\infty\int_{0}^\infty\diff{}{t}\ t^{s-1} e^{-t\left(\frac{\pi}{L_y}k_y\right)^2} \left(-\frac{1}{2}+\frac{L_x}{2\sqrt{\pi}}t^{-\frac{1}{2}}+\frac{L_x}{\sqrt{\pi}}t^{-\frac{1}{2}}\sum_{k_x=1}^\infty e^{-\frac{k_x^2 L_x^2}{t}}\right)\nonumber\\
&=-\frac{1}{2}\sum_{k_y=1}^\infty\left(\frac{\pi}{L_y}k_y\right)^{-2s}+\frac{L_x}{2\sqrt{\pi}}\frac{\Gamma(s-\frac{1}{2})}{\Gamma(s)}\sum_{k_y=1}^\infty\left(\frac{\pi}{L_y}k_y\right)^{-2s+1}+\nonumber\\
&\qquad\qquad\qquad\qquad\qquad\qquad \frac{2L_x^{s+\frac{1}{2}}}{\Gamma(s)\sqrt{\pi}}\sum_{k_x,k_y=1}^\infty\left(\frac{L_y k_x}{\pi k_y}\right)^{s-\frac{1}{2}}K_{s-\frac{1}{2}}\left(2\pi \frac{L_x}{L_y} k_x k_y\right)\nonumber\\
&=-\frac{1}{2}\left(\frac{L_y}{\pi}\right)^{2s}\zeta(2s)+\frac{L_x}{2\sqrt{\pi}}\frac{\Gamma(s-\frac{1}{2})}{\Gamma(s)}\left(\frac{L_y}{\pi}\right)^{2s-1}\zeta(2s-1)+\nonumber\\
&\qquad\qquad\qquad\qquad\qquad\qquad \frac{2L_x^{s+\frac{1}{2}}}{\Gamma(s)\sqrt{\pi}}\sum_{k_x,k_y=1}^\infty\left(\frac{L_y k_x}{\pi k_y}\right)^{s-\frac{1}{2}}K_{s-\frac{1}{2}}\left(2\pi \frac{L_x}{L_y} k_x k_y\right).
\end{align}
The derivative at $s=0$ of the first term is straight forward to calculate for the first term 
\begin{equation}
-\frac{d}{ds}\frac{1}{2}\left(\frac{L_y}{\pi}\right)^{2s}\zeta(2s)\vert_{s=0}=\frac{1}{2}\log(2L_y).
\end{equation}
In order to evaluate the derivatives at $s=0$ of the remaining two terms we need consider the following asymptotic behavior of the $\Gamma$- function. For $\varepsilon$ close to zero
\begin{gather}
\frac{1}{\Gamma(\varepsilon)}=\varepsilon+\ldots,\\
\frac{d}{ds}\frac{1}{\Gamma(s)}\vert_{s=\varepsilon}=1+\ldots.
\end{gather}
Since all other parts of the expressions are regular around $s=0$, this means that the only terms that survive are those where the the derivative hits the $\Gamma$-function. We can thus write for the second term in the spectral $\zeta$-function
\begin{equation}
\frac{d}{ds}\frac{L_x}{2\sqrt{\pi}}\frac{\Gamma(s-\frac{1}{2})}{\Gamma(s)}\left(\frac{L_y}{\pi}\right)^{2s-1}\zeta(2s-1)\Big\vert_{s=0} = \frac{\sqrt{\pi}}{2}\frac{L_x}{L_y}\Gamma(-\frac{1}{2})\zeta(-1)=\frac{\pi}{12}\frac{L_x}{L_y}
\end{equation}
and similarly for the third
\begin{multline}
\frac{d}{ds}\frac{2L_x^{s+\frac{1}{2}}}{\Gamma(s)\sqrt{\pi}}\sum_{k_x,k_y=1}^\infty\left(\frac{L_y k_x}{\pi k_y}\right)^{s-\frac{1}{2}}K_{s-\frac{1}{2}}\left(2\pi \frac{L_x}{L_y} k_x k_y\right)\Big\vert_{s=0}=\\
2\sqrt{\frac{L_x}{L_y}}\sum_{k_x,k_y=1}^\infty\sqrt{\frac{k_y}{k_x}}K_{-\frac{1}{2}}\left(2\pi \frac{L_x}{L_y} k_x k_y\right).
\end{multline}
Using the explicit form of $K_{-\frac{1}{2}}$ we can further simplify this to 
\begin{align}
2\sqrt{\frac{L_x}{L_y}}\sum_{k_x,k_y=1}^\infty\sqrt{\frac{k_y}{k_x}}K_{-\frac{1}{2}}\left(2\pi \frac{L_x}{L_y} k_x k_y\right)&=\sum_{k_x,k_y=1}^\infty \frac{1}{k_x}e^{-2\pi \frac{L_x}{L_y} k_x k_y}\nonumber\\
&=-\frac{\pi}{12}\frac{L_x}{L_y}-\log(\eta\left(i\frac{L_x}{L_y}\right)),
\end{align}
where $\eta$ is the Dedekind $\eta$-function. Putting everything together, we see that the derivative of the spectral $\zeta$-function at zero is
\begin{equation}
\zeta_S'(0)=\frac{1}{2}\log(2L_y)-\log(\eta\left(i\frac{L_x}{L_y}\right))
\end{equation}
and the determinant of the Laplacian on the rectangle is
\begin{equation}
\label{eq:det-rectangle-appendix}
\det \Delta=\frac{1}{\sqrt{2L_y}}\eta\left(i\frac{L_x}{L_y}\right).
\end{equation}
We note that due to the behavior of the Dedekind $\eta$-function under modular transformations the expression above is invariant under the exchange of $L_x$ and $L_y$.


\section{Correlation functions}
\label{sec:correlation-functions-appendix}
In this appendix we evaluate the correlation functions needed in the calculation of the entanglement entropy of the state with $m$ excitations in the $\lambda$ mode, see section \ref{sec:interesting-subcase}.
As a starting point, we can easily see that for integer $\beta$ and by linearity
\begin{align}
\label{eq:correl-one-field-appendix}
\left\langle(\varphi_X^\lambda)^{\beta}\right\rangle_X&=\delta_{\beta,\text{even}}(\text{\# of full Wick contractions})(G_X^\lambda)^\frac{\beta}{2}\nonumber\\
&=\delta_{\beta,\text{even}}(\beta-1)!!(G_X^\lambda)^\frac{\beta}{2},
\end{align}
where $X=A,B,M$ and $\delta_{\beta,\text{even}}:=\delta_{\beta,2n}$ for some integer $n$. Using the definition of Hermite polynomials \eqref{eq:Hermite-polynomials-def} and linearity we can rewrite the following correlation function as
\begin{align}
\frac{1}{2^m}&\sqrt{\frac{1}{m!}\binom{m}{k}\binom{m}{k'}}\left\langle H_{m-k}(\varphi_X^\lambda)H_{k'}(\varphi_X^\lambda)\right\rangle_X\nonumber\\
&=(m-k)!k'!\sqrt{\frac{1}{m!}\binom{m}{k}\binom{m}{k'}}\cross\nonumber\\
&\quad\sum_{\ell=0}^{\lfloor\frac{m-k}{2}\rfloor}\sum_{\ell'=0}^{\lfloor\frac{k'}{2}\rfloor}\frac{(-1)^{\ell+\ell'} 2^{k'-k-2(\ell+\ell')}}{\ell!\ell'!(m-k-2\ell)!(k'-2\ell')!}\left\langle(\varphi_X^\lambda)^{m-k+k'-2(\ell+\ell')}\right\rangle_X\nonumber\\
&=\delta_{m-k+k',\text{even}}(m-k)!k'!\sqrt{\frac{1}{m!}\binom{m}{k}\binom{m}{k'}}\cross\nonumber\\
&\quad\sum_{\ell=0}^{\lfloor\frac{m-k}{2}\rfloor}\sum_{\ell'=0}^{\lfloor\frac{k'}{2}\rfloor}\frac{(-1)^{\ell+\ell'} 2^{k'-k-2(\ell+\ell')} (m-k+k'-2(\ell+\ell')-1)!!}{\ell!\ell'!(m-k-2\ell)!(k'-2\ell')!}\left( G_X^\lambda\right)^\frac{m-k+k'-2(\ell+\ell')}{2}\nonumber\\
&=\delta_{m-k+k',\text{even}}\sum_{\ell=0}^{\lfloor\frac{m-k}{2}\rfloor}\sum_{\ell'=0}^{\lfloor\frac{k'}{2}\rfloor}f^m_{k,k'}(\ell,\ell')\left(G_X^\lambda\right)^\frac{m-k+k'-2(\ell+\ell')}{2},
\end{align}
where we included the prefactor that appears in front of the correlation functions that make up the tensors $\mathcal{A}$ and $\mathcal{B}$ in \eqref{eq:tensor-A-single-mode} and \eqref{eq:tensor-B-single-mode}, and in the last line we defined 
\begin{align}
\label{eq:coefficient-f-appendix}
f^m_{k,k';\ell,\ell'}&:=\frac{(-1)^{\ell+\ell'}}{2^{k-k'+2(\ell+\ell')}}\sqrt{\frac{1}{m!}\binom{m}{k}\binom{m}{k'}}\frac{ (m-k)!k'!(m-k+k'-2(\ell+\ell')-1)!!}{\ell!\ell'!(m-k-2\ell)!(k'-2\ell')!}.
\end{align}
If we define the following polynomial 
\begin{equation}
\label{eq:F-polynomial-definition-appendix}
F^m_{k,k'}(z):=\delta_{m-k+k',\text{even}}\sum_{\ell=0}^{\lfloor\frac{m-k}{2}\rfloor}\sum_{\ell'=0}^{\lfloor\frac{k'}{2}\rfloor}f^m_{k,k';\ell,\ell'}z^\frac{m-k+k'-2(\ell+\ell')}{2}
\end{equation}
then we can simply write
\begin{equation}
\frac{1}{2^m}\sqrt{\frac{1}{m!}\binom{m}{k}\binom{m}{k'}}\left\langle H_{m-k}(\varphi_X^\lambda)H_{k'}(\varphi_X^\lambda)\right\rangle_X=F^m_{k,k'}(G_X^\lambda).
\end{equation}
Next, we want to evaluate the correlation on $M$ appearing in the tensor $\mathcal{M}$ in equation \eqref{eq:tensor-M-single-mode}. Reminding ourselves that $\bar{\phi}^\lambda=\bar{\phi}^\lambda_A+\bar{\phi}_B^\lambda$ and using the identity of Hermite polynomials \eqref{eq:hermite-polynomial-identity-1} we can rewrite the correlation function as
\begin{align}
&\frac{1}{2^\frac{3m}{2}m!}\sqrt{\binom{m}{k}\binom{m}{k'}}\left\langle H_{m}\left(\frac{1}{\sqrt{2}}\bar{\phi}^{\lambda}\right) H_{k}\left(\bar{\phi}_{A}^{\lambda}\right)H_{m-k'}\left(\bar{\phi}_{B}^{\lambda}\right)\right\rangle_{M}\nonumber\\
&\quad=\frac{1}{2^{2m}m!}\sqrt{\binom{m}{k}\binom{m}{k'}}\sum_{n=0}^m \binom{m}{n}\left\langle H_{k}\left(\bar{\phi}_{A}^{\lambda}\right)H_{m-n}\left(\bar{\phi}_A^{\lambda}\right)H_{n}\left(\bar{\phi}_B^{\lambda}\right) H_{m-k'}\left(\bar{\phi}_{B}^{\lambda}\right)\right\rangle_{M}\nonumber\\
&\quad=\frac{1}{2^{2m}}\sqrt{\binom{m}{k}\binom{m}{k'}}\sum_{n=0}^m k!(m-k')!\cross\nonumber\\
&\qquad \sum_{r=0}^{\lfloor\frac{k}{2}\rfloor}\sum_{r'=0}^{\lfloor\frac{m-n}{2}\rfloor}\sum_{\ell=0}^{\lfloor\frac{n}{2}\rfloor}\sum_{\ell'=0}^{\lfloor\frac{m-k'}{2}\rfloor}\frac{(-1)^{r+r'+\ell+\ell'}2^{2m+k-k'-2(r+r'+\ell+\ell')}}{r!r'!\ell!\ell'!(k-2r)!(m-n-2r')!(n-2\ell)!(m-k'-2\ell')!}\cross\nonumber\\
&\qquad\qquad\qquad\qquad\qquad\qquad\qquad\qquad\quad \left\langle\left(\bar{\phi}_{A}^{\lambda}\right)^{m+k-n-2(r+r')}\left(\bar{\phi}_{B}^{\lambda}\right)^{m+n-k'-2(\ell+\ell')}\right\rangle_M
\end{align}
where we used the definition of Hermite polynomials in the last line. We will take care of the prefactors later, for now it is enough to see that the expression above is just a sum of correlation functions of the form $\langle A^\alpha B^\beta\rangle_M$, where we use the short-hand notation  $A\equiv \bar{\phi}^\lambda_A$ and $B\equiv\bar{\phi}_B^\lambda$. When we apply Wick's theorem to this type of correlation function, three distinct types of transformed two-point functions appear: $\langle A B\rangle_M=G^\lambda_{A,B}$, and $\langle X^2 \rangle_M=G^\lambda_{X,X}$ with $X=A,B$. Since we know from before what correlation functions of the form $\langle X^\alpha \rangle_M$ look like, we can evaluate $\langle A^\alpha B^\beta\rangle_M$ by successively taking mixed contractions creating an expansion of the form 
\begin{multline}
\left\langle A^\alpha B^\beta\right\rangle=a_0\left\langle A^\alpha\right\rangle\left\langle B^\beta\right\rangle+a_1\left\langle AB\right\rangle\left\langle A^{\alpha-1}\right\rangle\left\langle B^{\beta-1}\right\rangle+a_2\left\langle AB\right\rangle^2 \left\langle A^{\alpha-2}\right\rangle\left\langle B^{\beta-2}\right\rangle+\ldots,
\end{multline}
 where $a_\rho$ is a combinatorial factor counting the ways to select $\rho$ pairs $AB$ out of $A^\alpha B^\beta$ given by 
 \begin{equation}
 a_\rho=\frac{1}{\rho!}\frac{\alpha!}{(\alpha-\rho)!}\frac{\beta!}{(\beta-\rho)!}.
 \end{equation}
As a consistency check it isn't hard to see that this combinatorial factor leads to the expected amount of full Wick contractions: $(\alpha+\beta-1)!!$. Using equation \eqref{eq:correl-one-field-appendix} one can then evaluate the remaining correlation functions and write 
\begin{multline}
\left\langle A^\alpha B^\beta\right\rangle=
\sum_{\rho=0}^{\min(\alpha,\beta)}\delta_{\alpha-\rho,\text{even}}\delta_{\beta-\rho,\text{even}}\cross\\
\frac{1}{\rho!}\frac{\alpha!}{(\alpha-\rho)!!}\frac{\beta!}{(\beta-\rho)!!}\left(G^\lambda_{A,B}\right)^\rho \left( G_{A,A}^\lambda\right)^{(\alpha-\rho)/2}\left( G_{B,B}^\lambda\right)^{(\beta-\rho)/2},
\end{multline}
where we used the fact that $n!/(n-1)!!=n!!$. Let us thus define the following polynomial in three variables 
\begin{multline}
\label{eq:T-polynomial-def-appendix}
T^m_{k,k'}(x,y,z):=\delta_{k+k',\text{even}}\sqrt{\binom{m}{k}\binom{m}{k'}}\sum_{n=0}^m k!(m-k')!\cross \\
\sum_{r=0}^{\lfloor\frac{k}{2}\rfloor}\sum_{r'=0}^{\lfloor\frac{m-n}{2}\rfloor}\sum_{\ell=0}^{\lfloor\frac{n}{2}\rfloor}\sum_{\ell'=0}^{\lfloor\frac{m-k'}{2}\rfloor}\frac{(-1)^{r+r'+\ell+\ell'}2^{k-k'-2(r+r'+\ell+\ell')}}{r!r'!\ell!\ell'!(k-2r)!(m-n-2r')!(n-2\ell)!(m-k'-2\ell')!}\cross \\
\sum_{\rho=0}^{\min(m+k-n-2(r+r'),m+n-k'-2(\ell+\ell'))}\delta_{m+k-n-\rho,\text{even}}\frac{1}{\rho!}\frac{(m+k-n-2(r+r'))!}{(m+k-n-2(r+r')-\rho)!!}\cross \\
\frac{(m+n-k'-2(\ell+\ell'))!}{(m+n-k'-2(\ell+\ell')-\rho)!!}x^\rho y^{(m+k-n-2(r+r')-\rho)/2}z^{(m+n-k'-2(\ell+\ell')-\rho)/2}
\end{multline}
where we simplified the $\delta$-functions as  $\delta_{m+k-n-\rho,\text{even}}\delta_{m+n-k'-\rho,\text{even}}=\delta_{m+k-n-\rho,\text{even}}\delta_{k+k',\text{even}}$. 
In the end, we can write for the correlation function
\begin{multline}
\frac{1}{2^\frac{3m}{2}m!}\sqrt{\binom{m}{k}\binom{m}{k'}}\left\langle H_{m}\left(\frac{1}{\sqrt{2}}\bar{\phi}^{\lambda}\right) H_{k}\left(\bar{\phi}_{A}^{\lambda}\right)H_{m-k'}\left(\bar{\phi}_{B}^{\lambda}\right)\right\rangle_{M}\\
=T^m_{k,k'}(G^\lambda_{A,B},G^\lambda_{A,A},G^\lambda_{B,B}).
\end{multline}
Note that for a fixed surgery and eigenmode $G^\lambda_{A,B}$, $G^\lambda_{A,A}$, and $G^\lambda_{B,B}$ are just real numbers.


\bibliographystyle{utphys}
\bibliography{ES-EE-of-the-QLM}

\providecommand{\href}[2]{#2}\begingroup\raggedright\begin{thebibliography}{10}

\bibitem{Bennett1996}
C.~H. Bennett, D.~P. DiVincenzo, J.~A. Smolin, and W.~K. Wootters,
  ``Mixed-state entanglement and quantum error correction,''
  \href{http://dx.doi.org/10.1103/PhysRevA.54.3824}{{\em Phys. Rev. A}
  {\bfseries 54} (Nov, 1996) 3824--3851}.
  \url{https://link.aps.org/doi/10.1103/PhysRevA.54.3824}.

\bibitem{Plenio2007}
M.~B. Plenio and S.~Virmani, ``{An Introduction to entanglement measures},''
  {\em Quant. Inf. Comput.} {\bfseries 7} (2007) 1--51,
  \href{http://arxiv.org/abs/quant-ph/0504163}{{\ttfamily
  arXiv:quant-ph/0504163}}. \url{https://arxiv.org/abs/quant-ph/0504163}.

\bibitem{Calabrese2006}
P.~Calabrese and J.~L. Cardy, ``Entanglement entropy and quantum field theory:
  A non-technical introduction,''
  \href{http://dx.doi.org/10.1142/S021974990600192X}{{\em Int. J. Quant. Inf.}
  {\bfseries 4} (2006) 429},
  \href{http://arxiv.org/abs/quant-ph/0505193}{{\ttfamily
  arXiv:quant-ph/0505193 [quant-ph]}}.
  \url{https://arxiv.org/abs/quant-ph/0505193}.
Workshop on Quantum Entanglement in Physical and Information Sciences Pisa,
  Italy, December 14-18, 2004.

\bibitem{Amico2008}
L.~Amico, R.~Fazio, A.~Osterloh, and V.~Vedral, ``Entanglement in many-body
  systems,'' \href{http://dx.doi.org/10.1103/RevModPhys.80.517}{{\em Rev. Mod.
  Phys.} {\bfseries 80} (May, 2008) 517--576}.
  \url{https://link.aps.org/doi/10.1103/RevModPhys.80.517}.

\bibitem{Eisert2010}
J.~Eisert, M.~Cramer, and M.~B. Plenio, ``Area laws for the entanglement
  entropy - a review,'' \href{http://dx.doi.org/10.1103/RevModPhys.82.277}{{\em
  Rev. Mod. Phys.} {\bfseries 82} (2010) 277--306},
  \href{http://arxiv.org/abs/0808.3773}{{\ttfamily arXiv:0808.3773
  [quant-ph]}}.
\url{https://arxiv.org/abs/0808.3773}.

\bibitem{Laflorencie2016}
N.~Laflorencie, ``Quantum entanglement in condensed matter systems,''
  \href{http://dx.doi.org/https://doi.org/10.1016/j.physrep.2016.06.008}{{\em
  Physics Reports} {\bfseries 646} (2016) 1 -- 59}.
  \url{http://www.sciencedirect.com/science/article/pii/S0370157316301582}.
  Quantum entanglement in condensed matter systems.

\bibitem{Callan1994}
C.~Callan and F.~Wilczek, ``On geometric entropy,''
  \href{http://dx.doi.org/https://doi.org/10.1016/0370-2693(94)91007-3}{{\em
  Physics Letters B} {\bfseries 333} no.~1, (1994) 55 -- 61}.
  \url{http://www.sciencedirect.com/science/article/pii/0370269394910073}.

\bibitem{Holzhey1994}
C.~Holzhey, F.~Larsen, and F.~Wilczek, ``Geometric and renormalized entropy in
  conformal field theory,''
  \href{http://dx.doi.org/https://doi.org/10.1016/0550-3213(94)90402-2}{{\em
  Nuclear Physics B} {\bfseries 424} no.~3, (1994) 443 -- 467}.
  \url{http://www.sciencedirect.com/science/article/pii/0550321394904022}.

\bibitem{Calabrese2004}
P.~Calabrese and J.~Cardy, ``Entanglement entropy and quantum field theory,''
  \href{http://dx.doi.org/10.1088/1742-5468/2004/06/P06002}{{\em J. Stat.
  Mech.} (2004) }. \url{https://arxiv.org/abs/hep-th/0405152}.

\bibitem{Nishioka2009}
T.~Nishioka, S.~Ryu, and T.~Takayanagi, ``Holographic entanglement entropy: an
  overview,'' \href{http://dx.doi.org/10.1088/1751-8113/42/50/504008}{{\em
  Journal of Physics A: Mathematical and Theoretical} {\bfseries 42} no.~50,
  (Dec, 2009) 504008}.
  \url{https://doi.org/10.1088%2F1751-8113%2F42%2F50%2F504008}.

\bibitem{Rangamani2017}
M.~Rangamani and T.~Takayanagi, ``Holographic entanglement entropy,''
  \href{http://dx.doi.org/10.1007/978-3-319-52573-0}{{\em Lecture Notes in
  Physics} {\bfseries 931} (2017) }. \url{https://arxiv.org/abs/1609.01287}.

\bibitem{Srednicki1993}
M.~Srednicki, ``Entropy and area,''
  \href{http://dx.doi.org/10.1103/PhysRevLett.71.666}{{\em Phys. Rev. Lett.}
  {\bfseries 71} (Aug, 1993) 666--669}.
  \url{https://link.aps.org/doi/10.1103/PhysRevLett.71.666}.

\bibitem{Hastings2007}
M.~Hastings, ``{An area law for one-dimensional quantum systems},''
  \href{http://dx.doi.org/10.1088/1742-5468/2007/08/P08024}{{\em J. Stat.
  Mech.} {\bfseries 0708} (2007) P08024},
  \href{http://arxiv.org/abs/0705.2024}{{\ttfamily arXiv:0705.2024
  [quant-ph]}}.

\bibitem{Page1993}
D.~N. Page, ``Average entropy of a subsystem,''
  \href{http://dx.doi.org/10.1103/PhysRevLett.71.1291}{{\em Phys. Rev. Lett.}
  {\bfseries 71} (Aug, 1993) 1291--1294}.
  \url{https://link.aps.org/doi/10.1103/PhysRevLett.71.1291}.

\bibitem{Foong1994}
S.~K. Foong and S.~Kanno, ``Proof of page's conjecture on the average entropy
  of a subsystem,'' \href{http://dx.doi.org/10.1103/PhysRevLett.72.1148}{{\em
  Phys. Rev. Lett.} {\bfseries 72} (Feb, 1994) 1148--1151}.
  \url{https://link.aps.org/doi/10.1103/PhysRevLett.72.1148}.

\bibitem{Sen1996}
S.~Sen, ``Average entropy of a quantum subsystem,''
  \href{http://dx.doi.org/10.1103/PhysRevLett.77.1}{{\em Phys. Rev. Lett.}
  {\bfseries 77} (Jul, 1996) 1--3}.
  \url{https://link.aps.org/doi/10.1103/PhysRevLett.77.1}.

\bibitem{Ahmadi2006}
M.~Ahmadi, S.~Das, and S.~Shankaranarayanan, ``Is entanglement entropy
  proportional to area?,'' \href{http://dx.doi.org/10.1139/p06-002}{{\em
  Canadian Journal of Physics} {\bfseries 84} no.~6-7, (2006) 493--499},
  \href{http://arxiv.org/abs/https://doi.org/10.1139/p06-002}{{\ttfamily
  https://doi.org/10.1139/p06-002}}. \url{https://doi.org/10.1139/p06-002}.

\bibitem{Das2006}
S.~Das and S.~Shankaranarayanan, ``How robust is the entanglement entropy-area
  relation?,'' \href{http://dx.doi.org/10.1103/PhysRevD.73.121701}{{\em Phys.
  Rev. D} {\bfseries 73} (Jun, 2006) 121701}.
  \url{https://link.aps.org/doi/10.1103/PhysRevD.73.121701}.

\bibitem{Masanes2009}
L.~Masanes, ``Area law for the entropy of low-energy states,''
  \href{http://dx.doi.org/10.1103/PhysRevA.80.052104}{{\em Phys. Rev. A}
  {\bfseries 80} (Nov, 2009) 052104}.
  \url{https://link.aps.org/doi/10.1103/PhysRevA.80.052104}.

\bibitem{Alcaraz2011}
F.~C. Alcaraz, M.~I. Berganza, and G.~Sierra, ``Entanglement of low-energy
  excitations in conformal field theory,''
  \href{http://dx.doi.org/10.1103/PhysRevLett.106.201601}{{\em Phys. Rev.
  Lett.} {\bfseries 106} (May, 2011) 201601}.
  \url{https://link.aps.org/doi/10.1103/PhysRevLett.106.201601}.

\bibitem{Berganza2012}
M.~I. Berganza, F.~C. Alcaraz, and G.~Sierra, ``Entanglement of excited states
  in critical spin chains,''
  \href{http://dx.doi.org/10.1088/1742-5468/2012/01/p01016}{{\em Journal of
  Statistical Mechanics: Theory and Experiment} {\bfseries 2012} no.~01, (Jan,
  2012) P01016}.
  \url{https://doi.org/10.1088%2F1742-5468%2F2012%2F01%2Fp01016}.

\bibitem{CastroAlvaredo2018}
O.~A. Castro-Alvaredo, C.~De~Fazio, B.~Doyon, and I.~M. Sz\'ecs\'enyi,
  ``Entanglement content of quasiparticle excitations,''
  \href{http://dx.doi.org/10.1103/PhysRevLett.121.170602}{{\em Phys. Rev.
  Lett.} {\bfseries 121} (Oct, 2018) 170602}.
  \url{https://link.aps.org/doi/10.1103/PhysRevLett.121.170602}.

\bibitem{CastroAlvaredo2018a}
O.~A. Castro-Alvaredo, C.~De~Fazio, B.~Doyon, and I.~M. Sz{\'e}cs{\'e}nyi,
  ``Entanglement content of quantum particle excitations. part i. free field
  theory,'' \href{http://dx.doi.org/10.1007/JHEP10(2018)039}{{\em Journal of
  High Energy Physics} {\bfseries 2018} no.~10, (Oct, 2018) 39}.
  \url{https://doi.org/10.1007/JHEP10(2018)039}.

\bibitem{CastroAlvaredo2019}
O.~A. Castro-Alvaredo, C.~De~Fazio, B.~Doyon, and I.~M. Sz{\'e}cs{\'e}nyi,
  ``Entanglement content of quantum particle excitations. part ii. disconnected
  regions and logarithmic negativity,''
  \href{http://dx.doi.org/10.1007/JHEP11(2019)058}{{\em Journal of High Energy
  Physics} {\bfseries 2019} no.~11, (Nov, 2019) 58}.
  \url{https://doi.org/10.1007/JHEP11(2019)058}.

\bibitem{CastroAlvaredo2019a}
O.~A. Castro-Alvaredo, C.~De~Fazio, B.~Doyon, and I.~M. Sz{\'e}cs{\'e}nyi,
  ``Entanglement content of quantum particle excitations. part iii. graph
  partition functions,'' \href{http://dx.doi.org/10.1063/1.5098892}{{\em
  Journal of Mathematical Physics} {\bfseries 60} no.~8, (2019) 082301}.
  \url{https://doi.org/10.1063/1.5098892}.

\bibitem{Alcaraz2008}
F.~C. Alcaraz and M.~S. Sarandy, ``Finite-size corrections to entanglement in
  quantum critical systems,''
  \href{http://dx.doi.org/10.1103/PhysRevA.78.032319}{{\em Phys. Rev. A}
  {\bfseries 78} (Sep, 2008) 032319}.
  \url{https://link.aps.org/doi/10.1103/PhysRevA.78.032319}.

\bibitem{Alba2009}
V.~Alba, M.~Fagotti, and P.~Calabrese, ``Entanglement entropy of excited
  states,'' \href{http://dx.doi.org/10.1088/1742-5468/2009/10/p10020}{{\em
  Journal of Statistical Mechanics: Theory and Experiment} {\bfseries 2009}
  no.~10, (Oct, 2009) P10020}.
  \url{https://doi.org/10.1088%2F1742-5468%2F2009%2F10%2Fp10020}.

\bibitem{Moelter2014}
J.~{M\"olter}, T.~Barthel, U.~{Schollw\"ock}, and V.~Alba, ``Bound states and
  entanglement in the excited states of quantum spin chains,''
  \href{http://dx.doi.org/10.1088/1742-5468/2014/10/p10029}{{\em Journal of
  Statistical Mechanics: Theory and Experiment} {\bfseries 2014} no.~10, (Oct,
  2014) P10029}.
  \url{https://doi.org/10.1088%2F1742-5468%2F2014%2F10%2Fp10029}.

\bibitem{Ardonne2004}
E.~Ardonne, P.~Fendley, and E.~Fradkin, ``Topological order and conformal
  quantum critical points,''
  \href{http://dx.doi.org/10.1016/j.aop.2004.01.004}{{\em Annals of Physics}
  {\bfseries 310} no.~2, (Apr, 2004) 493–551}.
  \url{http://dx.doi.org/10.1016/j.aop.2004.01.004}.

\bibitem{Rokhsar1988}
D.~S. Rokhsar and S.~A. Kivelson, ``Superconductivity and the quantum hard-core
  dimer gas,'' \href{http://dx.doi.org/10.1103/PhysRevLett.61.2376}{{\em Phys.
  Rev. Lett.} {\bfseries 61} (Nov, 1988) 2376--2379}.
  \url{https://link.aps.org/doi/10.1103/PhysRevLett.61.2376}.

\bibitem{Henley1997}
C.~L. Henley, ``Relaxation time for a dimer covering with height
  representation,'' \href{http://dx.doi.org/10.1007/BF02765532}{{\em Journal of
  Statistical Physics} {\bfseries 89} no.~3, (Nov, 1997) 483--507}.
  \url{https://doi.org/10.1007/BF02765532}.

\bibitem{Moessner2001}
R.~Moessner, S.~L. Sondhi, and E.~Fradkin, ``Short-ranged resonating valence
  bond physics, quantum dimer models, and ising gauge theories,''
  \href{http://dx.doi.org/10.1103/PhysRevB.65.024504}{{\em Phys. Rev. B}
  {\bfseries 65} (Dec, 2001) 024504}.
  \url{https://link.aps.org/doi/10.1103/PhysRevB.65.024504}.

\bibitem{Castelnovo2005}
C.~Castelnovo, C.~Chamon, C.~Mudry, and P.~Pujol, ``From quantum mechanics to
  classical statistical physics: Generalized rokhsar–kivelson hamiltonians
  and the “stochastic matrix form” decomposition,''
  \href{http://dx.doi.org/https://doi.org/10.1016/j.aop.2005.01.006}{{\em
  Annals of Physics} {\bfseries 318} no.~2, (2005) 316 -- 344}.
  \url{http://www.sciencedirect.com/science/article/pii/S0003491605000096}.

\bibitem{Freedman2005}
M.~Freedman, C.~Nayak, and K.~Shtengel, ``Extended hubbard model with ring
  exchange: A route to a non-abelian topological phase,''
  \href{http://dx.doi.org/10.1103/PhysRevLett.94.066401}{{\em Phys. Rev. Lett.}
  {\bfseries 94} (Feb, 2005) 066401}.
  \url{https://link.aps.org/doi/10.1103/PhysRevLett.94.066401}.

\bibitem{Fendley2008}
P.~Fendley, ``Topological order from quantum loops and nets,''
  \href{http://dx.doi.org/https://doi.org/10.1016/j.aop.2008.04.011}{{\em
  Annals of Physics} {\bfseries 323} no.~12, (2008) 3113 -- 3136}.
  \url{http://www.sciencedirect.com/science/article/pii/S0003491608000614}.

\bibitem{Keraenen2017}
V.~Ker{\"a}nen, W.~Sybesma, P.~Szepietowski, and L.~Thorlacius, ``Correlation
  functions in theories with lifshitz scaling,''
  \href{http://dx.doi.org/10.1007/JHEP05(2017)033}{{\em Journal of High Energy
  Physics} {\bfseries 2017} no.~5, (May, 2017) 33}.
  \url{https://doi.org/10.1007/JHEP05(2017)033}.

\bibitem{Angel-Ramelli2019}
J.~Angel-Ramelli, V.~G.~M. Puletti, and L.~Thorlacius, ``Entanglement entropy
  in generalised quantum lifshitz models,''
  \href{http://dx.doi.org/10.1007/JHEP08(2019)072}{{\em Journal of High Energy
  Physics} {\bfseries 2019} no.~8, (Aug, 2019) 72}.
  \url{https://doi.org/10.1007/JHEP08(2019)072}.

\bibitem{Fradkin2006}
E.~Fradkin and J.~E. Moore, ``Entanglement entropy of 2d conformal quantum
  critical points: Hearing the shape of a quantum drum,''
  \href{http://dx.doi.org/10.1103/PhysRevLett.97.050404}{{\em Phys. Rev. Lett.}
  {\bfseries 97} (Aug, 2006) 050404}.
  \url{https://link.aps.org/doi/10.1103/PhysRevLett.97.050404}.

\bibitem{Hsu2009}
B.~Hsu, M.~Mulligan, E.~Fradkin, and E.-A. Kim, ``Universal entanglement
  entropy in two-dimensional conformal quantum critical points,''
  \href{http://dx.doi.org/10.1103/PhysRevB.79.115421}{{\em Phys. Rev. B}
  {\bfseries 79} (Mar, 2009) 115421}.
  \url{https://link.aps.org/doi/10.1103/PhysRevB.79.115421}.

\bibitem{Hsu2010}
B.~Hsu and E.~Fradkin, ``Universal behavior of entanglement in 2d quantum
  critical dimer models,''
  \href{http://dx.doi.org/10.1088/1742-5468/2010/09/p09004}{{\em Journal of
  Statistical Mechanics: Theory and Experiment} {\bfseries 2010} no.~09, (Sep,
  2010) P09004}.
  \url{https://doi.org/10.1088%2F1742-5468%2F2010%2F09%2Fp09004}.

\bibitem{Stephan2009}
J.-M. St\'ephan, S.~Furukawa, G.~Misguich, and V.~Pasquier, ``Shannon and
  entanglement entropies of one- and two-dimensional critical wave functions,''
  \href{http://dx.doi.org/10.1103/PhysRevB.80.184421}{{\em Phys. Rev. B}
  {\bfseries 80} (Nov, 2009) 184421}.
  \url{https://link.aps.org/doi/10.1103/PhysRevB.80.184421}.

\bibitem{Oshikawa2010}
M.~Oshikawa, ``Boundary conformal field theory and entanglement entropy in
  two-dimensional quantum lifshitz critical point,''
\href{http://arxiv.org/abs/1007.3739}{{\ttfamily arXiv:1007.3739
  [cond-mat.stat-mech]}}.

\bibitem{Zaletel2011}
M.~P. Zaletel, J.~H. Bardarson, and J.~E. Moore, ``Logarithmic terms in
  entanglement entropies of 2d quantum critical points and shannon entropies of
  spin chains,'' \href{http://dx.doi.org/10.1103/PhysRevLett.107.020402}{{\em
  Phys. Rev. Lett.} {\bfseries 107} (2011) 020402},
  \href{http://arxiv.org/abs/1103.5452}{{\ttfamily arXiv:1103.5452
  [cond-mat.str-el]}}.
\url{https://arxiv.org/abs/1103.5452}.

\bibitem{Zhou2016}
T.~Zhou, X.~Chen, T.~Faulkner, and E.~Fradkin, ``Entanglement entropy and
  mutual information of circular entangling surfaces in 2 + 1-dimensional
  quantum lifshitz model,''
  \href{http://dx.doi.org/0.1088/1742-5468/2016/09/093101}{{\em J. Stat. Mech.
  2016(9):093101, 2016} (2016) }. \url{https://arxiv.org/abs/1607.01771}.

\bibitem{AngelRamelli2020}
J.~Angel-Ramelli, C.~Berthiere, V.~G.~M. Puletti, and L.~Thorlacius,
  ``{Logarithmic Negativity in Quantum Lifshitz Theories},''
  \href{http://arxiv.org/abs/2002.05713}{{\ttfamily arXiv:2002.05713
  [hep-th]}}.

\bibitem{Zhou2016a}
T.~Zhou, ``Entanglement entropy of local operators in quantum lifshitz
  theory,'' \href{http://dx.doi.org/10.1088/1742-5468/2016/09/093106}{{\em
  Journal of Statistical Mechanics: Theory and Experiment} {\bfseries 2016}
  no.~9, (Sep, 2016) 093106}.
  \url{https://doi.org/10.1088%2F1742-5468%2F2016%2F09%2F093106}.

\bibitem{Parker2017}
D.~E. Parker, R.~Vasseur, and J.~E. Moore, ``Entanglement entropy in excited
  states of the quantum lifshitz model,''
  \href{http://dx.doi.org/10.1088/1751-8121/aa70b3}{{\em Journal of Physics A:
  Mathematical and Theoretical} {\bfseries 50} no.~25, (May, 2017) 254003}.
  \url{https://doi.org/10.1088%2F1751-8121%2Faa70b3}.

\bibitem{Bernien2017}
H.~Bernien, S.~Schwartz, A.~Keesling, H.~Levine, A.~Omran, H.~Pichler, S.~Choi,
  A.~S. Zibrov, M.~Endres, M.~Greiner, V.~Vuleti{\'{c}}, and M.~D. Lukin,
  ``Probing many-body dynamics on a 51-atom quantum simulator,''
  \href{http://dx.doi.org/10.1038/nature24622}{{\em Nature} {\bfseries 551}
  no.~7682, (Nov, 2017) 579--584}. \url{https://doi.org/10.1038/nature24622}.

\bibitem{Turner2018}
C.~J. Turner, A.~A. Michailidis, D.~A. Abanin, M.~Serbyn, and Z.~Papi{\'{c}},
  ``Weak ergodicity breaking from quantum many-body scars,''
  \href{http://dx.doi.org/10.1038/s41567-018-0137-5}{{\em Nature Physics}
  {\bfseries 14} no.~7, (Jul, 2018) 745--749}.
  \url{https://doi.org/10.1038/s41567-018-0137-5}.

\bibitem{Lan2017}
Z.~Lan and S.~Powell, ``Eigenstate thermalization hypothesis in quantum dimer
  models,'' \href{http://dx.doi.org/10.1103/PhysRevB.96.115140}{{\em Phys. Rev.
  B} {\bfseries 96} (Sep, 2017) 115140}.
  \url{https://link.aps.org/doi/10.1103/PhysRevB.96.115140}.

\bibitem{Wildeboer2020}
J.~Wildeboer, A.~Seidel, N.~S. Srivatsa, A.~E.~B. Nielsen, and O.~Erten,
  ``Topological quantum many-body scars in quantum dimer models on the kagome
  lattice,'' 2020.

\bibitem{Iadecola2020}
T.~Iadecola and M.~Schecter, ``Quantum many-body scar states with emergent
  kinetic constraints and finite-entanglement revivals,''
  \href{http://dx.doi.org/10.1103/PhysRevB.101.024306}{{\em Phys. Rev. B}
  {\bfseries 101} (Jan, 2020) 024306}.
  \url{https://link.aps.org/doi/10.1103/PhysRevB.101.024306}.

\bibitem{Moudgalya2018}
S.~Moudgalya, N.~Regnault, and B.~A. Bernevig, ``Entanglement of exact excited
  states of affleck-kennedy-lieb-tasaki models: Exact results, many-body scars,
  and violation of the strong eigenstate thermalization hypothesis,''
  \href{http://dx.doi.org/10.1103/PhysRevB.98.235156}{{\em Phys. Rev. B}
  {\bfseries 98} (Dec, 2018) 235156}.
  \url{https://link.aps.org/doi/10.1103/PhysRevB.98.235156}.

\bibitem{Moudgalya2018a}
S.~Moudgalya, S.~Rachel, B.~A. Bernevig, and N.~Regnault, ``Exact excited
  states of nonintegrable models,''
  \href{http://dx.doi.org/10.1103/PhysRevB.98.235155}{{\em Phys. Rev. B}
  {\bfseries 98} (Dec, 2018) 235155}.
  \url{https://link.aps.org/doi/10.1103/PhysRevB.98.235155}.

\bibitem{Calabrese2009}
P.~Calabrese and J.~Cardy, ``Entanglement entropy and conformal field theory,''
  \href{http://dx.doi.org/10.1088/1751-8113/42/50/504005}{{\em Journal of
  Physics A: Mathematical and Theoretical} {\bfseries 42} no.~50, (Dec, 2009)
  504005}. \url{https://doi.org/10.1088%2F1751-8113%2F42%2F50%2F504005}.

\bibitem{Ginsparg1988}
P.~H. Ginsparg, ``Applied conformal field theory,'' in {\em {Les Houches Summer
  School in Theoretical Physics: Fields, Strings, Critical Phenomena Les
  Houches, France, June 28-August 5, 1988}}, pp.~1--168.
\newblock 1988.
\newblock \href{http://arxiv.org/abs/hep-th/9108028}{{\ttfamily
  arXiv:hep-th/9108028 [hep-th]}}.
\newblock
\url{https://arxiv.org/abs/hep-th/9108028}.
\newblock

\bibitem{DiFrancesco1997}
P.~Di~Francesco, P.~Mathieu, and D.~Senechal,
  \href{http://dx.doi.org/10.1007/978-1-4612-2256-9}{{\em {Conformal Field
  Theory}}}.
\newblock Graduate Texts in Contemporary Physics. Springer-Verlag, New York,
  1997.
\newblock
\url{http://www-spires.fnal.gov/spires/find/books/www?cl=QC174.52.C66D5::1997}.
\newblock

\bibitem{Duplantier1988}
B.~Duplantier and F.~David, ``Exact partition functions and correlation
  functions of multiple hamiltonian walks on the manhattan lattice,''
  \href{http://dx.doi.org/10.1007/BF01028464}{{\em Journal of Statistical
  Physics} {\bfseries 51} no.~3, (May, 1988) 327--434}.
  \url{https://doi.org/10.1007/BF01028464}.

\end{thebibliography}\endgroup

\end{document}